# Dynamic Chemical Changes of $Ti_3C_2T_x$ MXenes in Halide Perovskite Precursor Environments

Kirill Sobolev[1,*], Nagaprasad Reddy Samala[2], Mariela J. Pavan[3], Sharon Hazan[3], Vladimir Ezersky[3], Shakked Schwartz[4], Omer Bartov[1], Adi Yakir[1], Michal Leskes[4], Tamar Goldzak Mizrachi[2,*], Yevgeny Rakita[1,3,*]

[1] Department of Materials Engineering, Ben Gurion University of the Negev, Beer-Sheva 8410501, Israel
[2] The Alexander Kofkin Faculty of Engineering, Bar Ilan University, Ramat Gan 5290002, Israel
[3] Ilse Katz Institute for Nanoscale Science and Technology, Ben-Gurion University of the Negev, Beer-Sheva 8410501, Israel
[4] Department of Molecular Chemistry and Materials Science, Weizmann Institute of Science, Rehovot 7610001, Israel

[*] corresponding authors: kirillso@bgu.ac.il, tamar.goldzak@biu.ac.il, rakita@bgu.ac.il

## Abstract

The integration of two-dimensional $Ti_3C_2T_x$ MXenes into Halide Perovskite (HaP) optoelectronics has shown immense potential for enhancing device performance, yet the fundamental chemical interactions within the precursor "ink" – a critical stage determining the final material state – remain largely unexplored. This study presents a comprehensive investigation into the chemical stability and surface evolution of $Ti_3C_2T_x$ MXenes when exposed to the individual and combined components of a model methylammonium lead iodide ($MAPbI_3$) solution. By combining detailed X-ray Photoelectron Spectroscopy (XPS) analysis with Density Functional Theory (DFT) calculations, we decouple the complex interactions between MXenes and the ink constituents. We demonstrate that elemental iodine ($I_2$) and polyiodides ($I_n^-$) act as aggressive oxidizers, triggering rapid degradation and dissolution of the MXene flakes. Conversely, lead iodide ($PbI_2$) induces a slower oxidation process driven by the strong adsorption of $Pb^{2+}$ ions onto oxygen-terminated sites. In stark contrast, methylammonium iodide (MAI) exhibits a stabilizing effect: it adsorbs onto the MXene surface via hydrogen bonding, effectively passivating the flakes and retarding degradation. In addition, our results reveal that the sequence of precursor addition governs the MXene surface state, requiring extra caution, but also enabling further tunability. These findings provide a critical view on how MXene integration with materials, in general, and HaP in particular, should be perceived, providing a fundamental for their rational engineering.

## 1. Introduction

MXenes are a family of two-dimensional (2D) materials sharing the common formula $M_{n+1}X_nT_x$, where M stands for a transition metal (Ti, V, Nb, Mo, etc.), X – for C or/and N, and $T_x$ is a surface termination group, with –F, =O and –OH being the most common ones[1,2]. After their introduction in 2011 by Gogotsi, Naguib et al.[3], MXenes attracted vast attention from a very broad scientific community. For example, they showed significant improvements in electrochemical cells for energy storage[4–6], demonstrated environmental remediation[7,8], and the ability to be integrated into flexible and printable electronics[9,10], sensors[11,12], biomedicine[13,14], and more.

The origin behind what renders MXenes with such wide applicability is, at their core, their chemical versatility – particularly the M metal and the chemically-active and dynamically-changing surface chemistry of the $T_x$ group[1,2]. Even without changing the M metal backbone, changes in the $T_x$ group show, for example, the ability to change MXene's electrical conductivity or chemical potential[15], stabilize colloidal solutions via surface modification[16], or make them hydrophilic or hydrophobic on demand[17]. All of these lead to a pronounced integrability of MXenes with different classes of materials, including polymers, nanomaterials, organic molecules[18–24], and so on.

A more recent rising interest within the scientific and technological communities, which will form the focus of this work, is the fusion between MXenes and Halide Perovskite (HaP) systems[25]. HaPs are mostly famous due to their outstanding photovoltaic performance as light-harvesting materials, demonstrating results that show a potential alternative or an enhancement to Si-based solar cells[26,27], with a light-to-electron conversion efficiency of over 26% alone or 30% in tandem with Si[28,29]. Besides photovoltaics, HaPs show success in other technological fields, such as photo-sensors, hard-radiation detectors, light-emitting diodes, and more[30].

HaPs are $ABX_3$ structures, where A is a monovalent organic or inorganic cation like methylammonium ($MA^+$), formamidinium ($FA^+$) or $Cs^+$, B is a divalent cation such as $Pb^{2+}$ or $Sn^{2+}$, and X is a monovalent anion, such as $I^-$ or $Br^-$[31]. Their technological superiority is attributed to a number of coexisting properties, such as long exciton recombination lifetime[32,33], formation of electronically non-active defects ("defect-tolerance")[34], and/or the capability for spontaneous self-healing[35,36]. These features are, to the great extent, attributed to high structural and ionic dynamics[37], or to bulk point defects that are rather benign (in contrast to classical semiconductors, such as Si, GaAs)[34]. Unlike bulk defects, surface defects do play a major role and are used in tuning HaPs' properties[38]. Therefore, two-dimensional defects, especially those with chemical versatility, such as MXenes, can play a crucial role in actively controlling the functional properties of HaP-based systems.

Both MXenes and HaPs are prone to changes in response to variations in their chemical environment. HaPs require only solution chemistry to form high-quality films[39,40], such as spin coating of an ink-like solution, which offers an advantage in simplicity and fabrication costs. Low temperature rapid structural arrangement, however, also presents several drawbacks associated with the high chemical sensitivity of the materials during both fabrication and operation[41,42]. Unencapsulated devices tend to rapidly degrade due to ambient oxygen and humidity[43], and when encapsulated, hysteresis indicates severe ion migration and internal chemical dissociation, which may lead to contact degradation as a result of interaction with chemically active species, such as polyhalides[44]. The ability to manipulate HaPs' chemical environment, especially surface/interface, will make HaPs more versatile and chemically robust.

This is exactly the motivation that ignited the fusion between HaPs and MXenes. Since the pioneering work by Guo Z. et al.[45], increasing number of studies have explored the integration of MXenes into HaP-based solar-cell architectures, demonstrating improvements in the overall device performance. MXenes were integrated as either passivation layers, when the flakes are located at the interface between the HaP and charge transport layers (CTLs), or reinforcements, when MXenes are directly mixed with HaP, CTLs, or both[46–52]. Since MXenes have highly tunable surface chemistry[53,54], while the surface chemistry of HaPs is known to strongly affect their work function[38], their integration allows tuning the band alignment with the CTLs, thus directly impacting the power conversion efficiency of the respective devices[55–58]. Other reports claim that MXenes help reduce trap state densities, improve charge carrier mobilities, or improve morphological features in the HaP layers[59,60]. For example, MXenes were found to interact with undercoordinated $Pb^{2+}$ ions in Pb-based HaPs[61,62] and, supposedly, act as defect-passivation agents. MXenes were shown to impact the morphology of the HaP films via promoting nucleation, leading to films with larger grains[63–65]. Since surfaces are the source of functionally-meaningful defects[38], larger grains will directly reduce the density of such defects. Overall, MXenes act as promising additives for enhancing the efficiency and shelf life of HaP-based devices[66,67].

However, unlike the functional aspects of the MXene–HaP integration, the chemical aspects arising from their interaction have received very little attention. Given that both HaPs and MXenes are strongly influenced by their chemical environment, a fundamental understanding of the MXene-HaP interaction mechanisms and binding chemistry is necessary, and will likely explain the observed, yet poorly-understood, functionality changes. HaPs' formation, degradation, and self-healing are closely related to their subtle chemical nature[41], so it is likely to assume there will be an active interplay between MXenes and HaPs. The chemistry occurring between MXenes and HaPs may impose positively or negatively on the perovskite system, more specifically, impact on ion-migration related hysteresis, degradation of contacts[44], impose electronically-active (surface) defects[38], and/or form intermediate structural states that promote nucleation during film formation.

While there are assumptions on why MXenes may affect HaPs, until now, there is actually no knowledge about the state in which MXenes are integrated with HaPs, in particular, their termination composition and oxidation state of the M metal. The latter is a critical measure for degradation and loss of structural integrity of MXene flakes, which typically occurs through increasing the oxidation state of the M metal[68]. It is well-known that the MXenes' termination groups, $T_x$, can be exchanged or modified when exposed to different chemical environments[69,70]. Therefore, it is reasonable to assume that when MXenes are exposed to a HaP-precursor solution, it will alter the MXenes from their initial as-cast form. With the chemical changes of the MXenes, the MXene-HaP system's behavior must, consequently, also change. At these interphases, local changes in the chemical potential may lead to doping and/or trapping defect states, trap impurities such as oxygen, or limit ion migration. Therefore, understanding the state of MXenes after their exposure to HaP-related precursors is among the first things we must ask ourselves to establish a predictive understanding of how MXene integration influences the properties and performance of HaP-MXene systems.

HaP-precursor solution (or as will be denoted later - a "HaP-ink") is where the interaction with MXenes is most likely to be immediately important. HaP film preparation techniques[71,72] include a broad set of precursors[31] and solvents[73–75], which create a variety of different potential chemical environments for MXenes before an actual crystallization of the HaP in a film. Solvated HaP components may contain corrosive and aggressive species, especially halides and polyhalides, like $I_n^-$ [76,77]. While MXenes are highly active and rather prone to oxidation and other chemical reactions[68,78,79], the impact of these surrounding chemical species on them is unknown and must be carefully investigated. Thus, one may expect a large scope of possible processes to happen to MXenes in a HaP-ink: from standard oxidation and decomposition to functional group exchange or complex electrostatic and chemical adsorption. Similar processes have already been observed when MXenes were exposed to different aqueous and organic solutions[68,78–80].

**This paper focuses on** providing a clear picture of **the chemistry and composition of MXenes** just before they are integrated into a HaP solid matrix, meaning **within the precursor solution.** We **explicitly** focus on understanding the chemistry of MXenes in HaP-inks, which we believe to be a critical gap that hinders further development of MXene-HaP integration technology. In particular, this study will cover the most abundantly studied $Ti_3C_2T_x$ MXene and a $MAPbI_3$[81] HaP-ink as a model system. While this study does not cover all the possible chemical specificities of different HaP-MXene systems, it provides a pioneering and essential step towards a more complete understanding of MXene-HaP composite films.

In particular, well-characterized $Ti_3C_2T_x$ MXene flakes were exposed to complete or partial $MAPbI_3$-inks solutions, where a combinatorial set of MAI, $PbI_2$, and $I_2$ precursors was dissolved in DMSO. Supported by various probing methods, we use X-ray Photoelectron Spectroscopy (XPS) as the main tool to examine what happens to the state of MXenes as they are extracted from the HaP-inks. Together with a detailed experimental investigation, we include a critical first-principles Density Functional Theory (DFT) study to support and explain our experimental findings. Our results show that the chemical nature of the termination groups and MXenes' stability are time-dependent in some solutions, while more stable in others. We show that different species composing the HaP-precursor solution, in particular MAI, $PbI_2$ and $I_n^-$, have each a unique role in the overall stability of MXenes in the MXene-HaP suspension. Using the data gathered from the combinatorial set of precursors, we will demonstrate experimentally, and support theoretically: the possitive protective effect of MAI; the mild but significant degrative effect of $PbI_2$, showing Pb cations adsorption alongside slow degradation of the MXenes; the fast and detrimental effect of $I_2$, or polyiodides, on on the stability of MXenes. We note that while MAI and $PbI_2$ are abundantly present in $MAPbI_3$ as its binaries, $I_2$ is typically not added to HaP-inks intentionally, but it is known as a product that may spontaneously form inside halide solutions[76,77] or films[82–84], thus understanding its role around MXenes is of non-negligible importance.

We note that the current study focuses solely on the MXene side, which was previously neglected, and aims to provide a fundamental understanding for explaining and engineering functional behavior in working HaP–MXene–based devices. Our deduced fundamental insights broaden the core understanding of MXene chemistry in a hybrid organic-inorganic environment. While not directly targeting the incorporation of MXenes with HaPs for improving them on a device level, this study poses a critical milestone for a greater, more general understanding of MXenes and optoelectronic composites made with them.

## 2. Materials and Methods

Given our use case-study focuses on $Ti_3C_2T_x$ MXenes and $MAPbI_3$ HaP, we started with a careful in-house preparation of $Ti_3C_2T_x$ MXenes using well-known practices[85,86], followed by their detailed characterization. Next, we integrated the freshly prepared MXenes with a combinatorial set of $MAPbI_3$-related precursors, dissolved in DMSO. We note that we majorly focused on DMSO in our use-case, but we later demonstrated the use of other solvents, such as DMF. To track the state of the immersed MXenes, we then analyzed them after one hour and one week of immersion. We were focusing on the changes in MXenes' chemistry after their exposure to the inks, and targeted the factors that led to these changes. While the major part of this work focused on suspended MXenes that were extracted via filtering, in some cases, where we found it necessary, we also characterized precipitants that were observed at the bottom of the mixed inks, and the inks themselves. In this paper we frequently use the term "ink" to label HaP-precursor solutions with or without MXenes, although we would like to underscore that we do not target ink formulation for industrial purposes, although it may be used for that regardless. A 'block-scheme', summarizing the conducted research, is given in **Figure 1Figure *1***; the following parts of the manuscript are organized in accordance with this skeleton.

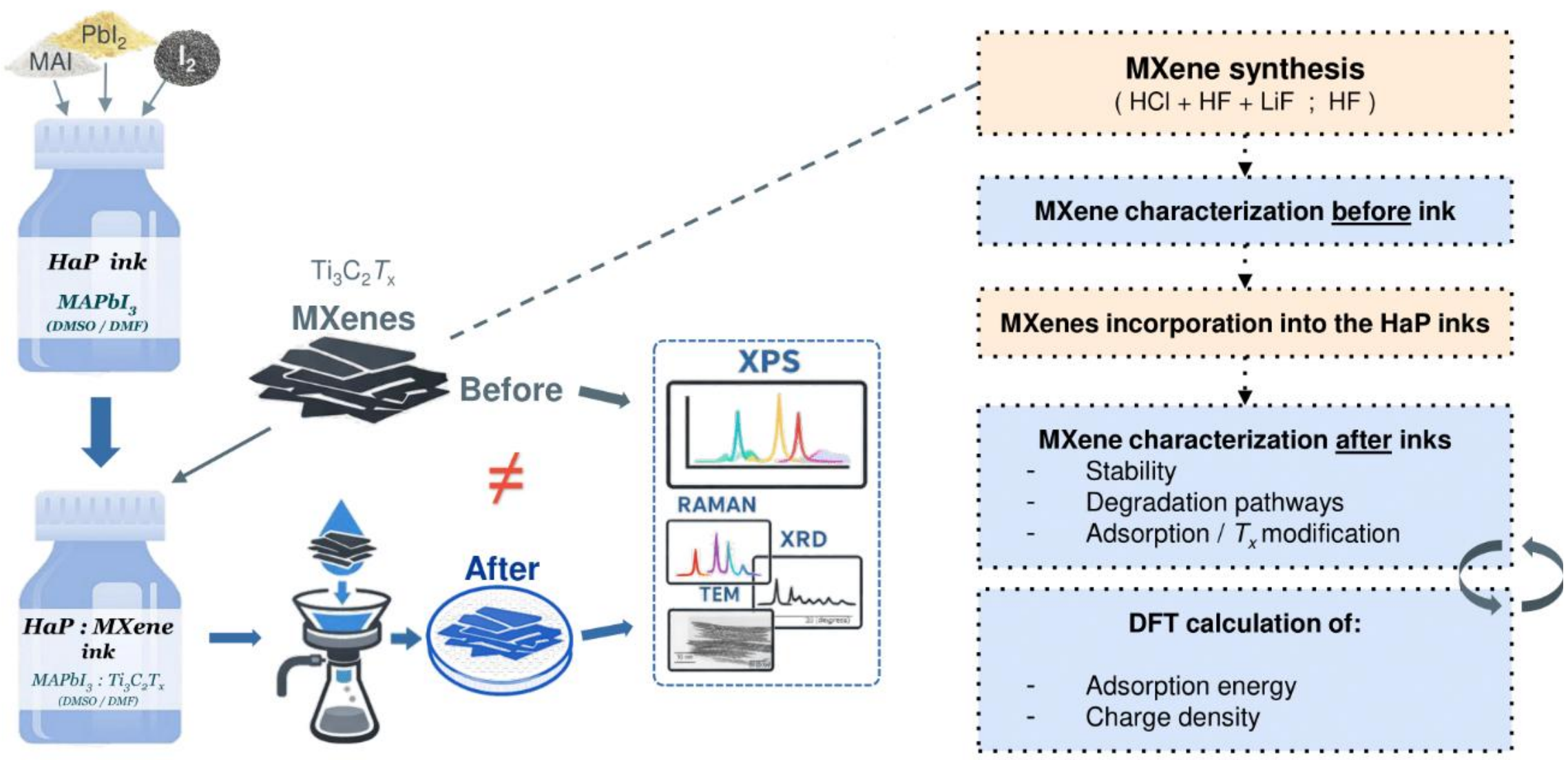

**Figure 1**. Illustrative block-schema of the conducted study: HaP-ink and $Ti_3C_2T_x$ MXene preparation, MXene+HaP ink formation, handling and characterization.

**Table 1.** Denotation and composition of $Ti_3C_2T_x$ MXene-based HaP-ink samples used in this work. All noted additives are of 1 M (mol/L); MXene concentration was targeted to be $2.5 \cdot 10^{-3}$ M (500 μg mL$^{-1}$); the actual volume of each solvent (DMSO or DMF) was 2 mL.

| Label | MAI [1.0 M] | $PbI_2$ [1.0 M] | $I_2$ [1.0 M] |
|---|---|---|---|
| 1-sol | - | - | - |
| 2-M | X | - | - |
| 3-MP | X | X | - |
| 4-MPI | X | X | X |
| 5-I | - | - | X |
| 6-PI | - | X | X |
| 7-P | - | X | - |
| 8-MI | X | - | X |

### 2.1. *$Ti_3C_2T_x$ MXenes synthesis*

For the synthesis of high quality $Ti_3C_2T_x$ MXenes we etched a $Ti_3AlC_2$ MAX phase precursor with a mixture of HCl and HF acids, followed by delamination with LiF[85]. The supporting batch of $Ti_3C_2T_x$ MXenes was synthesized using common HF-based etching protocol[86].

The precursor $Ti_3AlC_2$ MAX phase was prepared through a high temperature synthesis method, according to ref. [87]. The MAX phase was ground and passed through a sieving mesh of 45 $\mu m$. DDW, HCl and HF were intermixed in a volume ratio of 3 : 6 : 1, reaching the final volume of 20 mL. 1 g of the meshed $Ti_3AlC_2$ MAX powder was slowly immersed into the acidic solution; the solution temperature was brought up to 35°C, and the etching reaction was sustained for 24 hours under permanent magnetic stirring at 550 rpm.

After the completion of the reaction, pH of the solution was neutralized (reaching pH > 5) to separate the stacked MXene sheets and to remove unreacted slurry using repeated centrifugation. After each centrifugation cycle, the clear supernatant was discarded into waste, and the black precipitate was dispersed in fresh DDW. In total, we performed 12 centrifugation cycles at 3500 rpm for 5 minutes each. After the final centrifugation step, the obtained product was again re-dispersed in 50 mL of fresh

DDW, and then admixed with 1 g of LiF to delaminate MXene sheets. The solution was left to stir at 40°C for another 24 hours. One more centrifugation cycle at 5000 rpm for 5 minutes was performed to separate MXenes from LiF leftovers, after which the final centrifugation at 3500 rpm for 15 min segregated single-flake suspension from non-delaminated particles. After that, the black supernatant was frozen with the liquid nitrogen and freeze-dried (lyophilized) for 3 days to eliminate water and preserve re-dispersibility. The obtained spongy powder was stored in a vacuumized desiccator.

We investigated MXenes with two different combinations of initial functional groups (provided from two different synthesis routes) and followed their modification after being immersed into the HaP-inks. As a second synthesis route, we chose to use the classical HF etching reaction[86], which is described in more detail in Section SA in the SI. Overall, MXenes prepared using this method are denoted as HF-MXenes (or HF-$Ti_3C_2T_x$), and used as a supplementary batch of material. The reason we consider two different MXene preparation methods is to at least slightly control the $T_x$ termination group identity, where HCl+HF+LiF and HF etching methods are expected to result in a greater portion of either =O, or –F functional groups, respectively. We note that we know both approaches end up with a mixed =O, –F, and –OH termination, and what we attempted is to modify the ratio between these species. The quality of the prepared MXenes was validated prior to the following experiments (see **Figure 2** in the Results section and **Figure S1** in the SI). We note that HCl+HF+LiF-etched MXenes are found to be of higher quality, i.e., almost non-oxidized and naturally single-flake.

### *2.2. MXene-based HaP-inks preparation*

To study the interaction chemistry between $Ti_3C_2T_x$ MXenes and HaP precursor species, we prepared solutions with different precursor combinations of the widely used $MAPbI_3$ perovskite[81], meaning MAI, $PbI_2$ and $I_2$. The overall description of the studied set of samples is given in **Table 1**, the block-scheme of their preparation is given in **Figure 1**. In more detail, following **Table 1**, the inks with different combinations of components were prepared by adding only a sub-set of $MAPbI_3$ precursors into the solvent, followed by stirring and sonication until complete dissolution. Then, $Ti_3C_2T_x$ MXenes were added into the solutions, followed by a 30-minute sonication in an ice bath under argon bubbling to achieve homogeneous dispersions. The inks were stored in the sealed vials in a desiccator with a controlled-atmosphere (with desiccant and vacuumized).

To study the occurring chemistry, we analyzed the suspended MXenes in the solution, the separated dispersed MXenes, the residual solutions, and, in some cases, undesired precipitants. In order to test the chemistry across different timescales and track the evolution of the system, we examined the MXenes either after a short exposure (~1 hour) or a long (~1 week) storage. The 1-hour exposure is relevant to represent the use-case when MXenes are being dispersed in a freshly prepared HaP-ink, while 1 week scale is relevant to simulate the long-term storage conditions. For a solvent, we primarily used DMSO, but also had a comparison with DMF – both solvents are commonly employed in the HaP preparation processes separately or mixed[73,74]. For simplification, when noting ‘solvent’, we refer to DMSO; when DMF is used we will explicitly mention so within the text.

To extract dispersed $Ti_3C_2T_x$ MXenes, we passed the content of the reaction vial through a 0.4 µm pore-size PVDF filter membrane using vacuum-assisted filtration in a Buchner flask. When testing DMF as a solvent, we used a polyethylene filter instead (as DMF dissolves PVDF). The filtered solid was dried, and the residual solution of the ink was collected and analyzed separately. In cases when we observed precipitated solids forming in our reaction vials, they were collected and further analyzed. Any long-term storage was done in a vacuumized desiccator.

### *2.3. Characterization methods*

To study the type of chemical bonds in as-cast $Ti_3C_2T_x$ MXenes and MXenes reacted with HaP-precursor solutions, we used XPS. We employed ESCALAB photoelectron / Auger electron spectroscope (Thermo Fisher Scientific, Massachusetts, USA) in the XPS mode, with Al Kα X-ray source and 500 µm spot size. The analysis of the MXenes that were extracted from the precursor inks was performed directly on PVDF membranes, due to the limited amount of these materials. The substrate was not probed since the spot size was smaller than the MXenes-covered area. All the scans were obtained in a CAE analyzer mode with a pass energy of 20 eV and a step size of 0.1 eV. To gain a sufficient signal-to-noise ratio, the total number of scan repetitions was between 5 to 15, depending on the obtained intensity. For an extra set of experiments, we used the same facility in the UPS mode, with a HeII photon source, -5 V sample biasing, and 3 repetitive scans for each measurement.

To analyze the structural nature of as-cast MXenes, reaction products (filtered MXenes from HaP-inks) and precipitants, we used Panalytical Empyrean X-ray diffractometer (Malvern Panalytical, Malvern, UK) with Cu Kα radiation, operating at 45 V and 40 mA. All the scans were taken in a Bragg-Brentano mode with a 2θ range of $3-70°$, with a step size of 0.026°, and a counting time of 60 sec per step. The raw MXenes were measured as free powders on a zero-background substrate. The reacted MXenes were measured on a filtration membrane and background-corrected with a blank filter paper.

To analyze the chemical nature of dry MXenes (in a solid form on a glass substrate) and selected inks (in a liquid form inside a sealed quartz cuvette), we used Raman spectroscopy measurements. Spectra were collected with a confocal Horiba LabRam HR Evolution micro-Raman system equipped with a Syncerity Open Electrode CCD detector (deep-cooled to 60°C, 1024 × 256 pixels) and Synapse FIVIS EM-CCD. The light source was a 532 nm laser at 0.36 and 1.15 mW for liquid and solid samples, respectively. The laser was focused onto the surface of the solid or liquid with a 50x LWD objective (Olympus LMPlanFL-N, NA = 0.5), the measurements were taken using a 600 grooves/mm grating, and 100-120 µm confocal hole; exposure time varied between 0.5 - 60 seconds at the range of 56-3500 $cm^{-1}$. Several spots on each sample were recorded for the sake of gaining reliable statistics. LabSpec software (version 6.5.2.11) was used to operate the equipment and process the data.

We employed Transmission Electron Microscopy (TEM), using FEI Tecnai T12 G2 TWIN (FEI Company, Oregon, USA) and JEOL JEM-2100F (Jeol Ltd., Tokyo, Japan) TEMs, operating at 120 kV and 200 kV accelerating voltages, respectively, to visualize the morphology of as-cast $Ti_3C_2T_x$ MXenes, and MXenes reacted with selected HaP precursor inks, as well as of the precipitants from selected inks. We also performed SAED and EDS analysis of the respective products, using JEOL 50 $mm^2$ Si(Li) EDS detector equipped on JEM-2100F TEM. For as-cast MXene material analysis, low-concentration MXene-DDW suspensions (diluted until a pronounced light-green color) were drop-cast onto lacey carbon TEM grids that were pre-cleaned using plasma.

We used Cary 5000 UV-Vis-NIR Spectrophotometer (Agilent Technologies, California, USA) to study optical transmission spectra of MXene-containing and reference inks, to track their evolution upon aging. All the UV-Vis scans were measured in the wavelength range from 500 to 920 nm, with the step size of 1 nm and a counting time of 0.3 sec per step, all of them were background-corrected to a vial with pure solvent.

For a number of supplementary studies, we also used methods such as AFM, Confocal Microscopy, FTIR, and NMR. The detailed description of the conducted tests is given in the **Supporting Methods** section in the SI.

### *2.4 DFT calculations*

To support our experimental observations, we investigated a monolayer $Ti_3C_2T_x$ MXene with various surface terminations using DFT calculations. All the computations were performed with a plane-wave (PW) basis set as implemented in the Quantum ESPRESSO v7.2 software package[88]. The ion cores were modeled using ultrasoft pseudopotentials from the GBRV database[89]. The generalized gradient approximation of the Perdew–Burke–Ernzerhof exchange correlation functional was used in all calculations[90]. A kinetic energy cutoff of 60 Ry was employed for the wavefunctions, and 400 Ry – for the charge density, across all simulations. The Methfessel–Paxton smearing method was employed with a smearing width of 0.01 Ry, using the local-TF mixing scheme. Both atomic positions and lattice parameters were relaxed to preserve periodicity, with convergence criteria set to the default thresholds $10^{-4}$ a.u. for the energies and $10^{-4}$ a.u. for the forces in the geometry optimization. Target $Ti_3C_2T_x$ MXene structures were modeled as two-dimensional monolayers by isolating a single MXene layer and introducing a vacuum region ~20 Å along the z-direction to eliminate interlayer interactions. A two-dimensional Coulomb cutoff approach was applied to properly treat the charged species adsorption on slabs, which suppresses spurious periodic image interactions along the vacuum direction.

We considered 4*4*1 supercell with a 3*3*1 k-point sampling for all the studies. In general, surface terminations bind to the exposed titanium atoms on the outer surfaces of the $Ti_3C_2T_x$ monolayer. Depending on their chemical nature and size, these terminations preferentially adsorb at either on-top sites of surface Ti atoms or at hollow sites above the underlying atoms. We investigated a range of surface terminations, including =O, –F, –OH, and their mixtures. The results indicate that all these terminations exhibit a strong preference for hollow-site adsorption, which plays a critical role in stabilizing the monolayer structure and modulating the surface reactivity, and nicely correlates with the available literature[91]. Adsorption energies are determined by comparing the total energy of the surface with the adsorbed species to the sum of the energies of the clean surface and the isolated adsorbate.

## Results and Discussions

### *3.1 Preliminary validation of prepared $Ti_3C_2T_x$ MXenes*

Before moving forward to the MXene–HaP interaction experiments, we validate the quality and oxidation state of the as-cast $Ti_3C_2T_x$ MXenes, before they were exposed to the HaP-precursor solutions. In **Figure *2*** we summarize the characterization results of the produced MXenes. In **Figure *2*a** and **Figure *2*b** we validated the formation of the desired $Ti_3C_2T_x$ MXenes via XRD and Raman (see the detailed discussion in Section SB in the SI). Characteristic TEM image and the respective SAED pattern in **Figure *2*c** show high-quality single-layer flakes of $Ti_3C_2T_x$ MXenes with a typical hexagonal symmetry and lateral sizes of a few micrometers, which is rather large for MXenes and happens due to the optimal etching protocol[92]. AFM profiling revealed the thickness of ~2 nm, confirming the 2D flakes are mono-layered[93]. We also note that TEM does not reveal any undesired oxidation or damage. Overall, the conducted characterization confirms the formation of the high-quality $Ti_3C_2T_x$ product. Detailed analysis about HF-$Ti_3C_2T_x$ MXenes characterization is given in Section SC in the SI, and in **Figure S1**.

To further confirm the $Ti_3C_2T_x$ MXenes quality, and to form a baseline for comparing pristine MXenes with MXenes that are exposed to the HaP-precursor solutions, we probe their surface chemistry using XPS analysis, as shown in **Figure *2*d.** We note that while there are several notations for peak assignment in the MXene community for $Ti_3C_2T_x$[94–97], in this work we follow the comprehensive MXene XPS study by Natu, Barsoum, et al.[98].

**Ti2p** XPS spectrum can be deconvoluted into five doublets, positioned at Binding Energies (BE) of 455.4, 456.3, 457.1, 457.9 and 459.4 eV, where the first four correspond to $Ti^{2+}$ and $Ti^{3+}$ ions bonded to the carbon core and coordinated with =O/–OH and –F functional groups (C–Ti–*T*), and the fifth one – to some minority of $TiO_{2-x}F_{2x}$ oxofluoride contamination. **C1s** spectrum was deconvoluted into 3 peaks at BE of 282, 283, and 284.7 eV. The latter peak was used for charge compensation alignment, although we also kept an eye on the first one, whose positioning at the BE of 282 eV is rather universal for $Ti_3C_2T_x$ MXenes[98]. These peaks indicate Ti–C–Ti, C–Ti–O and C–C bonding, respectively. We would like to highlight the low amount of C–C carbon in our samples, which typically originates from over-etching or organic contaminants; this speaks in favor of the optimal etching conditions[99]. **O1s** spectrum was deconvoluted into 4 peaks at BE of 529.9, 530.9, 532.5 and 534.3 eV, referring to C–Ti–O, $TiO_{2-x}F_{2x}$, C–Ti–O/OH/F at the "bridge" site and adsorbed/intercalated $H_2O$, respectively. Since the lowest BE peak (529.9 eV) that refers to the oxygen termination is predominant, we conclude that =O population on the surface is significantly higher than that of –OH. However, in the following sections, we will raise some doubts regarding the accuracy of this notation, as some of our evidence suggests otherwise. **F1s** spectrum was deconvoluted into 2 peaks at BE of 685.3 and 686.5 eV, indicating C–Ti–F functionalization and the minor trace of $AlF_x$ secondary phase, respectively.

We clearly see the formation of a mixed-type =O, –F, and –OH surface termination groups, where, following Natu's analysis[98], the initial ratio between oxygen- and fluorine-based ones is close to 1 : 1, and their overall stoichiometry reaches $Ti_3C_2(O/OH)_{0.9}F_{1.1}$, demonstrating almost complete (~$Ti_3C_2T_2$) flake functionalization, i.e. the coverage of the MXene with termination groups.

### *3.2 General trends in the chemistry of MXenes in HaP-precursor solutions*

After validating the formation of high-quality $Ti_3C_2T_x$ MXenes, we exposed them to the HaP-precursor solutions with different precursor components (following **Table 1**), and followed the chemical state of both the MXenes and the hosting solutions. Our preliminary tests showed that some of the MXene-containing solutions (also denoted as "inks") became rather unstable, causing color-change of the solution, which indicated degradation processes. Among the tested inks, we found that some of them are significantly more stable than the others. To gain an in-depth understanding, a thorough study has been conducted after exposing MXenes to the HaP-inks, and observing them after 1 hour and after 1 week. Following **Table 1**, the study included a combinatorial set of DMSO-based solutions of the precursors, where MAI, $PbI_2$, and $I_2$ were added together or in part. This allowed us to isolate and identify (de)stabilization key factors and study the changes that happen to MXenes, including oxidation and surface chemistry changes in the suspended MXenes, providing critical fundamental understanding for the integration of MXenes with HaPs.

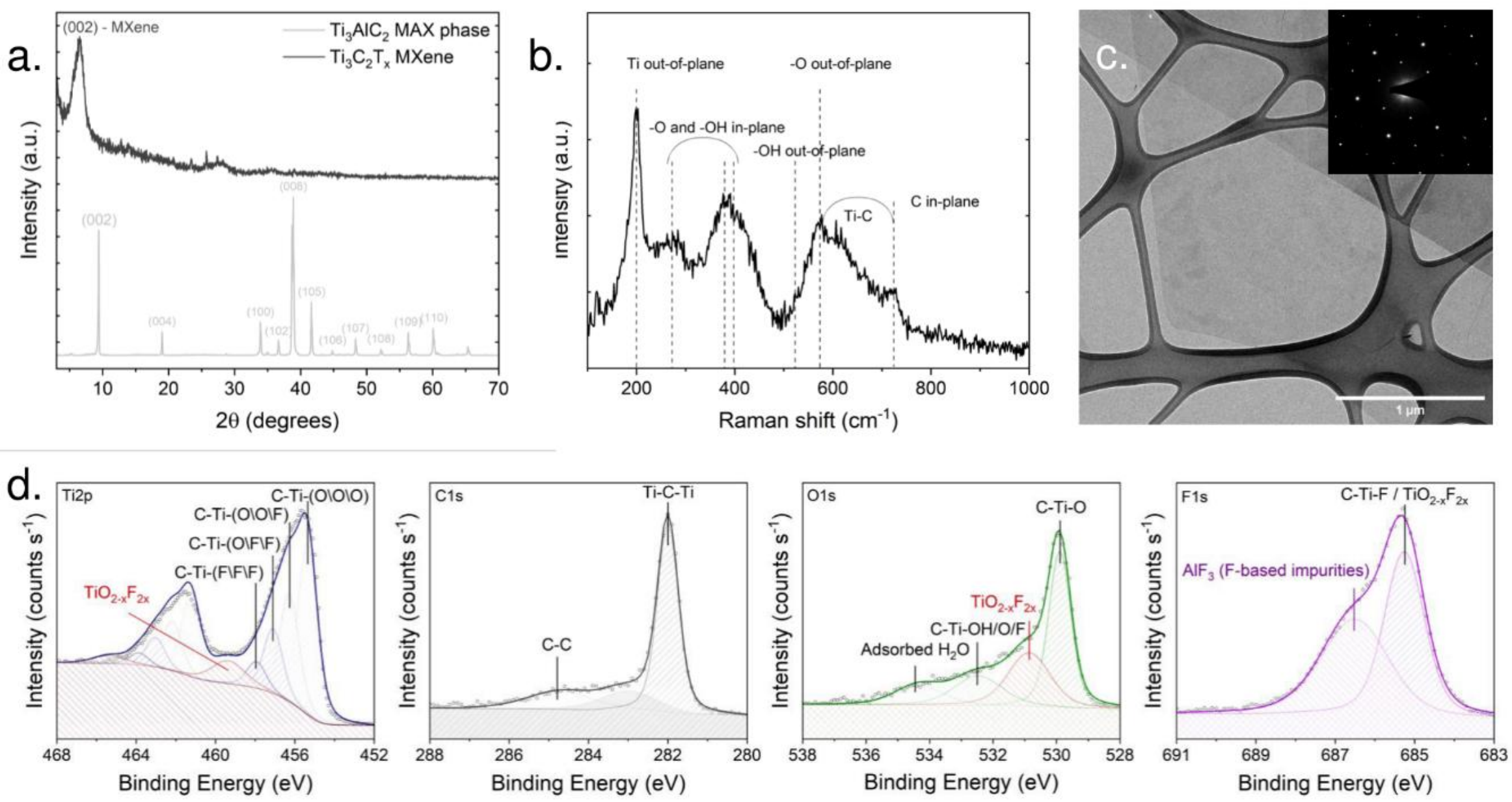

**Figure 2. Characterization of the as-cast $Ti_3C_2T_x$ MXenes.** (a) XRD pattern of as-cast $Ti_3C_2T_x$ MXenes and $Ti_3AlC_2$ MAX phase; (b) Raman spectrum, (c) TEM image of single-layer $Ti_3C_2T_x$ flakes with respective SAED pattern, and (d) - Ti2p, C1s, O1s, F1s XPS core spectra. The XPS peaks' notation follows ref. [98].

To provide insight into the changes in the chemical state and environment of the suspended MXenes, we used the unreacted MXenes as our benchmark and carried out XPS characterization on the remaining filtered solids. We found Ti2p and O1s core-level spectra to be the most chemically sensitive and indicative, as presented in **Figure *3***, while the complementary F1s core-level spectra are presented in **Figure S2**.

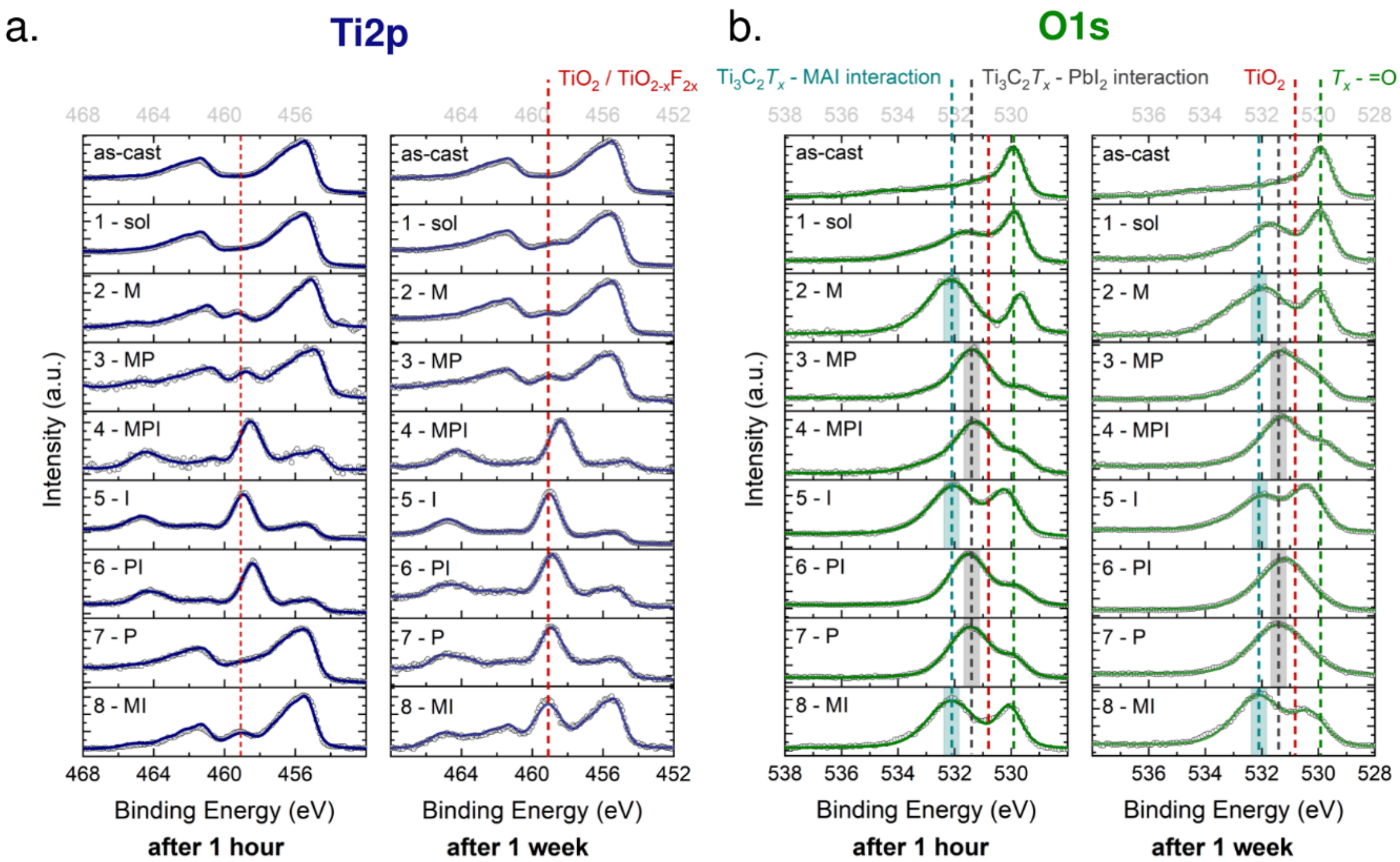

**Figure 3. Complete Ti and O XPS comparison between MXenes extracted from the HaP-precursor solutions.** (a) Ti2p, and (b) O1s XPS core spectra of the as-cast $Ti_3C_2T_x$ MXenes and MXenes extracted from solutions 1-8 after 1 hour and 1 week of exposure. In (a), the typical peak position that indicates MXene oxidation ($Ti^{3+} \rightarrow Ti^{4+}$; $Ti_3C_2T_x \rightarrow TiO_{2-x}F_{2x}$) is marked by a red vertical dashed line. In (b), the typical position referring to C – Ti – O termination group bonding is marked by the green vertical dashed line, the position referring to MXene oxidation is marked by a red line, while the positions marked with cyan and gray refer to new bonding types that we associate with $Ti_3C_2T_x$ reacting with MAI and $PbI_2$, respectively. The nature of this bonding is discussed in detail in the text.

We start with identifying inks that oxidize the MXenes ($Ti^{3+} \rightarrow Ti^{4+}$; $Ti_3C_2T_x \rightarrow TiO_2$ / $TiO_{2-x}F_{2x}$), and whether the oxidation is instantaneous or slow. The most informative XPS feature for the oxidation state of Ti in the $Ti_3C_2T_x$ MXenes is the Ti2p spectrum, and in particular the peak around 459.5 eV that indicates the presence of a $Ti^{4+}$ [98] (see the vertical red dashed line in **Figure *3*a**). First, we verify that the pure solvent (DMSO; ink 1-sol) does not oxidize MXenes, as no increase at ~459.5 eV is observed in the respective Ti2p spectra. A similar result is seen by the O1s edge (**Figure *3*b**), showing no evidence for an extra signal that could relate to the formation of a $TiO_2$ phase, which would appear at ~530.8 eV[98]; the broad peak at ~531.6 eV can be attributed to the intercalated solvent[100]. After verifying DMSO itself does not harm our MXenes, we focus on the interaction of the MXenes with MAI, $PbI_2$, and $I_2$. Initial XPS analysis identified the following phenomena:

***MAI containing ink*** – 1M MAI in DMSO ***(2-M)*** – Similar to the pure solvent, MAI does not lead to the oxidation of the $Ti_3C_2T_x$ MXenes even after 1 week. If we follow the $T_x$ termination groups in the O1s (**Figure *3*b**) and F1s (**Figure S2**) edges for ink 2-M, we find that after 1 hour the BEs are slightly down-shifted, proposing MAI is interacting with the MXene and acting as an electron donor. However, after 1 week the BEs return to their pure-MXene positions – a phenomenon that we will discuss later. In 2-M and other MAI- and/or $I_2$-containing inks (8-MI, 5-I), a characteristic broad peak in the O1s XPS core spectra appears, located around 532.1 eV. For ink **5-I** we use this characteristic peak with caution, though, since the presence of $I_2$ has an oxidizing and destructive effect on MXenes, as will be discussed later on in detail. We also note that this peak does not appear in all $PbI_2$-containing inks, even in the presence of MAI.

***$PbI_2$ containing ink*** - 1M $PbI_2$ in DMSO ***(7-P)*** – After a short exposure (1 hour), $PbI_2$ does not show a significant change in the $Ti_3C_2T_x$ MXene, the same as MAI, but after 1 week, the MXenes are found to be strongly oxidized, as indicated by the ~459.5 eV peak of Ti2p. The O1s edge spectra show a similar trend, where the O1s peak at ~530 eV, which is typical for the MXene oxygen termination group, decreases and broadens, while the signal close to 530.8 eV increases (red dashed line in **Figure 3b**), supporting the formation of $Ti^{4+}$. The fluorine edge spectra (**Figure S2**) show little shift in the peak position of the F1s peak, but also show a significant broadening with the oxidation of Ti, which we attribute to the weaker signal due to the lower amount of the surviving MXenes. We note that, unlike oxygen, which stays attached to Ti after oxidation, fluorine will deplete. Finally, we see from the O1s spectra that *all* $PbI_2$-containing systems (including: 3-MP, 4-MPI, 6-PI and 7-P) have a characteristic peak at the BE of ~531.4 eV.

***MAI-$PbI_2$ combined ink*** **–** (1:1) MAI:$PbI_2$ in DMSO ***(3-MP)*** **–** If we look at the combined effect of MAI and $PbI_2$ on the MXenes (ink **3-MP**), the system does not show severe oxidation of MXenes, and from the perspective of the titanium (Ti2p) and fluoride (F1s) spectra, it behaves similarly to the one in 2-M. Interestingly, from the perspective of oxygen (O1s), the system behaves similarly to 7-P. This suggests that the $Ti_3C_2T_x$ MXenes' =O termination groups will favor interaction with $PbI_2$-related species over MAI ones.

Overall, if we look at the O1s edge, we can suggest the type of interactions one may expect between =O and MAI / $PbI_2$. For the case of MAI, a =O---H-N hydrogen bond can be expected, which should increase the electron density around oxygen as a result of electron withdrawal from the hydrogen. The new =O---H-N structure may energetically resemble an –OH termination group, which is associated with a ~532 eV feature in O1s spectra[94,97,98], which thus can explain the characteristic feature at ~532.1 eV that appears in MAI-containing inks (2-M, 8-MI). $PbI_2$ shows a similar effect on the oxygen spectrum, with the appearance of a characteristic peak at ~531.4 eV, which can be explained by the chemical adsorption of Pb onto oxygen termination groups. We note that $PbI_2$ should be fully soluble in DMSO at the concentration we use (1 M), so we may expect to see the adsorption of $Pb^{2+}$ ions onto negatively-charged $Ti_3C_2T_x$ MXenes. The adsorption is likely followed by $Pb^{2+} \rightarrow Pb^0$ reduction and $Ti^{3+} \rightarrow Ti^{4+}$ oxidation – a mechanism closely resembling the one suggested in MXene-based water remediation studies[80,101]. We also note that O1s core spectra appear much stronger modified upon interaction with MAI and $PbI_2$ species, compared to F1s. This suggests that the chemical interactions in our systems mostly involve =O termination groups, while –F groups stay, to a large extent, inert.

***$I_2$ containing ink*** **–** 1M $I_2$ in DMSO ***(5-I)*** – not less importantly, we would like to briefly mention the strong and destructive effect of $I_2$ or $I_n^-$ polyiodides. In short, **Figure *3*** shows that any solution that contains elemental $I_2$, including: 4-MPI, 5-I, 6-PI, 8-MI, ends up having degraded $Ti_3C_2T_x$ MXenes, even just after 1 hour of exposure, as observed by the pronounced oxidation of Ti from the uprise in the signal at BEs of 458.5-459.5 eV. The slight shift of the $Ti^{4+}$-oxide peak towards lower BEs might be related to the formation of titanium oxo-iodides[102], as iodine has lower electronegativity compared to oxygen. If we look at the time scale of degradation, 8-MI is exceptional, showing significantly slower oxidation of Ti. This hints that MAI may possess a protective effect on $Ti_3C_2T_x$ from reacting with surrounding halides. The nature of $I_2$-assisted degradation and MAI-assisted preservation will be discussed in detail later on.

To summarize the main characteristic behavior when adding MXenes to MAI, $PbI_2$ and $I_2$ containing DMSO solutions, we suggest the following schematic chemical equations:

(1) $$[-\mathbf{Ti^{3+}} = \mathbf{O}]_{\mathbf{n}} \;+\; \mathbf{H - NH_2CH_3I} \Longrightarrow [-\mathbf{Ti^{3+}} = \mathbf{O}]_{\mathbf{n}} --- \mathbf{H - NH_2CH_3I}$$
$$\textit{hydrogen bonding}$$

(2) $$[-\mathbf{Ti^{3+}} = \mathbf{O}]_{\mathbf{n}} + [\mathbf{Pb^{2+}} \cdots \mathbf{2\ I^-}]_{(\mathbf{DMSO})} \Longrightarrow [-\mathbf{Ti^{3+}} = \mathbf{O}]_{\mathbf{n}} - \mathbf{Pb^{2+}} + \mathbf{2I^-} \Longrightarrow [-\mathbf{Ti^{4+}} = \mathbf{O}]_{\mathbf{n}} - \mathbf{Pb^0} + \mathbf{2I^-}$$
$$Pb^{2+}\ \textit{adsorption} \qquad \textit{MXene oxidation}$$

(3) $$[-\mathbf{Ti^{3+}} = \mathbf{O}]_{\mathbf{n}} \;+\; \frac{\mathbf{1}}{\mathbf{2}}\mathbf{I_2} \Longrightarrow [-\mathbf{Ti^{4+}} = \mathbf{O}]_{\mathbf{n}} + \mathbf{I^-}$$
$$\textit{MXene oxidation}$$

In the following section, we will further justify these proposed mechanisms using DFT calculations and N1s, Pb4f and I3d XPS core spectra. Moreover, we will follow the degradation kinetics, focusing on $I_2$ and $PbI_2$ oxidation effects, as well as the MAI protective effect.

### *3.2.1 DFT investigation of the MAI, $PbI_2$ ($Pb^{2+}$) and $I_2$ interaction with $Ti_3C_2T_x$*

To support the suggested equations, we studied the adsorption behavior of MAI, $PbI_2$, and $I_2$ on a bare, and on =O, –F, or mixed (=O/–F) terminated $Ti_3C_2T_x$ MXene surfaces by means of first-principles DFT methods. For the latter case, we considered 1 : 1 atomic ratio between O and F, as expected from the as-cast $Ti_3C_2T_x$ characterization given in Section 3.1. The snapshots of the examined systems after structural relaxation are given in **Figure 4** for the most realistic cases of MAI, $PbI_2$, and $Pb^{2+}$ on the mixed-terminated $Ti_3C_2T_x$ MXene, and in **Figure S3** for all the rest. The calculated adsorption energies are summarized in **Table 2**. We note that our calculations did not include the solvent, as it would majorly complicate the calculations. While we bear this fact in mind, the solvent, however, did not interact with the MXene (*cf.*, 1-sol), thus we can more confidently assume that the calculations represent the general trend of the interactions between the MXene and the different chemical species.

Starting with $CH_3NH_3I$ (MAI), on the bare $Ti_3C_2$ surface it undergoes dissociation, with $I^-$ ions preferentially occupying hollow adsorption sites acting as halide termination groups, while one of the hydrogen atoms bonded to the nitrogen in the methylammonium ($CH_3NH_3^+$) cation interacts directly with the $Ti_3C_2$ surface. However, we know that bare $Ti_3C_2$ MXene is rarely the case, and our material shows almost full termination coverage, as discussed in Section 3.1. In case of the terminated MXene surfaces, we found that with both =O and –F termination groups $CH_3NH_3I$ favor bonding via the amine ($-NH_3$), forming a hydrogen bond. If we compare the initial and the final states of the iodine ($I^-$) atom in $CH_3NH_3I$, we find it is somewhat dissociated, and closest to another $-NH_3$ hydrogen. Similar general behavior was observed for the surface with mixed terminations (=O/–F), whose adsorption characteristics closely resembled those of fully =O- or –F-functionalized surfaces. Although for both –F and =O terminations we see a negative adsorption energy, we found that bonding with the =O species is energetically more favorable. The adsorption energies, calculated with respect to the isolated terminated surfaces and a free $CH_3NH_3I$ molecule, are −0.82 eV, −0.49 eV, and −0.75 eV for =O, –F, and mixed-terminated surfaces, respectively.

Similar to MAI, on the bare $Ti_3C_2$ surface, $PbI_2$ does not show a molecular adsorption – it dissociates into Pb and I species that both bond chemically to the surface. In case of either =O, or –F termination (or in the mixed case), we see weak adsorption with characteristic energies of -0.144, and -0.075 eV, respectively. However, it is likely that during the dissolution process, $Pb^{2+}$ and $I^-$ ions will be solvated in DMSO; therefore, we studied the adsorption behavior of $Pb^{2+}$ ions in addition to $PbI_2$ molecule. Unlike $PbI_2$, $Pb^{2+}$ demonstrates drastically stronger adsorption with much lower free energies of -9.35, -8.49 and -8.68 eV for =O, –F and mixed-terminated surfaces, respectively. This clearly suggests the preference of reaction in Eq. 2 over the one in Eq. 1 when both MAI and $PbI_2$ are added to the system, which supports our experimental observations and related hypothesis for the preferential adsorption of Pb in ink 3-MP. Also, we see the tendency of the energetic preference of adsorption onto =O termination groups with respect to –F, similar to what was discussed earlier for MAI.

To gain better understanding about the behavior of $I_2$ around MXenes, we also performed DFT investigation. In case of bare $Ti_3C_2$ MXene surface, the calculations revealed that $I_2$ is *unstable* in its molecular form: it dissociates into $I^-$ ions that preferentially adsorb at unoccupied "top" sites above the surface Ti atoms. In contrast, adsorption studies on =O, –F, and mixed-terminated MXenes show that iodine molecules remain intact and do not dissociate. We note that $I_2$ does not exhibit stable adsorption and remains at distances greater than 3.5 Å from the functionalized surfaces, indicating weak or negligible interaction. The obtained results suggest that the observed destructive $Ti_3C_2T_x – I_2$ interaction happens on unoccupied termination sites, which are often regarded as structural defects at edges or pores. This may also hint that the impeded $I_2$-assisted degradation of the $Ti_3C_2T_x$ MXenes in presence of MAI or $PbI_2$ is related to them blocking the defect sites, providing an effective protective effect. Indeed, we saw that bare $Ti_3C_2$ possesses significant adsorption capacity towards MAI, $PbI_2$ and $Pb^{2+}$.

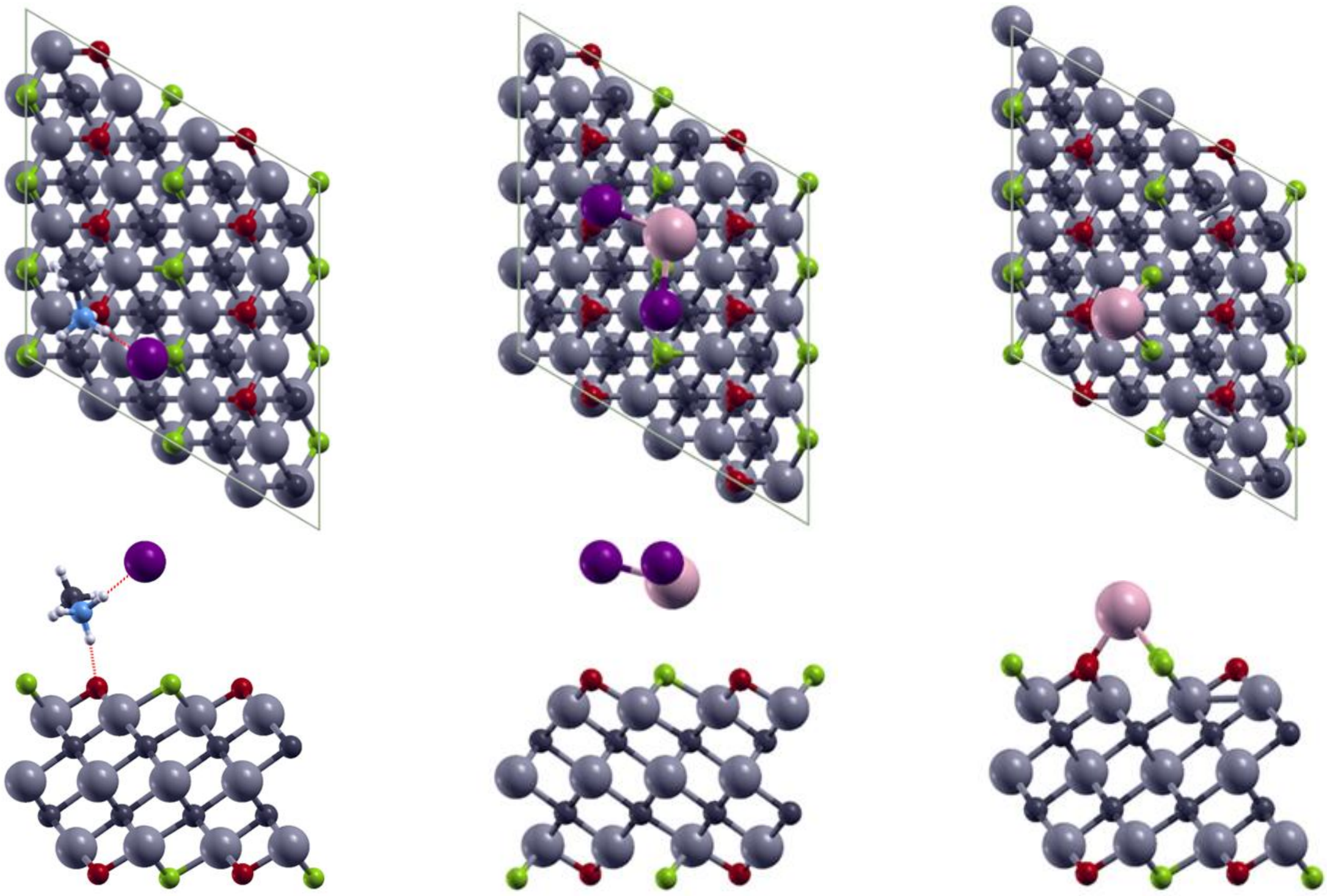

**Figure 4.** The snapshots showing DFT structurally-relaxed models for MAI, $PbI_2$, and $Pb^{2+}$ interaction with the realistic mixed-terminated (O/F) $Ti_3C_2T_x$ MXene surfaces - top and side views. Gray, black, red, green, blue, white, purple, and pink atoms represent Ti, C, O, F, N, H, I, and Pb, respectively.

**Table 2.** Calculated adsorption energies ($E_a$ in eV) for MAI, $PbI_2$, $Pb^{2+}$, $I^-$ and $I_2$ species on $Ti_3C_2T_x$ MXene surfaces with various terminations: bare (no termination groups attached), pure =O, pure –F, pure –OH, and mixed O/F (in 1 : 1 ratio).

| | **bare** | **=O** | **–F** | **mixed O/F** | **–OH** |
|---|---|---|---|---|---|
| **MAI** | -2.9917 | -0.8196 | -0.4917 | -0.7490 | -1.3918 |
| **$PbI_2$** | -6.3497 | -0.1437 | -0.0755 | -0.0632 | -2.4924 |
| **$I_2$** | -6.5633 | -0.0280 | -0.0188 | -0.0291 | -5.4798 |
| **$Pb^{2+}$** | -11.9035 | -9.3532 | -8.4894 | -8.6833 | -15.9649 |

Finally, we should note the potential contribution of an –OH termination group. Although we base our analysis regarding the ratio between =O and –OH on ref. [98], it is possible that we underestimate the real amount of –OH on our MXenes. If we follow the complete analysis of the DFT work that we show in **Figure S3** and **Table S1**, we see that the interaction of the different studied species with –OH termination group is dramatically different from =O, –F and mixed cases. For example, MAI interaction with the –OH group is energetically the strongest, and the configuration of the adsorbed molecule is completely different from other studied cases (see **Figure S3a**), with $I^-$ ions bonded to the termination hydrogen. The same implies to $I_2$ adsorption onto the –OH termination, which also appears to be the most energetically preferrable among the studied configurations, except for the bare $Ti_3C_2$ surface. This suggests that an –OH—I bond should exist, which would explain the appearence of the characteristic peak at ~532.1 eV that we see in **Figure 3** for 5-I, where we don't have $MA^+$ cations at all. The adsorption energies of $PbI_2$ and $Pb^{2+}$ on –OH are also the lowest with respect to the other examined terminations.

These critical DFT-based insights clearly support Eq. 1, and allow us to expand it to the following equations:

$$(1') \quad [-\mathbf{Ti^{3+}=O}]_\mathbf{n} \;+\; \mathbf{H-NH_2CH_3I} \Longrightarrow [-\mathbf{Ti^{3+}=O}]_\mathbf{n}\; ---\; \mathbf{H-NH_2CH_3I}$$

$$(1'') \quad [-\mathbf{Ti^{3+}=O-H}]_\mathbf{n} \;+ \mathbf{I\cdot NH_3CH_3} \Longrightarrow [-\mathbf{Ti^{3+}=O-H}]_\mathbf{n}\; ---\; \mathbf{I\cdot NH_3CH_3}$$

This explains what happens in ink 2-M upon $Ti_3C_2T_x$ MXene addition. MAI molecules tend to adsorb onto $Ti_3C_2T_x$ surface, preferring oxygen or, if available, –OH termination sites. Thus, the small down-shifts in both O1s and F1s spectra after one hour (see 2-M in **Figure *3*b** and **Figure S2**) may suggest the non-selective MAI adsorption onto various termination groups, while an extra peak at ~532.1 eV is more likely indicative of the –OH—I interaction as suggested in Eq. 1''.

We note that since the XPS notation reference we use[98] does not provide strong confidence about the ability to clearly distinguish between =O and –OH groups, we complemented it with an additional FTIR measurement. We compared the as-cast $Ti_3C_2T_x$ MXenes with those where we attempted to artificially convert the –OH groups into =O through annealing the MXenes in an Ar-filled quartz ampule following ref. [103]. The resulting spectra are given in **Figure S4**. We note the intensity of the peak associated with the –OH groups at ~1380 $cm^{-1}$ decreases after annealing, while the Ti-O-related signal at ~665 $cm^{-1}$, on the contrary, builds up[104]. This allows us to safely assume that the as-cast $Ti_3C_2T_x$ MXenes that we use in our study contain a certain, non-negligible amount of –OH terminations on their surface.

Overall, **Table 2** provides an interesting insight that may explain the mechanistic origin for the protective role of MAI, and, to some extent, $PbI_2$. Considering *bare* and –OH termination sites have the most significant adsorption energies, compared to =O and –F, one can conclude that the adsorption of MAI onto bare or –OH terminated sites is supposed to slow down, or even block, the potential oxidizing species, such as $I_2$ or $Pb^{2+}$. In case of the strongly oxidizing $I_2$, MAI seems to significantly impede the degradation of MXenes, as shown in ink 8-MI (and later further analyzed in Section 3.3.4). In case of the milder oxidation of MXenes via $PbI_2$, MAI seems to completely prevent $PbI_2$-related oxidation, as evident in ink 3-MP. While the interaction with –OH is more reasonable due to its potential natural abundance, protective effect can be attributed to MAI's interaction with bare/defective MXene sites, such as edges, which are known to be weak-links and highly reactive[105].

As for the $Pb^{+2}$ interaction, significantly higher adsorption energies on all possible functional groups suggest a strong, and likely covalent adsorption. The relatively lower energies for =O and –OH termination groups can explain an extra peak in the respective O1s XPS spectra, supporting the formation of the =O/–OH – Pb bond, as suggested in Eq. 2. In case of the competitive processes involving $Pb^{2+}$ and MAI, the former will likely get adsorbed onto MXenes due to much stronger adsorption energies, which clarifies the situation observed in ink 3-MP. If we look at the charges on the outer Ti atoms and termination groups before and after the studied interaction processes, as shown in **Table S-T1**, we see that in case of both MAI and $Pb^{2+}$ adsorption the $T_x$ slightly acquires charge, indicating electron withdrawal from the adsorbed species. Interestingly, similar trends in charge-density variations are evident in the BE chemical shifts in the XPS measurements, as we mentioned earlier, and as we will see in further detail in the following sections.

### *3.3 Comprehensive analysis of MXene interaction chemistry with HaP-precursor components*

Based on the results so far, we can summarize our findings in the following way:

1) MAI is adsorbing to MXenes, supposedly through selective adsorption at =O or, if available, –OH terminations, or active bare sites. Its adsorption is likely promoting a stabilization effect that protects MXenes from oxidation;
2) $PbI_2$, and more likely $Pb^{2+}$ as its binary, adsorbs onto $Ti_3C_2T_x$ MXenes selectively (preferentially at =O, or –OH terminations), and expectedly acts as a reducing agent that promotes the oxidation of the $Ti_3C_2T_x$ MXene;
3) Free $I_2$ or polyiodides degrade $Ti_3C_2T_x$ rapidly via a redox reaction, with a simultaneous formation of $Ti^{4+}$ and $I^-$ species. The reaction is expected to happen at bare surface sites.

To support the DFT prediction, suggesting MAI and $Pb^{2+}$ adsorption, we performed UPS measurements to evaluate whether the work function of the filtered MXenes has changed, as an indication for surface modifications; see **Figure S4**. Our starting work function (ink 1-sol) is 4.67 eV, which is in line with previous works[53,106,107]. When we measure MXenes extracted from inks 2-M and 7-P, which correspond to MAI and $Pb^{2+}$ adsorption, respectively, we observe clear shifts in the work function to 4.87 eV with MAI and 4.49 eV with $PbI_2$. We note that the ± 200 meV difference from the pristine state is of the order one will gain by exchanging the MXene surface from fully O-terminated to fully F-terminated[53], and comparable to other reported surface modification strategies[107]. This highlights the importance of knowing the exact surface chemistry of MXenes, prior to their incorporation in devices. In particular for HaPs, such changes in work function may lead to significant effects in optoelectronic devices, such as solar cells, LEDs, and others.

Overall, these conclusions provide us with an important framework and general guidelines for understanding the HaP-solution – MXene system. However, they still lack the evidence that can provide us with an in-depth understanding of the degradation mechanisms and kinetics, and the actual state of the MXenes in cases where the MXenes are not degraded. In this section, we would like to shed light on these three mechanisms, providing a deeper understanding of the nature of passivation and degradation of the $Ti_3C_2T_x$ MXene when exposed to MAI, $PbI_2$, and $I_2$. We will explore the XPS of MAI (N1s and I3d) and $PbI_2$ (Pb4f and I3d) components in the characteristic MAI- and $PbI_2$-containting inks, and will provide the degradation kinetics distinguishing between the specific role of $I_2$ and that of $PbI_2$.

#### *3.3.1 MXene passivation chemistry with MAI*

Starting with MAI – the only non-aggressive species in our set – we focus on solutions 2-M, 3-MP and 8-MI. 2-M and 3-MP show minimal degradation, and overall are significantly more stable than the rest of the inks. In 8-MI the degradation seems to be evident, but is rather mild compared to other $I_2$ containing inks. After focusing on the O1s and F1s XPS spectra in the Section **Error! Reference source not found.**, suggesting chemical adsorption of MAI in 2-M and 8-MI, and of $Pb^{2+}$ in 3-MP, here we further investigate the I3d and N1s core spectra, which are given in **Figure 5**.

After short exposure to ink 2-M, N1s peak is located at 402.1 eV, and the I3d consists of two doublets: one located at 619.3 eV, and the second one at 619.9 eV. All these peak positions are up-shifted with respect to what is expected for a free-standing MAI (401.9 eV for N and 619 eV for I)[108]. Coupled with the evidence on oxygen and fluorine termination group shifts discussed above (O1s: 529.9 → 529.6 eV; F1s: 685.3 → 685.0 eV), these minor effects may suggest non-selective adsorption of MAI onto the MXene, as suggested in Eq. 1’. After 1 week in 2-M, however, we see a strong effect showing that both N and I signals are significantly shifted towards lower BEs of 400.9 and 618 eV, respectively. Simultaneously, in **Figure 3b** and **Figure S2**, we can see that the terminal oxygen and fluorine peaks shift to their original positions. We attribute it to the transition from a non-selective bonding to a selective adsorption onto the –OH termination sites that follows the DFT calculated models (**Figure 4** and **Figure S3**) and Eq. 1'', where the –OH termination group acts as an electron donor[109,110].

As a complementary confirmation regarding the interaction of MAI with the $Ti_3C_2T_x$ MXenes, we performed $^1H$ solution NMR measurements. For this we used 1M ink 2-M-*type* solutions with deuterated DMSO (DMSO-d6) to avoid the interference between DMSO and MAI methyl signals in the $^1H$ spectrum, with and without MXenes. The resulting $^1H$ spectra, displaying the $CH_3$ and $NH_3$ signals, are given in **Figure S6**. Following the addition of MXenes, we observed a $5.8 \pm 0.6\%$ and $4.7 \pm 0.6\%$ signal loss for $CH_3$ and $NH_3$ respectively, indicative of MAI surface binding. Additionally, both signals exhibited a 0.02 ppm shift and broadening, most evidently for the $CH_3$ signal, as the $NH_3$ signal is likely motionally narrowed due to fast exchange of its labile protons. These observations strongly support the evidence we collected so far regarding the chemical interaction occurring between $Ti_3C_2T_x$ MXenes and MAI molecules. **Table S-T2** summarizes the $^1H$ relaxation rates for the $CH_3$ and $NH_3$ groups with and without MXenes in the solution. The relaxation rates showed very slight changes between the MXene-containing sample and the reference, with the transverse relaxation rate unchanged within error. As adsorbed states are characterized by rapid transverse relaxation $R_2$, measurable surface exchange would increase the observed $R_2$ in the MXene-containing sample[111].The unchanged $R_2$, alongside the equivalent signal loss for both signals, indicate that MAI is effectively irreversibly adsorbed on the NMR timescale and sequestered by the $Ti_3C_2T_x$ MXenes.

As for ink 8-MI, after 1 hour the trend is similar to what we observe for the material extracted from 2-M, where we associate the minor shifts to the adsorption of MAI on top of $Ti_3C_2T_x$ through a hydrogen bonding, where hydrogen donates electrons to the termination groups. In contrast to 2-M, however, after 1 week in 8-MI we do not see any significant down-shifts in the position of N1s and I3d peaks, only the increase in the relative intensity for the peak that closer resembles a free-standing MAI. Most likely, this is due to the propagating degradation of MXenes in this ink (see Ti2p core spectra in **Figure 3**), leaving us with some precipitated MAI on the degraded $Ti^{4+}O_x$.

As for the passivation effect of MAI, we see that the degradation in 3-MP and 8-MI by both $PbI_2$ and $I_2$ is practically stopped or strongly impeded with respect to the inks that do not contain MAI. Given the experimental and DFT-based evidence we collected, MAI should occupy bare MXene sites, which act as activated spots for degradation. Therefore, having MAI that will be attached to the bare sites (which can also be referred to as defect sites) is likely to provide protection from the aggressive $I_2$ attack. Since $Pb^{2+}$ competes with MAI on the adsorption on the =O functional groups, MAI is likely passivating the MXene, or reacts with $Pb^{2+}$ to form [MAI + $PbI_2$] complexes[112], thus effectively reducing the overall available $Pb^{2+}$ for adsorption and oxidation (*cf.* Eq. 2).

If we focus on a charge density in means of the XPS peak shift (**Figure 5**) for ink 3-MP, after 1 hour there is a minor down-shift in both N1s and I3d. Overall, such behavior is very different compared to that of inks 2-M and 8-MI. Following the previous discussion on the O1s spectrum in **Figure 3b**, and the energy values for adsorption of MAI and $Pb^{2+}$ (**Table 2**), we noticed that $Ti_3C_2T_x$ favors Pb-related adsorption (Eq. 2) rather than MAI-related adsorption (Eq. 1' or 1''). Overall, the assembly of the findings supports a negligible interaction of MAI with the MXene in ink 3-MP and a rather preferential adsorption of Pb as =O/–OH – Pb. Hence, we attribute the small chemical shifts in the positions of N1s and I3d peaks after 1 hour to the interaction of MAI with the adsorbed Pb as =O/–OH – Pb···MAI. We note that after 1 week both N1s and I3d signals are matching a free-standing MAI (see dashed vertical lines in **Figure 5**), suggesting weak MAI-surface interaction. In addition, we note that in ink 3-MP [MAI + $PbI_2$]-based complexes should form[113,114], so the observations for 3-MP ink may relate to the adsorption of these whole complexes onto $Ti_3C_2T_x$ MXene surface.

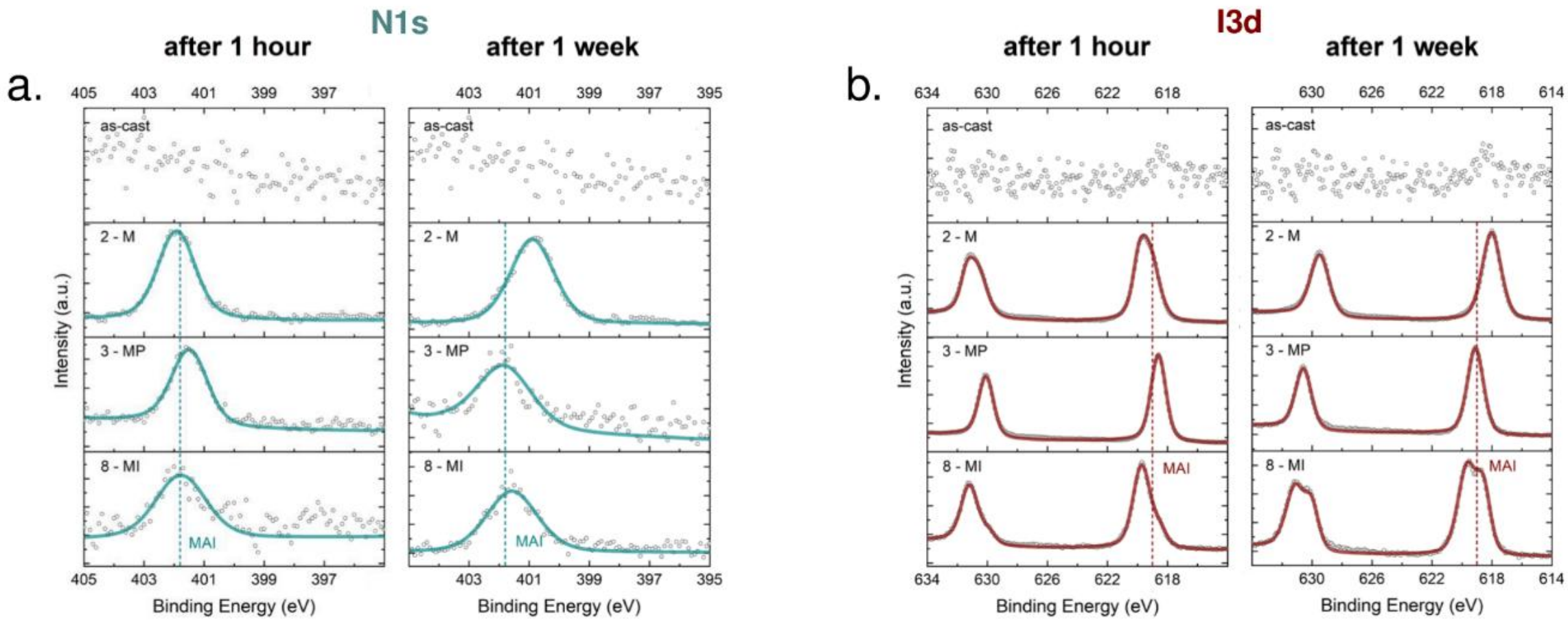

**Figure 5.** (a) N1s and (b) I3d XPS core spectra of the nitrogen- and iodine-based species from $Ti_3C_2T_x$ MXenes extracted from inks that focus on the effect of MAI: 2-M 3-MP, 8-MI after 1 hour, and 1 week of exposure. The typical N1s and I3d peak positions for nitrogen and iodine in MAI[108] are marked by cyan and dark-red vertical lines, respectively.

### *3.3.2 MXene* ***oxidation*** *chemistry with $PbI_2$*

Next, we shift our focus to inks 3-MP, 4-MPI, 6-PI, and 7-P, to better understand the chemistry around $PbI_2$ + $Ti_3C_2T_x$ in DMSO. $PbI_2$-containing inks are used for a two-step HaP films preparation[71], so knowing the chemistry of MXenes in these solutions has an immediate practical impact. Focusing on ink 7-P, we saw $PbI_2$ leads to an accelerated oxidation of the MXene (see **Figure 3**). To understand the nature of this degradation, we again turn to XPS, where now we focus on the Pb and I core spectra, namely Pb4f and I3d. In **Figure 6** we analyze these spectra for MXenes after exposure to inks 3-MP, 4-MPI, 6-PI and 7-P after 1 hour and 1 week.

Focusing on ink 7-P, where we can isolate the effect from $PbI_2$, in **Figure 6a** we see two distinct states of Pb. Following Keersmaecker et al. [115], the first, and more dominant, peak at 138.2 eV is attributed to a crystalline $PbI_2$, and the second up-shifted component at 138.9 eV can be attributed to a fully coordinated Pb. We associate the appearance of the strongly coordinated Pb to its adsorption onto $Ti_3C_2T_x$ MXene, as suggested by Eq. 2, and also supported by the DFT results. Moreover, it perfectly aligns with the available evidence of MXenes reacting with undercoordinated $Pb^{2+}$ defects in HaPs[61,62], and nicely supports this conception. We expect this adsorption to be =O (or –OH) group specific, following much more pronounced changes in O1s core spectra (see **Figure 3b**) with respect to F1s (**Figure S2**). Further support can be found in the fact that the location of the peak also perfectly agrees with the one reported for the study of $Ti_3C_2T_x$-mediated Pb extraction from aquatic solutions[116].

Generally, it has been reported that $Ti_3C_2T_x$ MXenes are efficient in adsorbers for Pb[101,117]; thus, it reasonable to assume that a similar behavior may happen in other solvents. To further understand the nature of the Pb adsorption on $Ti_3C_2T_x$ in DMSO, we measured in a TEM isolated flakes that were extracted from ink 7-P (see **Figure S7)**. EDS analysis from the thinnest region on a $Ti_3C_2T_x$ flake detected Pb atoms in a ratio of ~1 : 1 with respect to Ti, while almost no iodine was detected at the same positions, supporting that Pb signal does not come from $PbI_2$. When studying their diffraction via SAED (see **Figure S7**), we found only $Ti_3C_2T_x$ MXene diffraction peaks (typical for 001 zone-axis), without any evidence for additional peaks that could relate to a $PbI_2$ phase.

As a supporting test, we checked $Ti_3C_2T_x$ flakes extracted from ink 6-PI, and indeed much lower content of lead was observed with respect to Ti (Pb : Ti atomic ratio was ≤ 1 : 5). This suggests that in ink 6-PI the processes of Pb adsorption and $I_2$-mediated degradation are running in parallel, where iodine is oxidizing MXenes, leaving fewer active sites to adsorb Pb. To support this, we return to the XPS results in **Figure 6**. Interestingly, in ink 6-PI we see almost the same pattern after 1 hour exposure as in ink 7-P; however, it drastically changes after 1 week. After long exposure to 6-PI, the Pb4f spectrum reveals the absolute dominance of the pure $PbI_2$-related signal at 138.2 eV meaning that upon the $I_2$-assisted degradation of $Ti_3C_2T_x$, the adsorbed Pb transforms back to a $PbI_2$ phase, possibly acting as a nucleation site for further $PbI_2$ growth as the solvent dries.

Whenever MAI is added to the system (see **Figure 6a -** 3-MP and 4-MPI), we find that initially, after 1 hour, the peak related to the coordinated Pb is not as pronounced compared to 7-P and 6-PI. This supports our earlier suggestion of a competition between Pb and MAI adsorption on MXenes, or complex formation with MAI that reduce the activity of Pb to adsorb onto MXenes – both of which suggest impeded occupancy of Pb on the MXene. Moreover, the whole XPS core spectra are significantly shifted towards lower BEs, suggesting a very different and highly reducing chemical environment. The local formation of $MAPbI_3$ phase is also possible, which may lead to similar chemical shifts[115]. In addition, when we add MAI, we observe some metallic $Pb^0$, which actually speaks in favor of $MAPbI_3$ phase formation after filtration and solvent drying, and a consequent Pb reduction to $Pb^0$ by the probing X-rays – a phenomenon that is typical for XPS measurements of $MAPbI_3$[118]. After 1 week of exposure, however, the pattern in 3-MP almost perfectly resembles the one in 7-P, supporting that Pb adsorption indeed happens, or [MAI + $PbI_2$] complex reacts with the surface through Pb.

If we analyze I3d core spectra in **Figure 6b**, we see that in 7-P and 6-PI, iodine seems to be coming predominantly from $PbI_2$-like phase[115], while in 3-MP and 4-MPI it is shifted closer to what is expected for a free-standing MAI[108]. These positions support the previous discussion of favoring Pb-adsorption, followed by a weaker MAI attachment.

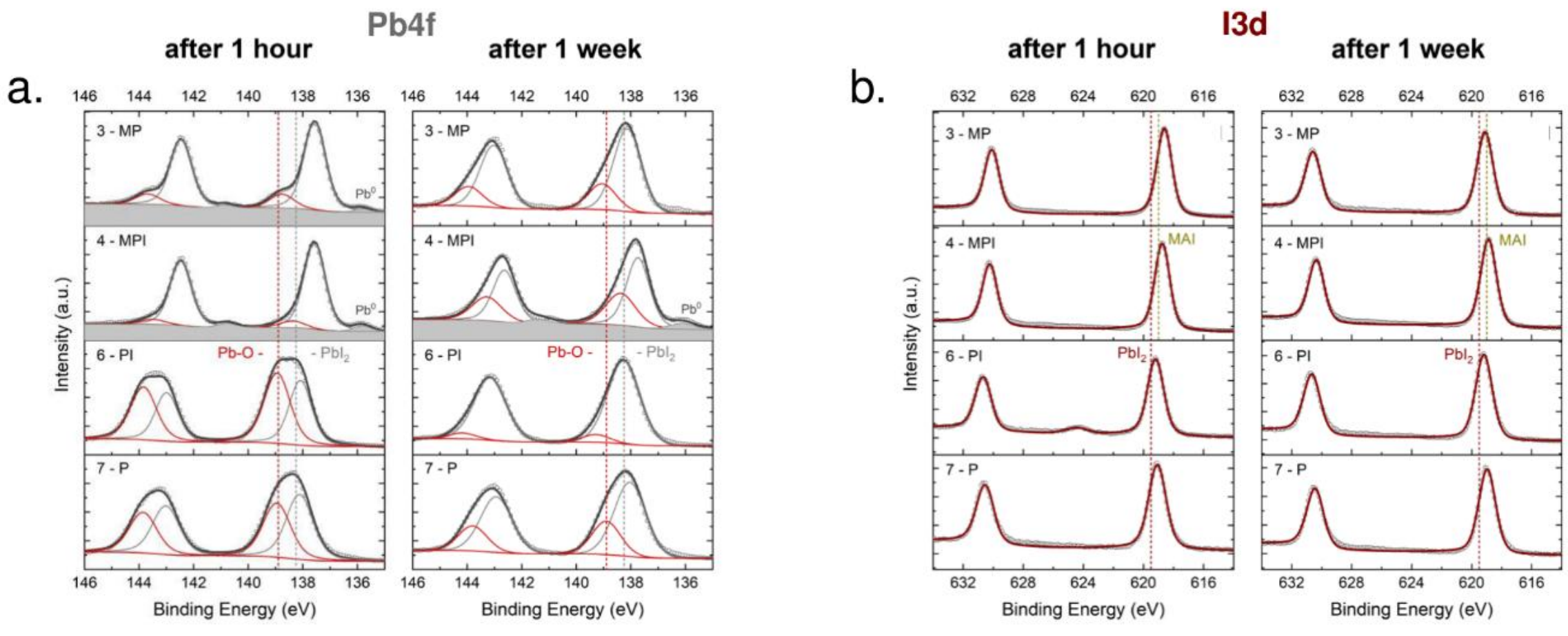

**Figure 6.** (a) Pb4f and (b) I3d XPS core spectra of the lead- and iodine-based species adsorbed at $Ti_3C_2T_x$ MXenes extracted from that focus on the effect of $PbI_2$: 3-MP, 4-MPI, 6-PI, and 7-P, after 1 hour, and 1 week of exposure. The typical peak positions for nitrogen and iodine in $PbI_2$ are marked by a gray vertical dashed line in (a), and dark-red line in (b), respectively. The typical peak position for iodine in MAI is marked by a yellow vertical dashed line in (b). The position marked with red in (a) refers to the bonding type that we associate with Pb selective adsorption onto $Ti_3C_2T_x$ oxygen termination groups.

To summarize, the demonstrated findings provide solid support to the first part of Eq. 2, meaning the adsorption of $Pb^{2+}$ onto $Ti_3C_2T_x$, while the evidence of MXene oxidation (**Figure *3*a**) supports the second part. The adsorption of metals onto MXenes was reported for several divalent ones, including $Pb^{2+}$ [80,101]. Having confirmed that the adsorption process is really happening, we expect the oxidation part in Eq. 2 to be a realistic concept of what is happening next, due to the observed oxidation of Ti shown in **Figure 3**. We note that $Pb^0$ was directly observed only in inks 3-MP and 4-MPI, as discussed above, but not in 7-P and 6-PI (see **Figure 6a**), where partial reduction of Pb was observed as a slight shift of $PbI_2$-related peak towards lower BEs. One possible reason can be a rapid oxidation of the adsorbed $Pb^0$ while solvent drying during the sample preparation for XPS. Thus, we re-write Eq. 2 in a more cautious form as follows:

$$(2') \quad [-\mathbf{Ti^{3+}} = \mathbf{O}]_{\mathbf{n}} + [\mathbf{Pb^{2+}}]_{(\mathbf{DMSO})} \underset{Pb^{2+}\ adsorption}{\Longrightarrow} [-\mathbf{Ti^{3+}} = \mathbf{O}]_{\mathbf{n}} --- \mathbf{Pb^{2+}} \underset{MXene\ oxidation}{\Longrightarrow} \left[-\mathbf{Ti^{\delta+}} = \mathbf{O}\right]_{\mathbf{n}} --- \mathbf{Pb^{\delta-}}$$

### 3.3.3 *Implications to the joint effects of **MAI + PbI₂** with $Ti_3C_2T_x$ MXene*

As discussed earlier, both $PbI_2$ and MAI get preferably adsorbed onto oxygen (or –OH) termination groups, but the oxidation of Ti in ink 3-MP is suppressed (unlike what we observe in ink 7-P), which suggests that Pb is further complexed with MAI, or MAI is complexed with the MXene. When we leave ink 3-MP intact at room temperature or cooled down, it transforms from dark-green transparent liquid to pale white-yellow suspension within several hours (see **Figure S8a**). If we heat up the vial above 70°C upon mixing, it reversibly turns back to its original state (a dark-green color is preserved even after several weeks), which indirectly supports complex formation.

To further distinguish the nature of MXenes after inks 2-M and 3-MP, we checked XRD patterns of $Ti_3C_2T_x$ MXenes extracted from inks 1-sol, 2-M and 3-MP (**Figure S8b**), using the same samples that we used for XPS tests. We see that after 1-sol and 2-M the material does not exhibit any extra reflections except for the (002) MXene peak. However, after ink 3-MP we no longer observe any MXene-related reflections, suggesting the MXene flakes are significantly separated and lose any periodicity. Instead, we start seeing characteristic peaks of $PbI_2$ and MAI phases. If we highlight low intensity peaks by plotting the pattern at the logarithmic scale (**Figure S8c**), we will see the plethora of them, proving large degree of complexation between the components in this system, as well as some partial nucleation of $PbI_2$ and MAI around MXenes. We note that it is unusual to see MAI / $PbI_2$ peaks without $MAPbI_3$, since $MAPbI_3$ often rapidly forms when these two binaries are present in the same space. This may indicate, though, that MAI and $PbI_2$ are spatially separated on the MXenes, or that the HaP decomposes into its binaries before the measurement. While it is out of the scope of the current work, we will be looking into it in the following studies.

When we focus on the position of (002) $Ti_3C_2T_x$ MXene reflection (see **Figure S8d**), we see that it shifts from the original 6.53° to 4.72° after exposure to ink 1-sol, due to intercalation of DMSO molecules[100,119], and to 4.95° after ink 2-M. This indicates a different intercalation state after ink 2-M, compared to ink 1-sol, and serves as extra indirect evidence for MXenes' interaction with MAI.

Noting that MAI and $PbI_2$ compete for adsorption on the MXene, we further evaluate the ability to control the MXene surface chemistry by a sequential exposure to these different elements. To do that, we soak the MXenes into a solution that contains only MAI (2-M like), wait for 1 hour, and only then add dissolved $PbI_2$ in DMSO. Following **Figure S9,** we do not see Pb adsorption, but rather MAI adsorption. This is evident by the characteristic O1s peaks at ~532.0 eV and ~531.5 eV, which are attributed to MAI (–OH—I like; Eq. 1'') and $PbI_2$ (=O/–OH — Pb like; Eq. 2') related adsorption, respectively. We found that if we start with MAI, even after the addition of $PbI_2$, the characteristic peak remained on the position associated with MAI adsorption, i.e., –OH—I like at ~532.1 eV. This is in contrast to the original 3-MP experiment shown in **Figure 3b**, where we show that when $PbI_2$ is there from the start, the surface of the MXene would be =O/–OH — Pb like. Moreover, we do not see the characteristic coordinated Pb peak at 138.9 eV in the respective Pb4f core spectra, which further confirms the absence of interaction between Pb ions and MXene surface groups.

These results clearly suggest that the functional control of MXenes can be achieved not only through chemical environment and time, but also by controlling the sequence chemicals are added. We should note that we are not aware of other publications that showed a similar attribute in MXenes, in general, and this work is likely to be the first to report it. Interestingly, such an attribute is rather common in coordination chemistry and the stabilization of nano-particles[120], where the sequence of precursor addition determines the coordinated species (the ligands) and their overall stabilization. Therefore, we should be able to design MXenes with specifically functionalized surfaces by simply controlling the sequence of the added components before implementation – an idea we are currently investigating and will report in a future work.

Functionally, if MAI or Pb+MAI are adsorbed onto the MXene, they both can act as nucleation centers – a statement that was hypothesized several times[48,63–65], but without a solid support. Particularly, we note a recent publication[121] that suggests that exposure to MAI prior to bringing $Ti_3C_2T_x$ MXenes in contact with $MAPbI_3$ is beneficial for the formation of a less-defective HaP-MXene interface. The authors suggested chemical adsorption of MAI based on indirect empirical evidence from the device functionality. Our results clearly shed light on many of these hypothesized statements, providing a solid ground for this result with additional critical insights – both empirical and theoretical – and providing a better platform for further engineering HaP integration with MXenes.

#### *3.3.4 MXene **degradation** chemistry with $I_2$*

Finally, we will focus on the effect of $I_2$ on MXenes when added to different inks (4-MPI, 5-I, 6-PI, and 8-MI), mostly on the degradation kinetics and the chemical origin for the degradation. In addition, we will elaborate on the differences between $I_2$- and $PbI_2$-based degradation.

As mentioned earlier, while MAI and $PbI_2$ are the natural binaries of $MAPbI_3$, $I_2$ is typically not intestinally added to HaP-inks, but it is known as a product that may spontaneously form inside halide solutions[76,77] or films[82–84]. While we used a very high concentration of 1M $I_2$, which is likely not something one will intentionally introduce in a realistic HaP-based ink, we wanted to compare the strength of its effect with that of MAI and $PbI_2$. This allowed us to observe the strong and detrimental effect of $I_2$ on the MXenes, as well as to see the clear protective nature of MAI on MXenes, and the different characteristic degradation mechanism compared to that of $PbI_2$.

To analyze the degradation kinetics, we employed optical transmission experiments of the various inks. The principle behind this experiment is the fact that optical opacity to transmitted light should decrease once MXenes' density decreases due to degradation. MXenes scatter light due to their metallic nature, but once dissolved or oxidized to a $TiO_2$ form, light transmission should increase (see the schematic illustration of this principle in **Figure 7a**). As a control, the impact of the solution without MXenes is shown in **Figure S10**, where we prepared inks without adding $Ti_3C_2T_x$ and measured their optical transmission spectra. It is expected that $I_2$ will lead to optical absorption, and indeed, in the solutions where we added $I_2$, we found an absorption band below 780 nm. Since above 780 nm all the solutions stop absorbing light, we could isolate the scattering effect due to the MXenes and use the transmitted light intensity as a metric to the $Ti_3C_2T_x$ MXenes concentration. Therefore, to follow their degradation we focused on the intensity changes at 920 nm – well above the iodine-related absorption (< 780 nm) or MXene-related plasmon-related absorption at 780-800 nm[122].

As a quasi-quantitative calibration to MXene's concentration, we prepared fresh DMSO solution in which we suspended known concentrations of $Ti_3C_2T_x$: 10, 50, 100, 200, 300, 400, and 500 $\mu g\ mL^{-1}$ ; right after suspending the MXenes in the solvent, we measured their optical transmission spectra. The resulting spectra are given in **Figure S11a-b**, and the derived calibration curve (intensity values at 920 nm) is given in **Figure S11c**. We see a perfect logarithmic decay in the light transmission intensity with increasing $Ti_3C_2T_x$ concentration. A similar calibration test was performed with a DMSO+MAI solution, to show that the effect on the transmission intensity depends on the MXene concentration and not the solution itself. As we will see below, the relevant calibration concentration for our decay experiments is marked by a green zone in **Figure S11c**.

Following the calibration and control experiments, to investigate the aging kinetics due to the addition of $I_2$, we followed the optical transmission spectra for inks 2-M, 5-I and 8-MI after 1, 2, 3, 7 and 9 days (**Figure 7b**). It is easy to see that between day 1 and day 9 in inks 5-I and 8-MI, there is an increase in the transmission intensity along the spectrum, where in ink 5-I it is much faster. In **Figure 7c** we show the changing intensity at 920 nm, so we can follow the degradation kinetics. Overall, by plotting and comparing the degradation kinetics curves from all the inks: 2,5,8 in **Figure 7c**; 4,6,7 in **Figure 7d ;** and 1,2,3 in **Figure S12**, we found that the degradation kinetics in 5-I is the fastest, while inks 1-sol, 2-M, 3-MP show no degradation. Inks 6-PI and 8-MI show somewhat slower degradation with respect to that of 5-I, while 4-MPI and 7-P show strongly impeded degradation.

We clearly see that with MAI and $PbI_2$ (compare 5-I, 8-MI, and 6-PI in **Figure 7c,e**), the MXene dissolution is impeded, where their combined effect can effectively stop MXene dissolution (**Figure 7e**), but not their oxidation (**Figure 3**). Following **Figure S12**, indeed, we do not see inks 1-sol, 2-M, 3-MP degrade, as expected from the earlier XPS study (**Figure 3**), confirming the high stability of $Ti_3C_2T_x$ in these media. We note that the reverse trend of the transmittance decrease with time for inks 3-MP and 2-M might be associated with flake agglomeration or complex formation.

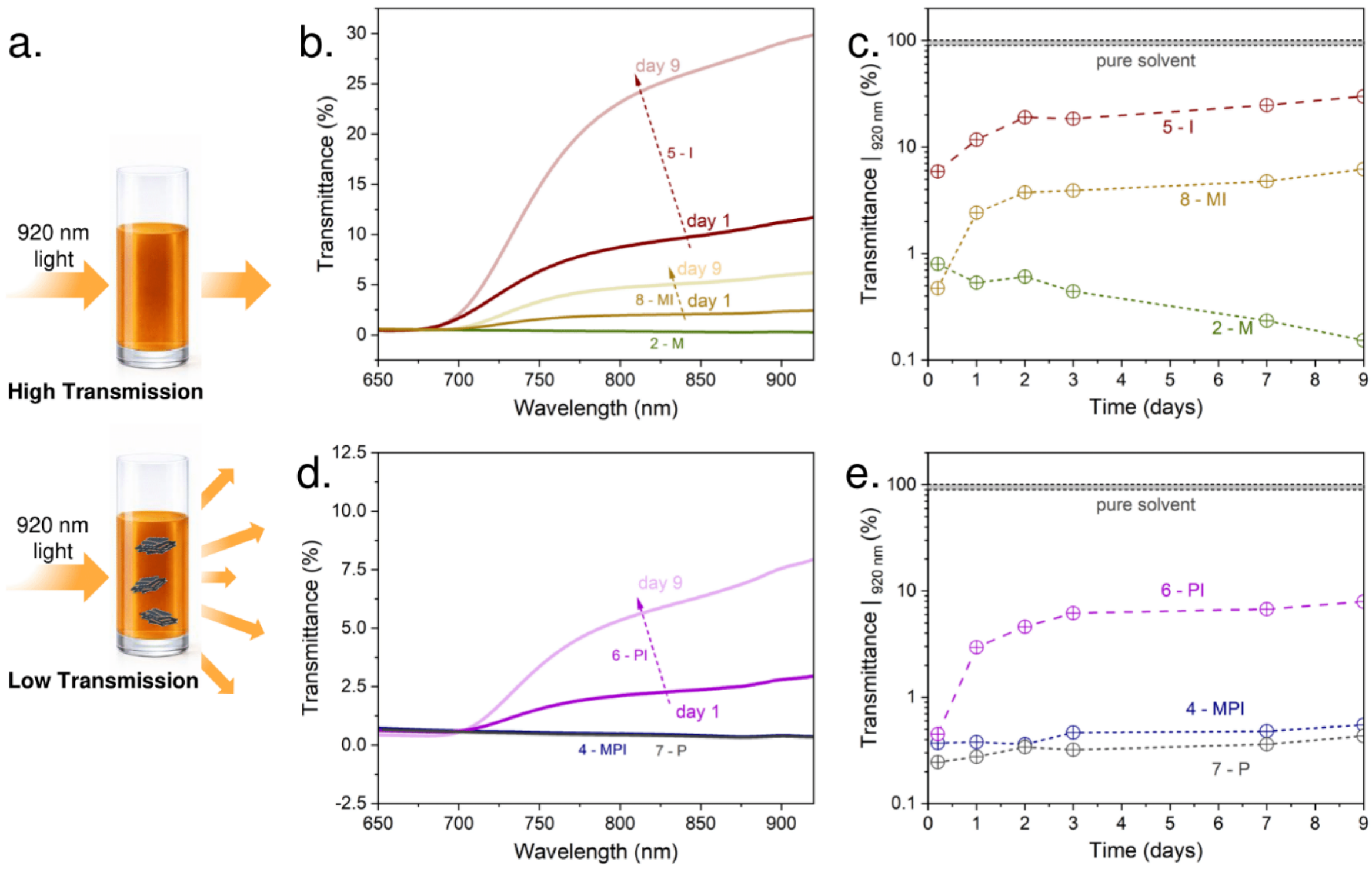

**Figure 7. Optical transmission study of MXene degradation**. (a) schematics of the conducted light transmission experiment that follows MXene dissolution – as MXenes dissolve, more light passes through. (b,d) Representative optical transmission spectra of MXene-containing solutions for studying (b) the effect of $I_2$ and MAI (5-I , 2-M, and 8-MI) and (d) the effect $PbI_2$ (7-P, 6-PI and 4-MPI ), after day 1 and day 9; 2-M, 7-P, and 4-MPI show no significant change in their spectrum with time; (c,e) time-dependent transmittance at 920 nm following the dissolution of MXenes with time. The results are respective to the solutions presented in (b,d); a reference solution (without $Ti_3C_2T_x$ MXene addition) is added as a benchmark. The thickness of the "pure solvent" line in (c,e) estimates the error following **Figure S8**.

Focusing on the effect of $I_2$, if we quantify the degradation due to $I_2$ in ink 5-I using the calibration curve in **Figure S11c**, we can estimate that after 1 hour we already lose more than 50% of the initial MXenes, and after a week, more than 80%. Keeping $Ti_3C_2T_x$ MXenes in ink 5-I for 2 weeks led to no material on PVDF membrane upon vacuum filtration, which supports the assumption the **$I_2$ leads to the dissolution** of MXenes. We note that degradation due to $I_2$ starts with a quick process that spans about 1-2 days, then follows a slower degradation step, as clearly shown for inks 5-I, 6-PI and 8-MI in **Figure 7c,e**. From this simple qualitative analysis, and following DFT-related discussion in Section 3.2.1, we suggest that halide species attack defective sites of $Ti_3C_2T_x$ MXene flakes, break them down to smaller pieces, leading to a fast reduction in the scattering intensity due to particle size. The slow step, then, is a dissolution that occurs from the MXene edges until complete disappearance.

As for $PbI_2$-contaitning inks, we do not see an increase in transmittance for inks 7-P and 4-MPI (see **Figure 7d,e**), suggesting that in these inks the degradation of $Ti_3C_2T_x$ MXenes is mostly via their oxidation but not dissolution. Regardless the reason, this process is much slower compared to the rapid oxidation by $I_2$. As for the possible mechanistic explanation, our suggestion is that $PbI_2$-containing solutions are expected to lead to degradation following Eq. 2', where we suppose Pb adsorption followed by Ti oxidation that leads to the flake's oxidation without dissolution, keeping it structurally intact. Alternatively, the oxidation process happens slowly only when the MXenes are exposed to air. The dissolution process when $I_2$ is added may be related to the formation of $TiI_4$, which is known to be soluble in DMSO[123]. These alternative mechanisms need to be investigated separately.

We would like to briefly mention a rather unique phenomenon we observed in ink 4-MPI, where we noticed the formation of a precipitants within the solution, even without MXenes (see **Figure S13a**). As a side note to this work, we studied the precipitants that formed in the body of ink 4-MPI, and compared this solution to the stable ink 3-MP (see SI, Section SD). We observed these precipitants right inside the volume of the ink (without the filtration or solvent drying) using a confocal microscope, and compared inks with and without $Ti_3C_2T_x$ MXene addition; the respective characteristic images are given in **Figure S13a**. In the absence of MXenes, we observed the formation of large, elongated rods with lengths above 100 μm. $Ti_3C_2T_x$ addition led to their shortening and disoriented "hedgehog"-like morphology, which supports the idea that MXenes play a role in nucleation functionality, as previously suggested[48,63–65]. Further details can be found in Section SD in the SI.

To understand the chemistry behind the $I_2$-based degradation of MXenes, we studied the Raman spectra of the solutions, in which we suspended the MXenes. **Figure 8a** shows the Raman spectra of the freshly prepared inks 3-MP, 4-MPI and 5-I, with and without the addition of $Ti_3C_2T_x$.

In ink 4-MPI we see one dominant peak at 110 $cm^{-1}$ that is indicative of $I^-_3$ stretching mode; the broad peak at 140-146 $cm^{-1}$ originates from the asymmetric $I^-_3$ vibration. Since MAI is a source for $I^-$, $I_n^-$ polyiodides will naturally form when exposed to $I_2$,[124,125] and if the MAI : $I_2$ stoichiometry is ~1 : 1, the most dominant polyiodide will be $I^-_3$. A supplementary Raman study on ink 8-MI shows a similar result (see **Figure S14a**), confirming the formation of $I_3^-$. We note that $I^-_3$ peak intensity is very strong, partially because it may fall into Raman resonance conditions, making [$PbI_2$]-related features and DMSO features absent. The most aggressive ink, 5-I, also demonstrated the existence of polyiodides, but with richer iodine chemistry. Except for the $I^-_3$ feature at 110 $cm^{-1}$ and 140 $cm^{-1}$, we also revealed peaks at 168 and 187 $cm^{-1}$, referring to $I^-_5$ and $I_2$ species, following:

$$(4) \quad I_2 + I^- \leftrightarrow I_3^-$$
$$(5) \quad I_3^- + I_2 \leftrightarrow I_5^-$$

With no MAI influence, which provides a source of $I^-$, we have an excess of $I_2$ with respect to a minority of the spontaneously forming $I^-$ [126], leading to the formation of $I_5^-$ and $I_3^-$ species. The iodine and polyiodide modes in all the spectra were assigned in accordance with refs [126–129]. We note that the broad feature at 114 $cm^{-1}$ that we see in ink 3-MP, and also observed in 7-P (see **Figure S14b**), can be attributed to $PbI_2$-DMSO or [MAI + $PbI_2$]-DMSO complexes[130,131].

When adding MXenes (full lines in **Figure 8**), 3-MP and 4-MPI show no changes in the chemical environment, as an identical shape appears for both spectra, with and without MXenes. When adding MXenes to ink 5-I, however, we do notice a slight increase in the relative intensity of $I_3^-$ and $I_5^-$ related peaks with respect to the $I_2$ feature; however, the effect is rather weak. Since this effect was a result of MXene addition, to enhance it, we increased MXene concentration to 15 mmol $L^{-1}$ (or 3 mg $mL^{-1}$) – 6 times higher than in our standard studied inks (see **Figure 8b)**. With a much more pronounced effect, we now see that after 1 week of storage in darkness (so no light-induced polyiodide conversion occurred), MXene-containing inks show a very clear increase in $I_3^-$ and $I_5^-$ features with respect to $I_2$, while fresh inks with and without MXenes barely differ from each other. We note that with time, even without MXenes, we see spontaneous formation of $I_3^-$ and $I_5^-$, likely due to residual exposure to light, but it is clear that the presence of MXenes accelerates their formation.

The origin for the spontaneous formation of $I^-_3$ and $I^-_5$ is clear: since $I_2$ is in excess, a spontaneous reduction of $I_2$ to $I^-$ leads to an increase in the concentration of $I^-$, which is the source for $I_n^-$ polyhalides. $I^-_5$ also forms since $I_2$ is in excess, and stays in excess even after the reduction of $I_2$ within the time frame of the experiment. We note that if the ratio between $I_2$ and $I^-$ had increased closer to equimolarity (1 : 1), we would have measured a spectrum similar to 4-MPI or 8-MI.

Overall, polyiodides are known to be extremely aggressive and corrosive species[76,77], which suggests a redox reaction between MXenes and polyiodides. If we look on the electrochemical potentials for the oxidation of $Ti_3C_2T_x$ MXenes to $TiO_2$ ($E_{ox}$ = -0.54 eV) and the reduction of iodine ($E_{red}$ = +0.54 eV)[132] in aqueous solutions, we see that energetically it is easy to explain the redox reaction. If we add the fact that higher polyhalides have slightly more positive reduction potential than that of free iodine ($I_5^- > I_3^- > I_2$)[133], we can explain our finding that ink 5-I is the most reactive towards $Ti_3C_2T_x$ (**Figure 7b,d**), since in **Figure 8** we show that it contains the highest concentration of $I_3^-$ and $I_5^-$ species. Following the above, we can explain the degradation reactivity of MXenes with $I_n^-$, shown in **Figure 3a**, by expanding Eq. 3 as follows:

$$(3') \quad 2\mathrm{Ti}^{3+} + \mathrm{I}_2 \rightarrow 2\mathrm{Ti}^{4+} + 2\mathrm{I}^-$$

$$(3'') \quad (n-1)\mathrm{Ti}^{3+} + \mathrm{I}_\mathrm{n}^- \rightarrow (\mathrm{n}-1)\mathrm{Ti}^{4+} + n\mathrm{I}^-$$

We remind that the pronounced formation of $I^-$ in presence of MXenes also nicely supports our DFT-based assumption discussed in Section 3.2.1. As for whether it follows the formation of $TiI_x$ species that can further dissolve in DMSO[123] – it is a matter to be studied.

We underscore that a very minor addition of MXenes (two to three orders of magnitude smaller molar ratio with respect to iodine) is already manifested in a pronounced change of the inks' chemistry. When a stabilizing agent comes into play, such as MAI in our case, degradation slows down, as shown when comparing 8-MI with 5-I (see **Figure 3**, **Figure 4**, **Figure 7b,d**). The inhibition of MXene degradation by MAI is likely to be affected by the interaction of MAI with polyiodides (like in inks 4-MPI and 8-MI)[124], but also due to the adsorption reaction with MXenes, as was discussed in the previous sections.

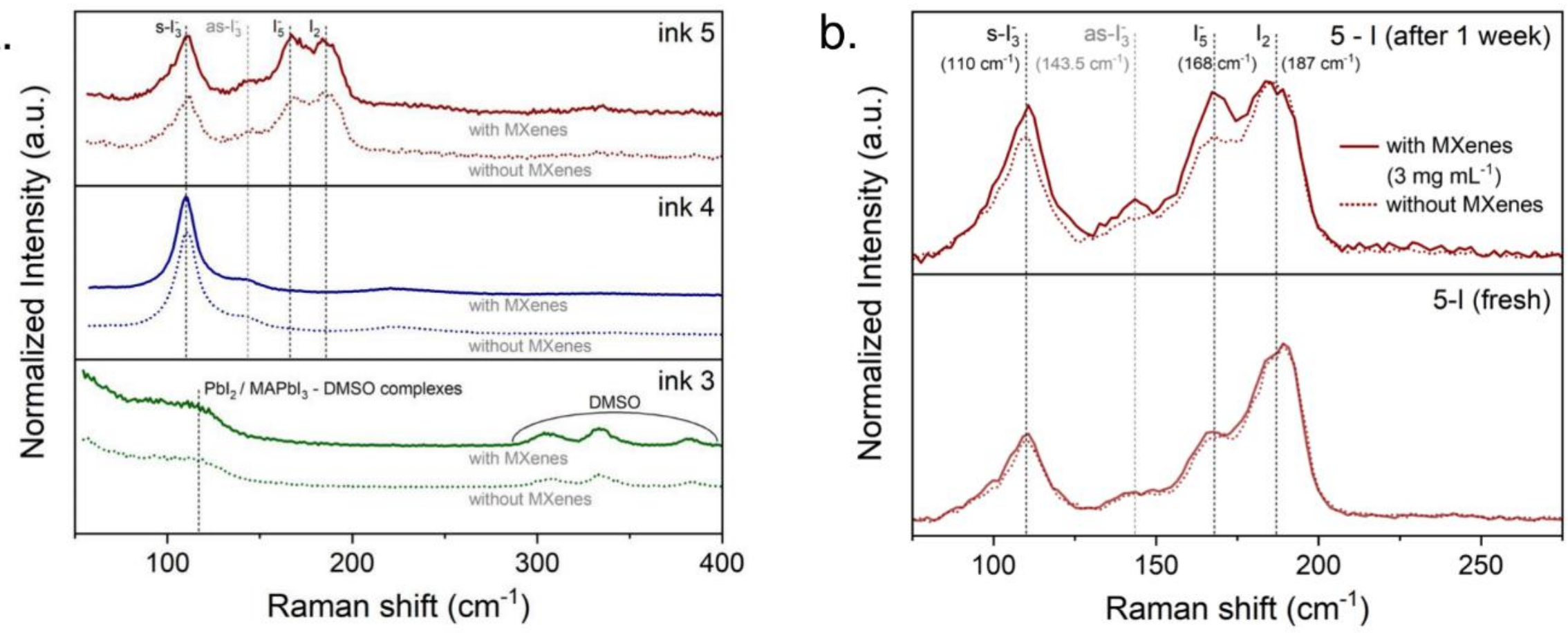

**Figure 8.** (a) Raman spectra of the solutions 3-MP, 4-MPI, and 5-I with and without $Ti_3C_2T_x$ MXenes addition; (b) Raman spectra of 5-I with and without an excessive addition of $Ti_3C_2T_x$ MXenes (3 mg $mL^{-1}$ concentration compared to 1 mg $mL^{-1}$ for regular inks): as-prepared (fresh) and after 1 week storage in a dark place.

### *3.4 Exploring the broader applicability of our findings*

To understand whether what we observed are the general trends, we carried out a number of supplementary studies, changing the type of MXenes, the solvent, or the halide salts.

**Type of MXene:** First, we studied the behavior of a supplementary HF-$Ti_3C_2T_x$ MXene batch upon their exposure to inks 1-3, the respective XPS spectra for both the Ti core and the termination groups, =O and –F, are given in **Figure S15**. As we noted before, the initial state of the HF-MXenes is somewhat more oxidized, and the further oxidation due to $PbI_2$ (in ink 3-MP) is more obvious. This can be related to higher natural defectivity of MXenes prepared using solely HF[86,99]. The evolution of functional groups is similar to what we observed for our main batch, although we see an unexpected modification of F1s core spectra in parallel with O1s after inks 2-M and 3-MP. Likely since HF-$Ti_3C_2T_x$ has more –F terminations, the reaction with MAI appears on fluorine sites as well, because, as we showed in Section 3.2.1, the formation of –F—I bonds is also energetically favorable. As a rather obvious result, to avoid accelerated degradation we suggest using the least defective MXenes possible for integration with HaPs , and using a passivating molecules or cations (MAI or $MA^+$), as was also suggested elsewhere[121]. As we realize that pre-exisiting defects on MXenes may play an important practical role, an important conclusion is that our, and others', theoretical studies are limited to a high-quality isolated flakes of $Ti_3C_2T_x$, while an important complementary work is needed to understand the interaction between the defect states of MXenes with its surrounding. Overall, although the observed trends could remain similar, the details on the termination group population and defects will affect the MXene – HaP-precursor interactions.

**Solvent:** To check the effect on the exploited solvent, we investigated the similar set of inks, but with DMF being a solvent instead of DMSO. Ti2p XPS core spectra of $Ti_3C_2T_x$ MXenes extracted from these inks are presented in **Figure S16a**, while **Figure S14b** provides the respective O1s, F1s, I3d and N1s spectra. Here we see some clear differences with respect to when using DMSO as a solvent. First of all, the oxidation of MXenes in DMF-based inks is significantly faster than in DMSO. Second, we do not actually see the interaction between MAI and $Ti_3C_2T_x$ in this case, neither from O1s and F1s, nor from I3d and N1s core spectra, as no characteristic chemical shifts or extra peaks occur, unlike what was discussed in the previous sections. These results mean that the solvent has an impact on the MXene – HaP-precursor behavior, possibly due to differences in the coordination chemistry between the solvents with the different species in the ink, and require a complementary study of the behavior of MXenes in the hosting solutions with tools similar to what we provide in this work.

**Precursors**: Finally, we checked if similar effects happen in the presence of alternative organic and metal salts, i.e. formamidinium (FAI) instead of MAI, and CuI instead of $PbI_2$; all using DMSO as a solvent. If we change MAI with FAI, the termination group XPS core spectra modification also takes place, as evidenced in **Figure S17a**, suggesting similar hydrogen-bond mediated adsorption processes. O1s core spectra evolution looks similar to the case of MAI, but we also observe changes in F1s spectra (an extra peak arising at higher BEs), suggesting that FAI may be less selective between different termination groups. **Figure S17b** provides Ti2p, Cu2p and I3d XPS spectra after $Ti_3C_2T_x$ MXenes interaction with DMSO+CuI ink. Similar to Pb-salt, we detect copper in more richer chemical state than what is expected for pure copper iodide[134], suggesting the interaction of CuI with $Ti_3C_2T_x$. The pronounced shift of Ti2p core spectrum towards higher BEs also speaks in favor of the strong interaction between the MXene and the metal halide, which nicely correlates with what was shown in one of the earlier works on copper adsorption for water remediation[135].

Overall, we see that the MXene-HaP system is very much chemistry specific. While the use-case we chose draw the general trends in the interaction between MXenes and HaP-related precursors, the details may matter. In particular we found that the solvent and the state of the MXene itself play a significant role, but also that clear similarities between very different chemical environments (e.g., when changing the precursors) do exist. Concluding the supplementary set of tests, we underscore that what we provide in this work is a foundation, as well as an established experimental infrastructure, for further studies in different environments and with different MXenes. That would allow use-case specific understanding of MXenes chemistry and stability in chemically-active media, not necessarily HaP-related.

## 4 Summary and Outlook

As an overall conclusive remarks, the thorough analysis of $Ti_3C_2T_x$ MXene stability upon interaction with MAI, $PbI_2$ and $I_2$ in DMSO-based HaP-inks allowed us to identify the following phenomena:

1. MAI, and likely other amino-halide salts, are not harmful to MXenes even after long-term exposure. They tend to adsorb onto the $Ti_3C_2T_x$ surface through the formation of hydrogen bonds of –N-H with =O and –F termination atoms, or –I with –OH. We note that the XPS quantification analysis that we use here[98] does not allow us to properly estimate the population of –OH termination groups on our MXenes, however, we can be confident about their presence on the surface based on the FTIR analysis, as well as overall knowledge about MXenes. These groups are expected to strongly affect the MAI adsorption behavior, and MAI shows much higher affinity towards –OH. We associate at least one of the observed XPS features to the formation of an –OH—I bond, where adsorbed iodine act as an electron acceptor. Moreover, MAI was found to provide the overall passivation effect through blocking the bare and defective sites, inhibiting, for example, free-halide assisted degradation.
2. Metal halide salts, like $PbI_2$, when being solvated in DMSO, donate $Pb^{2+}$ ions to adsorb onto $Ti_3C_2T_x$ surface through =O (or –OH) terminations, and promote the oxidation of MXenes. These important finding suggests, for example, that it is not recommended to use the two-step fabrication process for HaP[71], where $PbI_2$ is deposited first from a polar aprotic solvent (incl. DMSO), followed by exposure to MAI. While HaP preparation immediately after admixing MXenes into the metal halide precursor can still be implemented, long-term storage of such inks will likely change the content of the ink over time, and with it the quality of the deposited film. Since the proposed mechanism closely resembles MXene-mediated heavy metal adsorption from the aquatic environment, we expect similar mechanisms when other metal salts are being used, such as Cu- or Sn-based ones. We predict similar behavior also for monovalent metals, such as Cs, which can be used as an A-site cation for some HaPs, although this hypothesis is left to be proved.

3. Clearly, the addition of free halides, such as $I_2$, into MXene-containing HaP-inks triggers the rapid degradation of the $Ti_3C_2T_x$ phase. Halide molecules attack the defect sites on MXene basal planes of edges, causing fast redox reaction with the formation of $I^-$, and further polyiodides, and Ti oxidation to $Ti^{4+}$. Although competitive species, such as MAI or $PbI_2$, may impede this process through blocking available sites, the reaction inevitably propagates, leading to severe $Ti_3C_2T_x$ MXene degradation. Thus, further researchers should avoid preparing such solutions in their work, and should take into account the possible formation of polyhalides due to improper storage conditions (the access of light should be minimized).
4. Despite the potential degradation, it was encouraging to find that $Ti_3C_2T_x$ MXene surface chemistry strongly depends on a sequence of HaP-precursor addition within a HaP-ink. For example, a simultaneous exposure to both MAI and $PbI_2$ will result in the preferable adsorption of Pb species onto $Ti_3C_2T_x$ surface, while preliminary exposure to MAI will enrich the adsorption sites with MAI species, blocking further Pb adsorption. The ability to construct chemically-defined $Ti_3C_2T_x$-Pb and $Ti_3C_2T_x$-MAI surfaces prior to actual HaP crystallization allows us to effectively control and tune MXene-HaP interaction. The effect of such chemical control over MXene-HaP interfaces on morphological, chemical and optoelectronic properties of the resulting bulk composites is the main topic of our ongoing research.
5. Last, but not least our evidence shows that –F, =O or –OH termination groups have a different interaction nature or interaction power with the additives, and play an important role in both the activity of MXenes and their role in MXene-HaP composites. By understanding this, now one may, for example, engineer MXene-HaP interaction from very strong, with a predominant =O or –OH termination of $Ti_3C_2T_x$ surface, to relatively weak, if the MXenes are enriched with –F terminations. Preparing such MXenes using known protocols[103,136], and studying their properties upon HaP-ink exposure is our plan for the future research.

Given this fundamental understanding, we can suggest that commonly applied steps, such as storing the inks refrigerated, and in low oxygen/humidity conditions (e.g., glovebox)[137,138], should minimize their degradation, although the above mentioned reactions will still take place. We strictly suggest using the most well-optimized MXene synthesis protocols[85,86], to minimize the defectiveness of the obtained flakes, which act as active sites and increase their susceptibility to degradation. We found that the reported processes are strongly solvent-dependent, so a follow-up study focused on $Ti_3C_2T_x$ MXene stability and surface chemistry evolution in various HaP-relevant solvents (DMF, GBL, acetonitrile, etc.) is a necessity.

This research establishes the understanding that MXenes have a variety of degrees of freedom that directly impact their state, thus cannot be overlooked or can be harnessed to engineer their integration. Overall, it is clear that when it comes to the integration of MXenes with HaPs, we have to take into account that what we add is not what we finally have in our layer – not because of the solvent, but because of the precursors. To this end, also the sequence at which we add MXenes and the duration we hold them in the solution play a role in defining their chemical state – both at the level of the Ti-oxidation and the identity of the termination groups chemistry. These findings put in doubt past research that completely overlooked these points, and provide a critical view on how MXene integration, in general, should be perceived.

**Acknowledgments:**

K.S. thanks Dr. Natalia Froumin for the assistance in XPS data analysis and interpretation.

Y.R. and T.G.M. acknowledge the Ministry of Energy, Israel, Grant #223-11-057, for their support in this project.

T.G.M. acknowledges funding from the Ministry of Innovation, Science and Technology, Israel, Grant #5802.

# *Supporting Information*

## Dynamic Chemical Changes of $Ti_3C_2T_x$ MXenes in Halide Perovskite Precursor Environments

Kirill Sobolev[1,*], Nagaprasad Reddy Samala[2], Mariela J. Pavan[3], Sharon Hazan[3], Vladimir Ezersky[3], Shakked Schwartz[4], Omer Bartov[1], Adi Yakir[1], Michal Leskes[4], Tamar Goldzak Mizrachi[2,*], Yevgeny Rakita[1,3,*]

[1] Department of Materials Engineering, Ben Gurion University of the Negev, Beer-Sheva 8410501, Israel
[2] The Alexander Kofkin Faculty of Engineering, Bar Ilan University, Ramat Gan 5290002, Israel
[3] Ilse Katz Institute for Nanoscale Science and Technology, Ben-Gurion University of the Negev, Beer-Sheva 8410501, Israel
[4] Department of Molecular Chemistry and Materials Science, Weizmann Institute of Science, Rehovot 7610001, Israel
[*] corresponding authors: kirillso@bgu.ac.il, tamar.goldzak@biu.ac.il, rakita@bgu.ac.il

---

### Supplementary Methods

To analyze the thicknesses and lateral dimensions of as-cast $Ti_3C_2T_x$ MXene flakes we used CoreAFM Atomic Force Microscope (Nanosurf, Liestal, Switzerland). For this, small amounts of MXene powders were dispersed in the AR-grade dehydrated ethanol (30 minutes sonication in an ice bath at 37 kHz was used to achieve homogeneous colloidal dispersions), then drop-cased onto flat Si substrates and analyzed in a semi-contact mode after drying.

We exploited IKI Zeiss LSM880 Confocal Laser Scanning Microscope (Zeiss, Oberkochen, Germany) as a specific tool to observe the volume of the inks that were drop-casted onto the glass slide and sandwiched between two slides, without the necessity to filter them or do any other modification.

FTIR spectra were acquired from KBr + $Ti_3C_2T_x$ MXene (or annealed $Ti_3C_2T_x$) pellets using a Nicolet Impact 410 spectrophotometer (Nicolet, USA). The measurements were performed in the spectral range from 400 to 4000 $cm^{-1}$ with a step size of 7.7 $cm^{-1}$, all the spectra were automatically background-corrected.

NMR measurements were performed on a 11.7 Tesla, Bruker AVANCE III spectrometer using a 5 mm BBFO probe with a 500.08 MHz $^1H$ Larmor frequency. The samples were measured under deuterium locking and shimmed manually on the reference sample, which subsequently related any residual inhomogeneity to the particle organization in the $Ti_3C_2T_x$ MXene-containing sample. The $^1H$ longitudinal and transverse relaxation rates for each sample were measured using inversion recovery (IR) and Carr-Purcell-Meiboom-Gill (CPMG) experiments, respectively. The spectra were acquired with a relaxation delay equal to 5 times $T_1$ to ensure steady state quantitative measurements, with a 9 μs pulse (calibrated with respect to a 90° flip angle) and 16 scans. Bruker Topspin software was used for initial spectral processing, followed by signal integration and plotting in MATLAB (R2023B).

For NMR measurements, the sample containing 1M methylammonium iodide (MAI) in 0.5 mg $mL^{-1}$ $Ti_3C_2T_x$ MXene dispersion in DMSO-d6 was prepared by mixing equal parts of 2M MAI stock solution and 1 mg $mL^{-1}$ MXene dispersion in DMSO-d6, the mixing continued overnight for stabilization. For the reference sample, the 2M MAI stock solution was diluted with the equal part of DMSO-d6. Sample preparation and packing of the NMR tube were performed in an Ar-filled glovebox ($O_2$, $H_2O$ < 0.5 ppm). A sealed capillary containing Benzene-D6 (Sigma Aldrich, 99.5%) was placed inside the NMR tube as a non-interacting chemical shift reference.

**Section SA**

*HF-MXenes synthesis*

For the HF-based etching reaction[1], 1 g of the precursor $Ti_3AlC_2$ was added into 5 wt.% solution of HF in DDW and stirred for 24 hours at room temperature. When the reaction was finished, the pH was neutralized (pH > 5) through 8 centrifugation cycles at 3500 rpm for 5 min each. After washing, the supernatant was vacuum-filtered through the polyvinylidene difluoride (PVDF) membrane with 0.22 µm pore size to collect $Ti_3C_2T_x$ multilayered particles, natural for this preparation method. The membrane with the material dried and stored in a vacuumized desiccator.

**Section SB**

*Validation of as-cast $Ti_3C_2T_x$ MXenes: XRD ana Raman analysis*

Powder XRD and Raman validated the synthesis of a desired $Ti_3C_2T_x$ MXene phase. In Error! Reference source not found.**a** we see the XRD pattern showing the characteristic (002) peak at 6.53°, which is shifted from the (002) $Ti_3AlC_2$ MAX phase at 9.5°. The shift to smaller angles and broadening indicates the expansion of the interlayer spacing during etching, delamination, and functional group attachment[2]. Raman spectrum, provided in **Error! Reference source not found.b**, is also rather typical for $Ti_3C_2T_x$ MXenes[3], showing the common vibrations for delaminated MXenes. The vibrations around 200 $cm^{-1}$ consist of *Eg* (Ti, C, O) and *A1g* (Ti, C, O), which are the in-plane and out-of-plane vibrations of titanium atoms at the outer interfaces of the flakes, as well as carbon and surface termination groups, respectively. The region between 580-730 $cm^{-1}$ is assigned mainly to carbon vibrations (both *Eg* and *A1g*)[4]. The region of 230-470 $cm^{-1}$ represents in-plane vibration (*Eg*) of functional groups attached to titanium atoms.

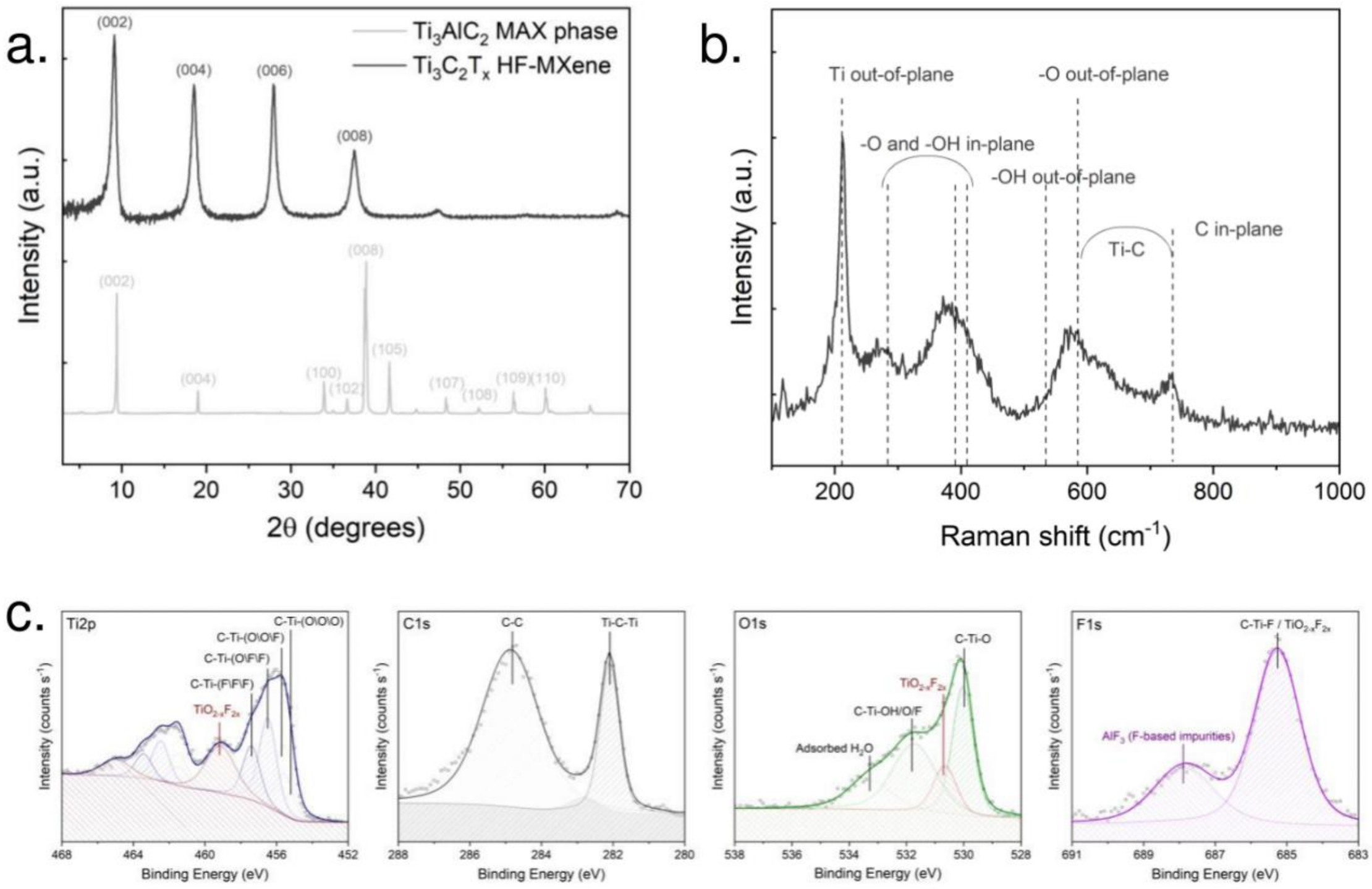

**Figure S1.** (a) XRD patterns of as-cast HF-$Ti_3C_2T_x$ MXenes and the respective $Ti_3AlC_2$ MAX phase precursor; (b) Raman spectra, and (c) Ti2p, C1s, O1s, F1s XPS core spectra of as-cast HF-$Ti_3C_2T_x$ MXenes.

## Section SC

*Validation of as-cast HF-$Ti_3C_2T_x$ MXenes*

Results of the complex initial characterization of as-cast HF-$Ti_3C_2T_x$ MXenes[1] are given in **Figure S1**. Briefly, the provided characterization proves sufficient quality of the prepared materials. XRD pattern of HF-MXenes (**Figure S1a**) possesses repeating (002)-(004)-(006)-(008) reflections, because, unlike milder protocols that yield single- or few-flake MXene clays (see **Figure 2a**[5]), HF etching yields stacked multilayer MXene particles[1]. Raman spectrum (**Figure S1b**) is very similar to what is provided in **Figure 2b**[3,4], and does not indicate any extra phases. A shift for the *A1g* of titanium indicates increased stiffness for the titanium environment and can result from a different assembly or oxidation[3,4]. Ti2p, C1s, O1s and F1s XPS spectra (**Figures S1c,d**) were deconvoluted similarly to what is described in Section 3.1 of the main text[6]. We can note slightly richer –F functionalization of HF-MXenes with respect to the regular ones, as well as the slight over-oxidation of the product (C-C carbon and $TiO_{2-x}F_{2x}$ signals arising), resulting from the harsher etching environment.

**Section SD**

*XRD and TEM-SAED analysis of precipitates from ink 4-MPI*

XRD analysis of precipitants from ink 4-MPI, given in **Figure S13b**, reveals that initially they possess very complex structure, which most likely contains solvent molecules involved[7]. Thus, after drying the precipitate at 140°C for several minutes, we observed only reflections originating from $PbI_2$ and $MAPbI_3$ phases. Precipitants from ink 4-MPI were also deposited onto a grid and analyzed by TEM, the respective images and the varying-tilt SAED patterns are given in **Figure S13c**. Precipitant particles show quite pronounced hexagon-shape core with randomly oriented needles at the periphery. We used varying-tilt ED measurements to confirm the crystal structure of the core, which turned out to be $PbI_2$. The same experiment for the needles turned out to be rather challenging, as they were found to rapidly decompose and change morphology under the electron beam. We assume them to be $MAPbI_3$ perovskite due to the initial composition of the ink, particle shape, and the illumination-sensitive behavior[8], which confirms our assumptions from the XRD analysis The reason behind spontaneous precipitant formation in ink 4-MPI could be that in its complex chemical environment MAI is partially blocked by iodine, as suggested from Raman analysis, which promotes $PbI_2$ nucleation. The interior of $PbI_2$ particles reacts with MAI, leading to $MAPbI_3$ formation and growth. $Ti_3C_2T_x$ presence here simply provides more available sites for lead iodide nucleation, as MXenes were shown to strongly adsorb Pb. As this solution does not transmit light above 780 nm, unlike its $Ti_3C_2T_x$-free counterpart (see **Figure S10** and **Figure 7d**), we assume MXene flakes survive in the core of the observed particles, as they are protected from the aggressive halide environment by the surrounding particle.

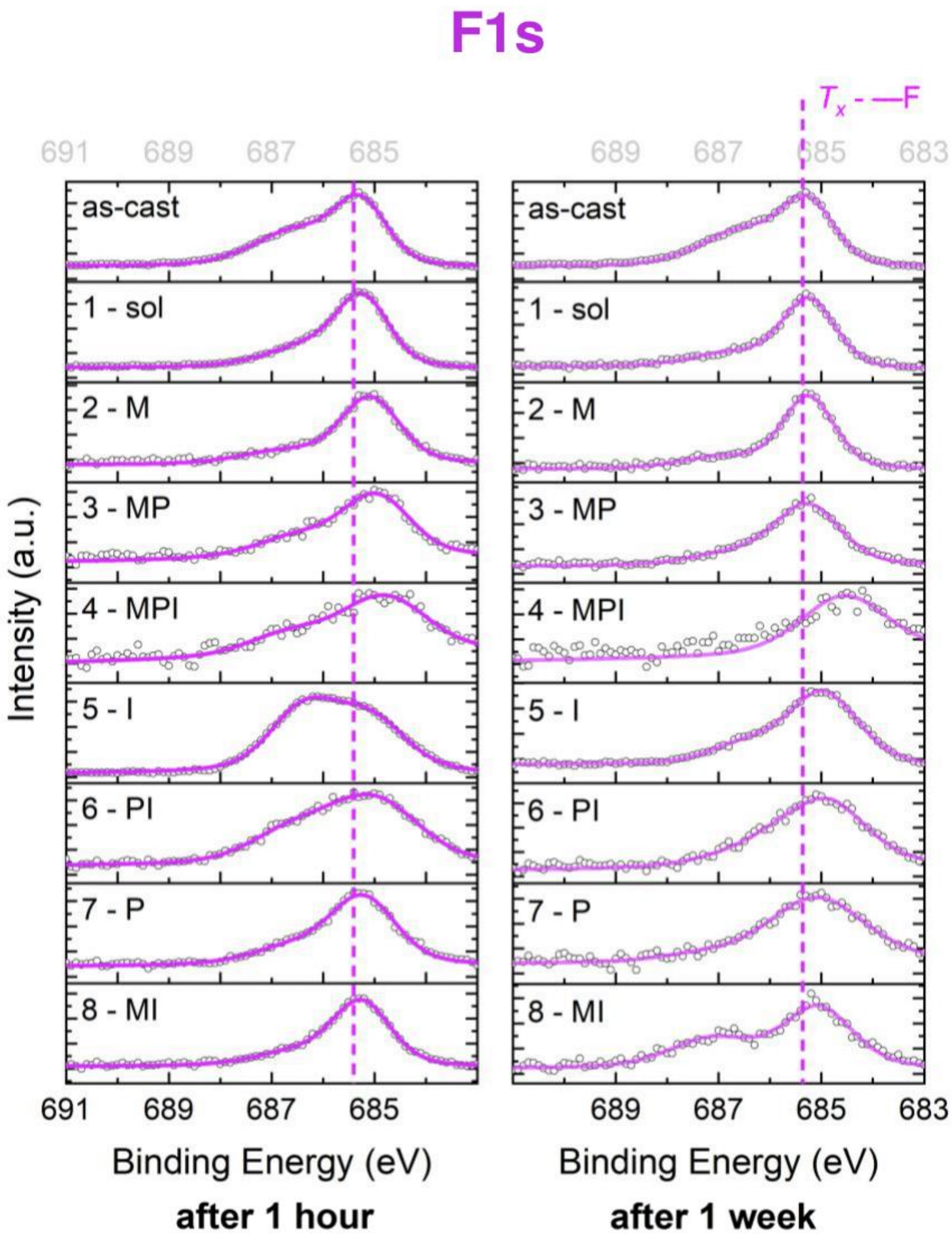

**Figure S2.** F1s XPS core spectra of the as-cast $Ti_3C_2T_x$ MXenes and MXenes extracted from solutions 1-8 after 1 hour and 1 week of exposure. The typical position referring to C – Ti – F termination group bonding is marked by the magenta vertical dashed line.

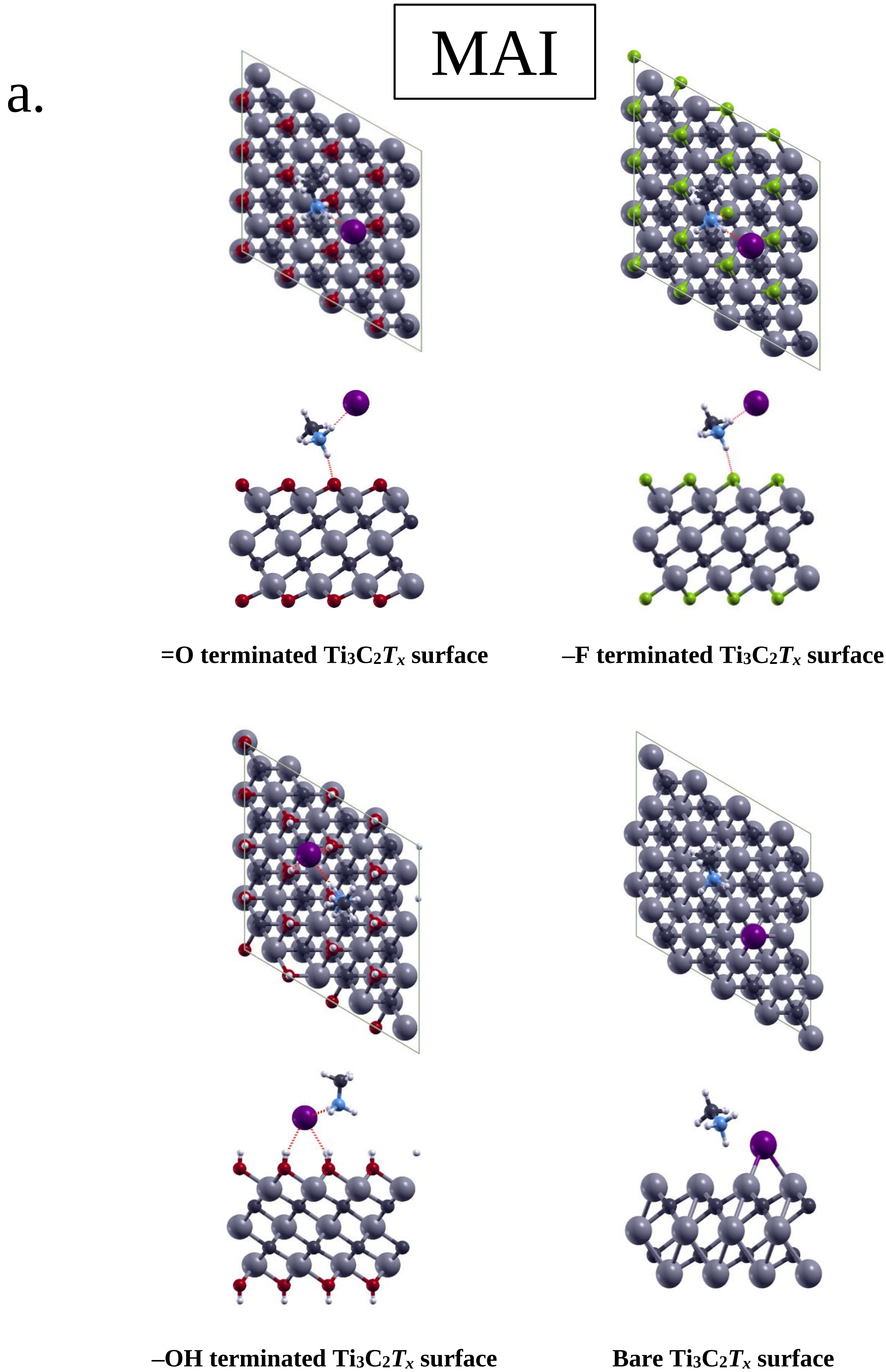
a.
MAI
=O terminated Ti3C2Tx surface
–F terminated Ti3C2Tx surface
–OH terminated Ti3C2Tx surface
Bare Ti3C2Tx surface

b.

$PbI_2$

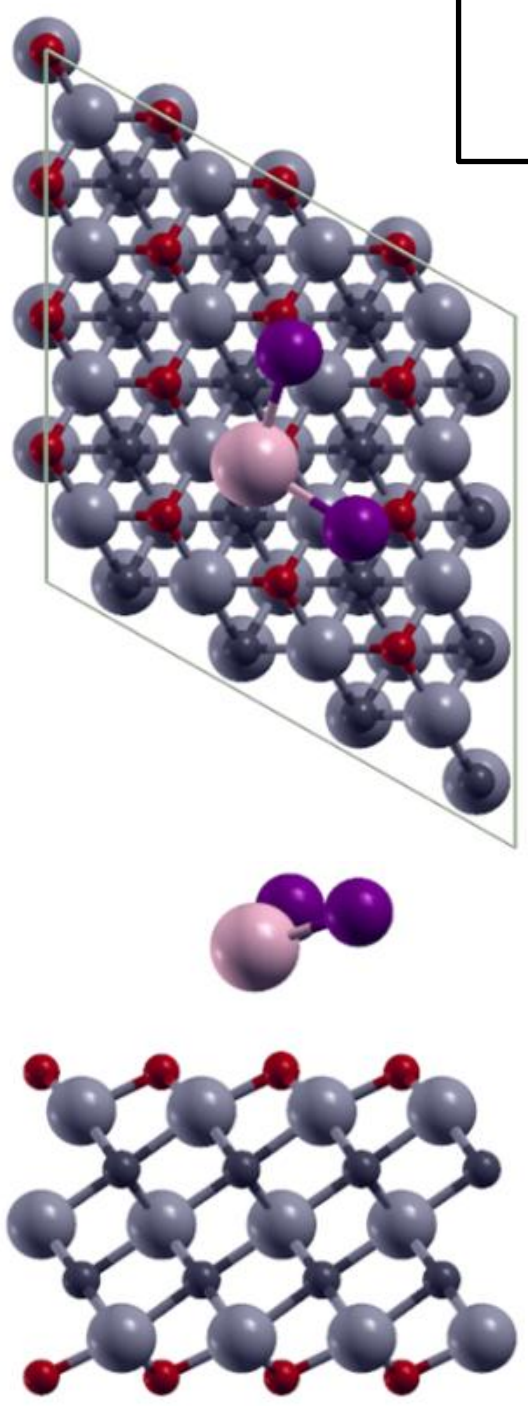

**=O terminated $Ti_3C_2T_x$ surface**

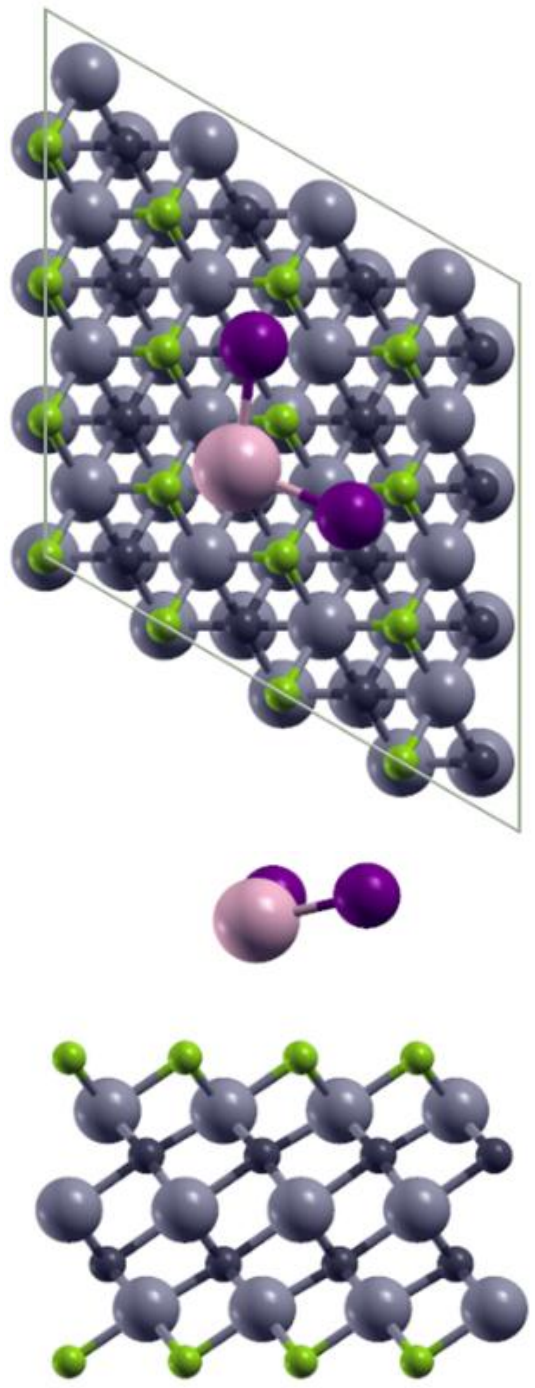

**–F terminated $Ti_3C_2T_x$ surface**

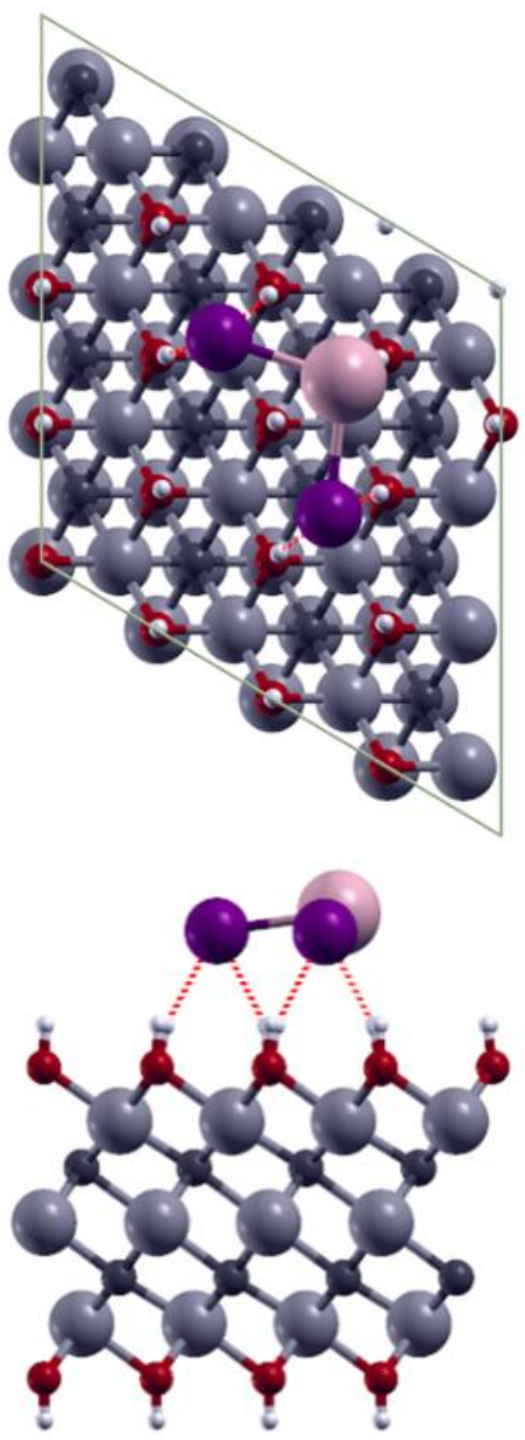

**–OH terminated $Ti_3C_2T_x$ surface**

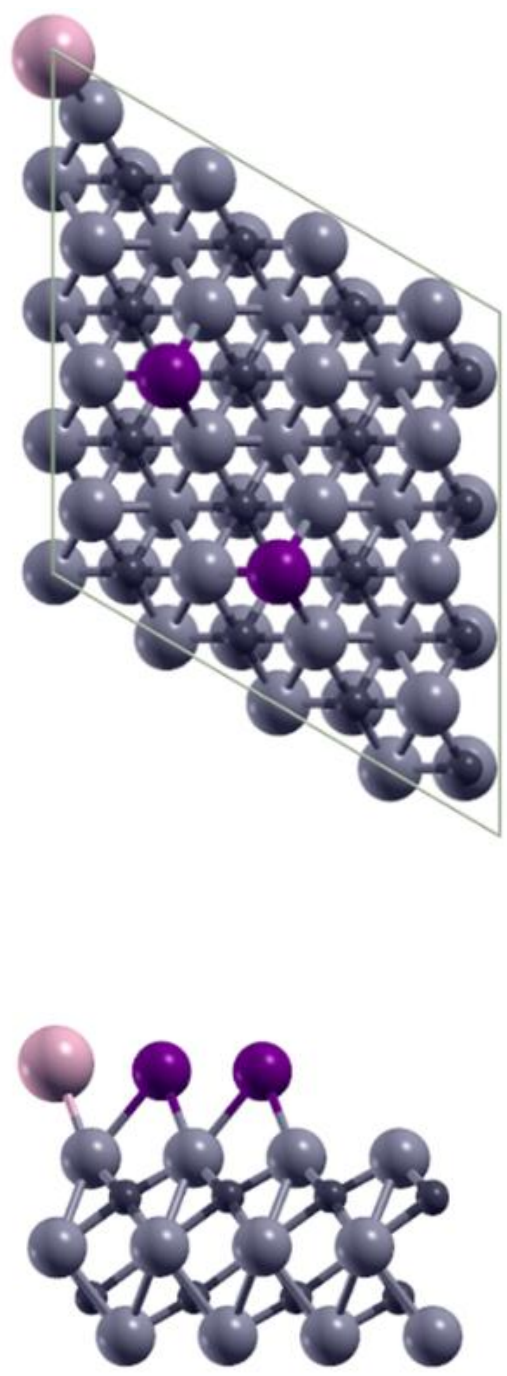

**Bare $Ti_3C_2T_x$ surface**

c.

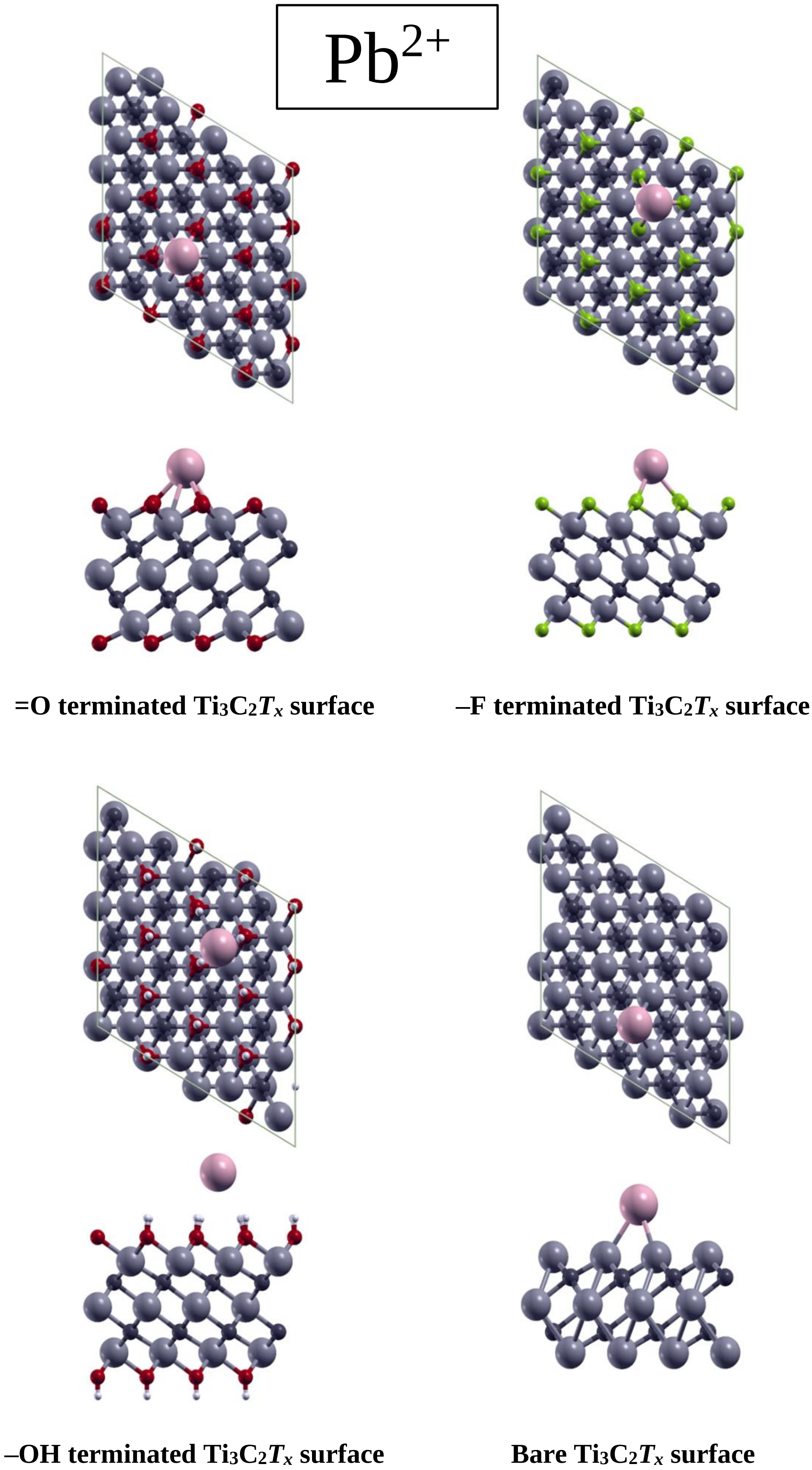

$Pb^{2+}$
=O terminated $Ti_3C_2T_x$ surface
–F terminated $Ti_3C_2T_x$ surface
–OH terminated $Ti_3C_2T_x$ surface
Bare $Ti_3C_2T_x$ surface

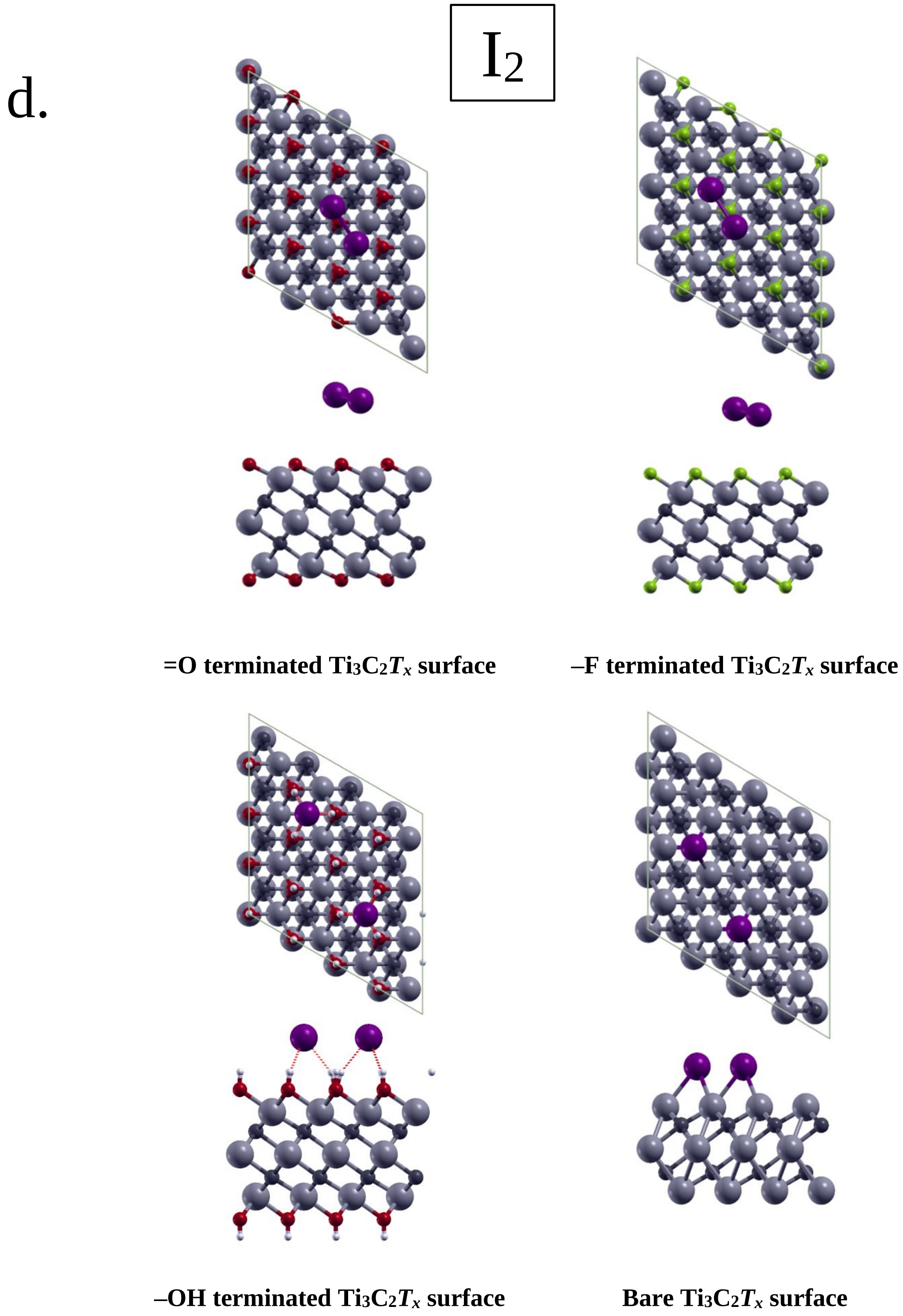

**Figure S3.** The snapshots showing DFT structurally-relaxed models for a) MAI, b) $PbI_2$, c) $Pb^{2+}$, and d) $I_2$ interaction with $Ti_3C_2T_x$ MXene surfaces - top and side views. For each studied species, four different MXene termination states are presented: =O, –F, –OH terminated and bare surfaces. The snapshots for the most realistic mixed-terminated (O/F) $Ti_3C_2T_x$ surfaces are given in **Figure 4** in the main text. Gray, black, red, green, blue, white, purple, and pink atoms represent Ti, C, O, F, N, H, I, and Pb, respectively.

**Table S-T1.** Calculated electrical charge values (in the units of elemental charge) for outer Ti atoms (atoms bonded to termination groups) and $T_x$ groups of: bare (no termination groups attached), pure =O, pure –F, pure –OH, and mixed O/F (in 1 : 1 ratio) terminated $Ti_3C_2T_x$ MXenes before and after MAI, and $Pb^{2+}$ adsorption. The difference in charge due to adsorption is given below each respective line, where a negative sign corresponds to electron acceptance, and a positive – to electron donation. Significant values are highlighted in **bold**. Values for the inner Ti and C atoms are not given here, as they were found unaffected by the adsorption.

| | **bare** | **=O** | **–F** | **–OH** | **mixed O/F** |
|---|---|---|---|---|---|
| **Outer Ti** | 11.1256 | 10.5812 | 10.7238 | 10.7571 | 10.6442 |
| **$T_x$ group** | - | 6.7468 | 7.5535 | 0.6261 | 6.753 |
| | **after MAI adsorption** | | | | |
| **Outer Ti** | 11.0688 | 10.5831 | 10.7254 | 10.7529 | 10.6468 |
| *difference* | **0.0568** | -0.0019 | -0.0016 | 0.0042 | -0.0026 |
| **$T_x$ group** | - | 6.8065 | 7.5724 | 0.6433 | 6.8099 |
| *difference* | - | **-0.0597** | **-0.0189** | **-0.0172** | **-0.0569** |
| | **after $Pb^{2+}$ adsorption** | | | | |
| **Outer Ti** | 11.048 | 10.5785 | 10.7168 | 10.75 | 10.6414 |
| *difference* | **0.0776** | 0.0027 | 0.007 | 0.0071 | 0.0028 |
| **$T_x$ group** | - | 6.8191 | 7.5653 | 0.6822 | 6.8458 |
| *difference* | - | **-0.0723** | **-0.0118** | **-0.0561** | **-0.0928** |

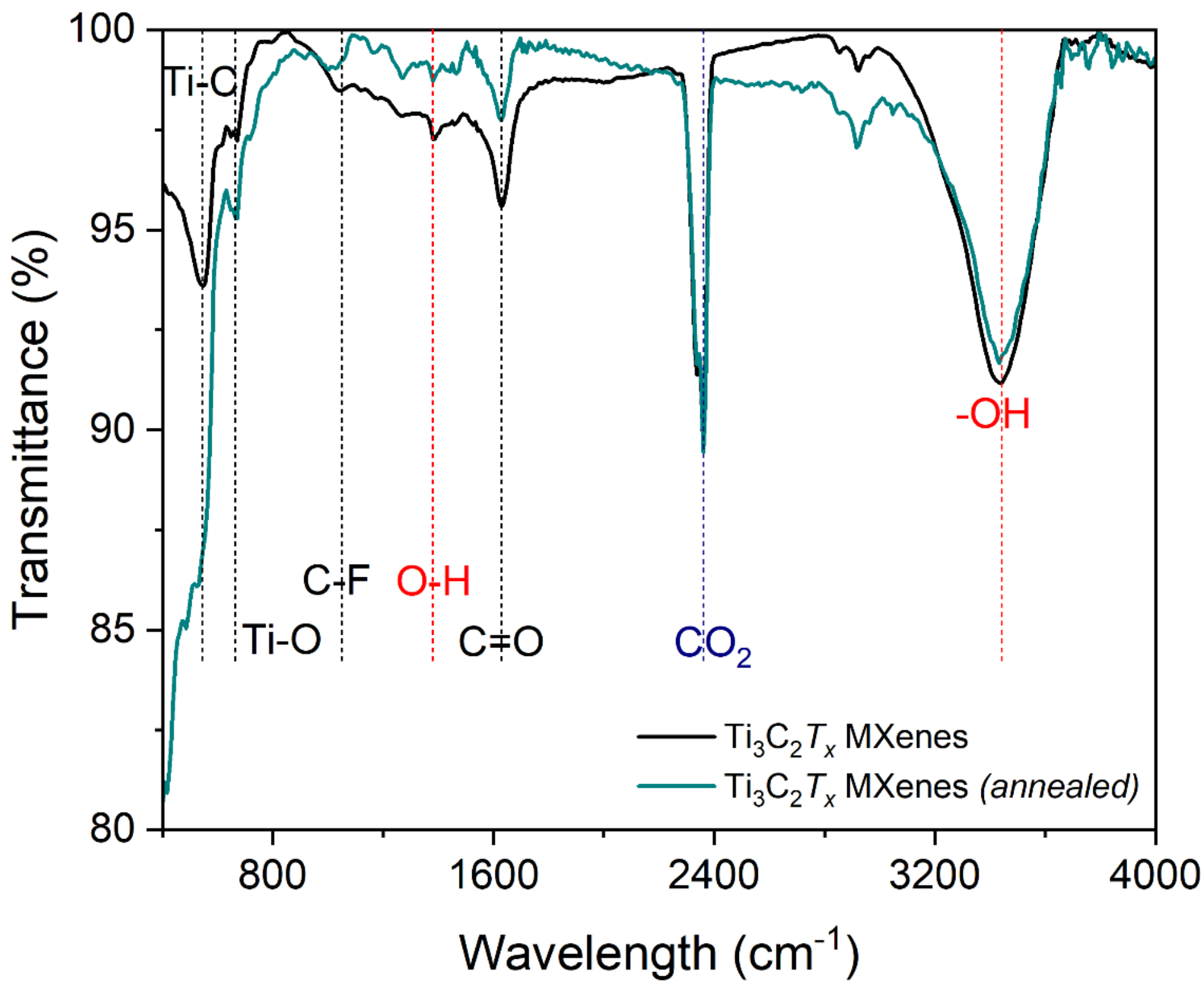

**Figure S4.** Baseline-corrected **FTIR spectra** for as-cast $Ti_3C_2T_x$ MXenes and MXenes calcinated at 300°C for 1 hour in the quartz ampule filled with Ar through 3 sequential pump-purge cycles. According to ref. [9], this treatment is expected to convert –OH surface termination groups into =O ones. Peak assignment was done according to ref. [10].

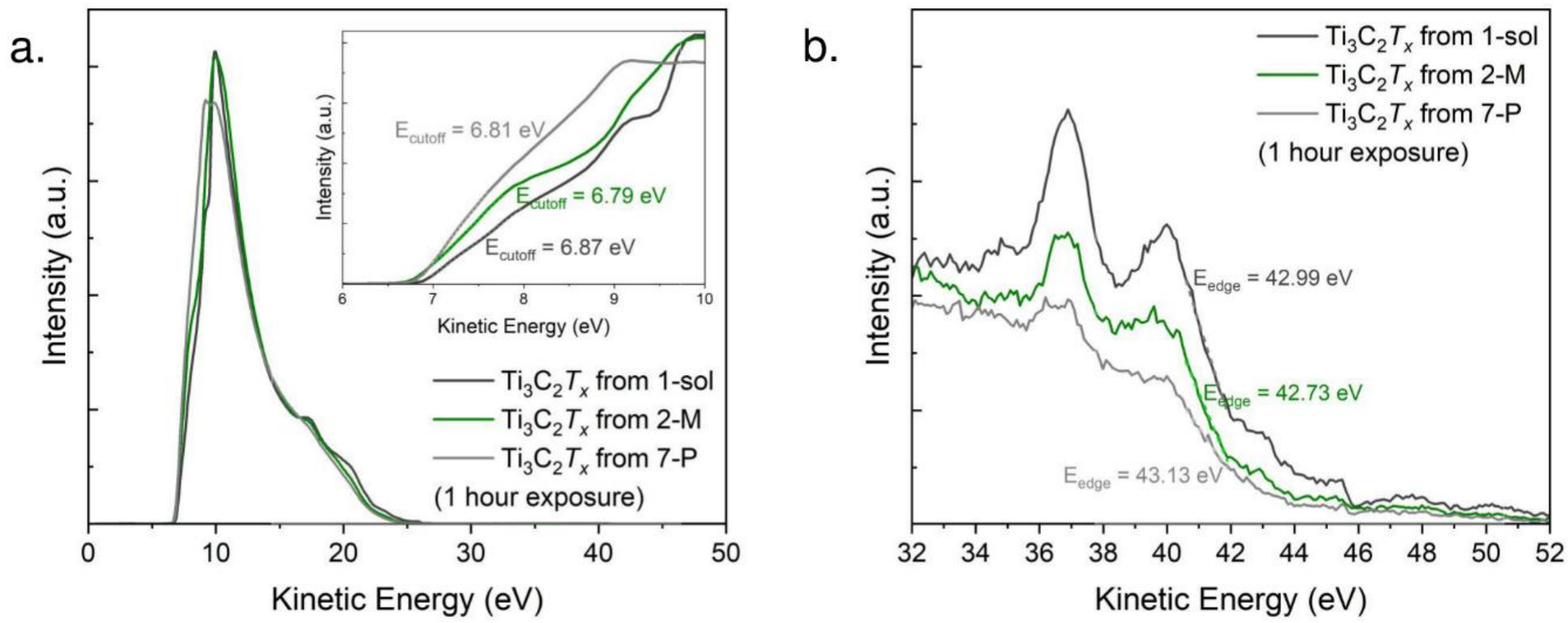

**Figure S5. UPS work function evaluation** for $Ti_3C_2T_x$ MXenes with MAI- and $PbI_2$-based species adsorbed on their surface: (a) complete UPS spectra for $Ti_3C_2T_x$ MXenes extracted from inks 1-sol, 2-M and 7-P after 1 hour of exposure; the secondary electron cutoff region and its linear fitting is highlighted in the insert; (b) zoomed-in valence band region and its linear fitting for the UPS spectra given in (a).

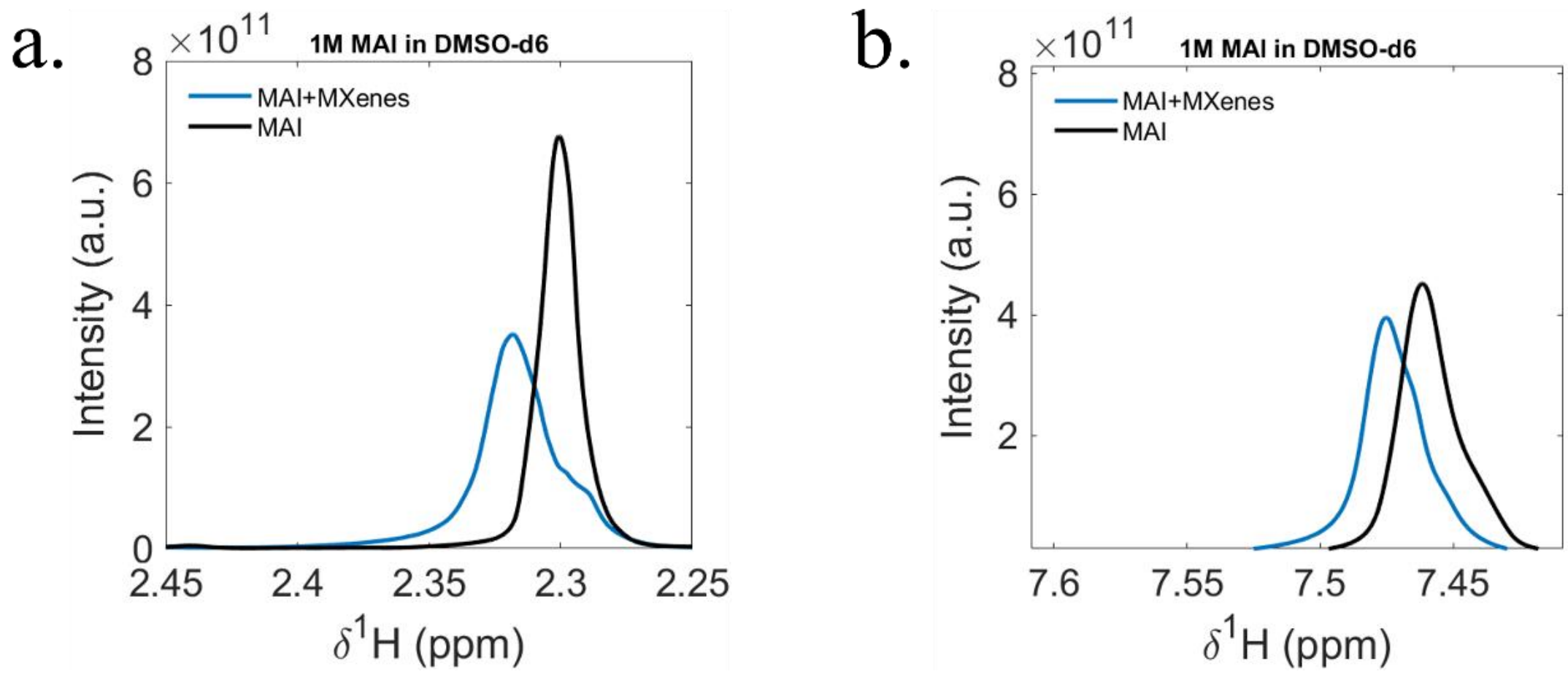

**Figure S6.** **$^{1}$H** (a) **$CH_3$ and** (b) **$NH_3$ NMR spectra** of 1M ink 2-M-*type* solutions with deuterated DMSO (DMSO-d6), with and without $Ti_3C_2T_x$ MXenes.

**Table S-T2.** $^{1}$H relaxation rates for $CH_3$ and $NH_4$ NMR spectra of 1M ink 2-M-*type* solutions with deuterated DMSO (DMSO-d6), with and without $Ti_3C_2T_x$ MXenes.

| **$^{1}$H relaxation rates** | **$R_1$ ($s^{-1}$)** | | **$R_2$ ($s^{-1}$)** | |
|---|---|---|---|---|
| | **$CH_3$** | **$NH_3$** | **$CH_3$** | **$NH_3$** |
| **1M MAI in DMSO-d6** | 0.394 ± 0.003 | 1.05 ± 0.01 | 0.43 ± 0.02 | 7.1 ± 0.1 |
| **1M MAI in DMSO-d6 + $Ti_3C_2T_x$** | 0.368 ± 0.004 | 0.99 ± 0.01 | 0.45 ± 0.02 | 7.1± 0.2 |
| **$\Delta R_2$** (with $Ti_3C_2T_x$ MXenes – without) | - | - | 0.02 ± 0.03 | 0 ± 0.2 |

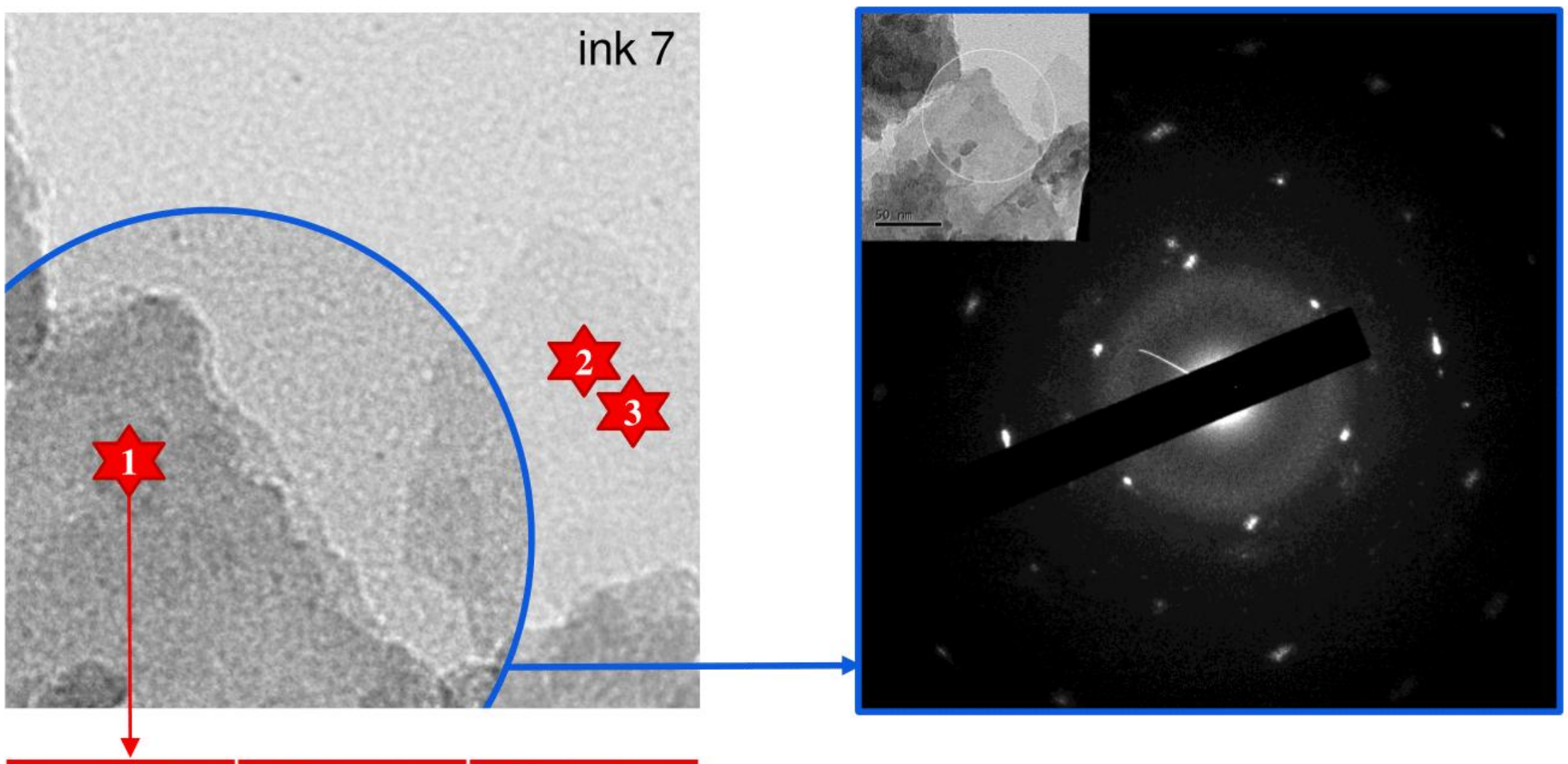

| Area Number | Ti (at.%) | Pb (at.%) |
|---|---|---|
| 1 | 66.5 | 33.5 |
| 2 | 50.1 | 49.9 |
| 3 | 50.2 | 49.8 |

**Figure S7.** Characteristic TEM image with respective SAED pattern and EDS analysis results for $Ti_3C_2T_x$ MXene flakes extracted from ink 7-P. EDS analysis in three given positions (marked with red stars) revealed high amount of the adsorbed Pb (discussed in the main text). The selected area, from which the SAED pattern was taken, is highlighted in TEM image by the blue circle. SAED analysis supported the TEM observations, where no $PbI_2$ crystals were detected in the studied region.

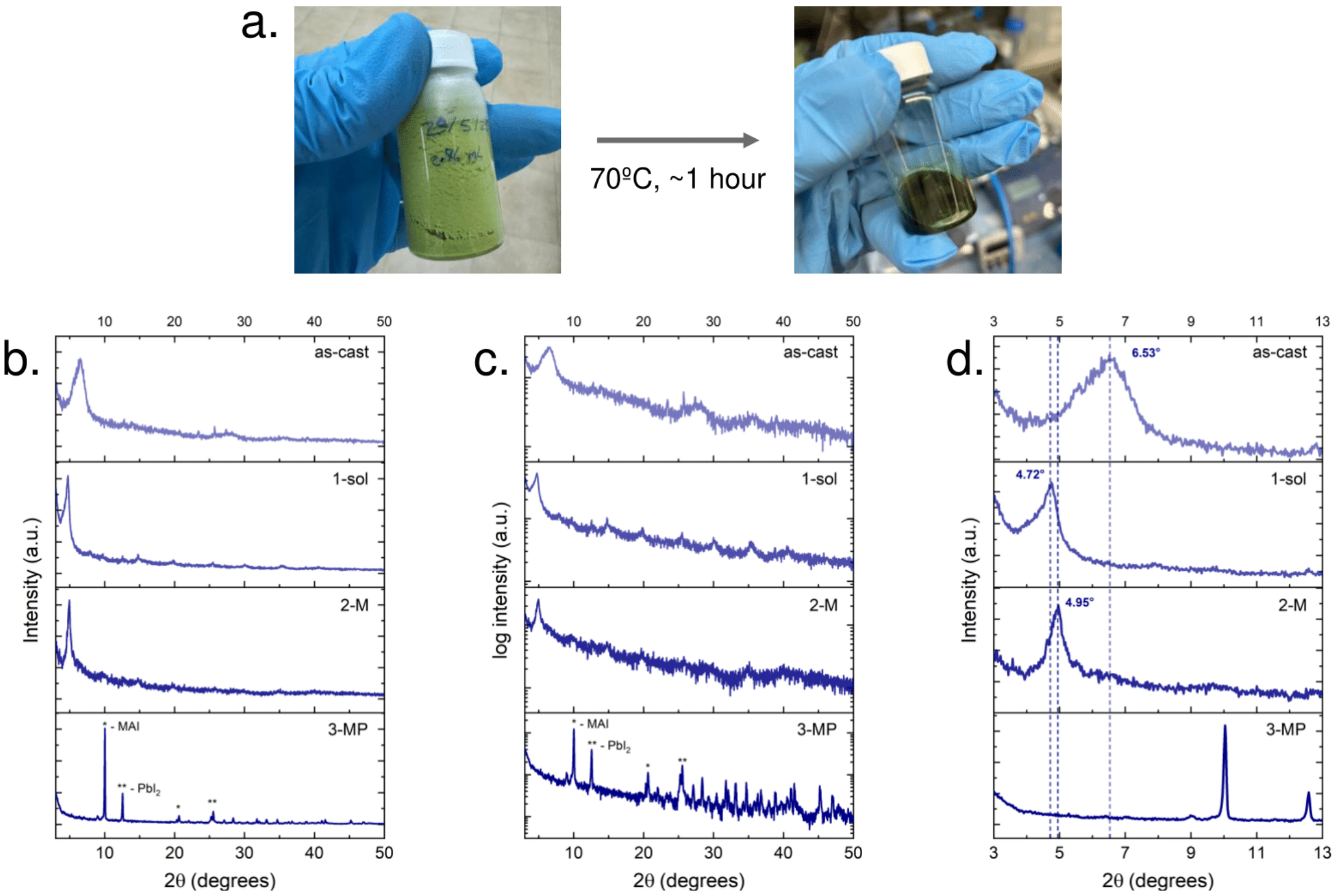

**Figure S8.** (a) photos of the vial with ink 3-MP after several days of storage in the ambient conditions (*left*), and after "remediation" by heating at 70°C for an hour (*right*); (b) XRD patterns of as-cast $Ti_3C_2T_x$ MXenes and materials (MXenes + adsorbed or surrounding species) extracted from solutions 1-3 after 1 week exposure; (c) supplementary representation of the same XRD patterns, but 3-MP plotted at the logarithmic scale to highlight the lower intensity peaks; (d) - supplementary representation of the same XRD patterns, but in low 2θ range of 3-13° to highlight the shifts in characteristic (002) MXene reflection (2θ positions for each case are given at the graphs).

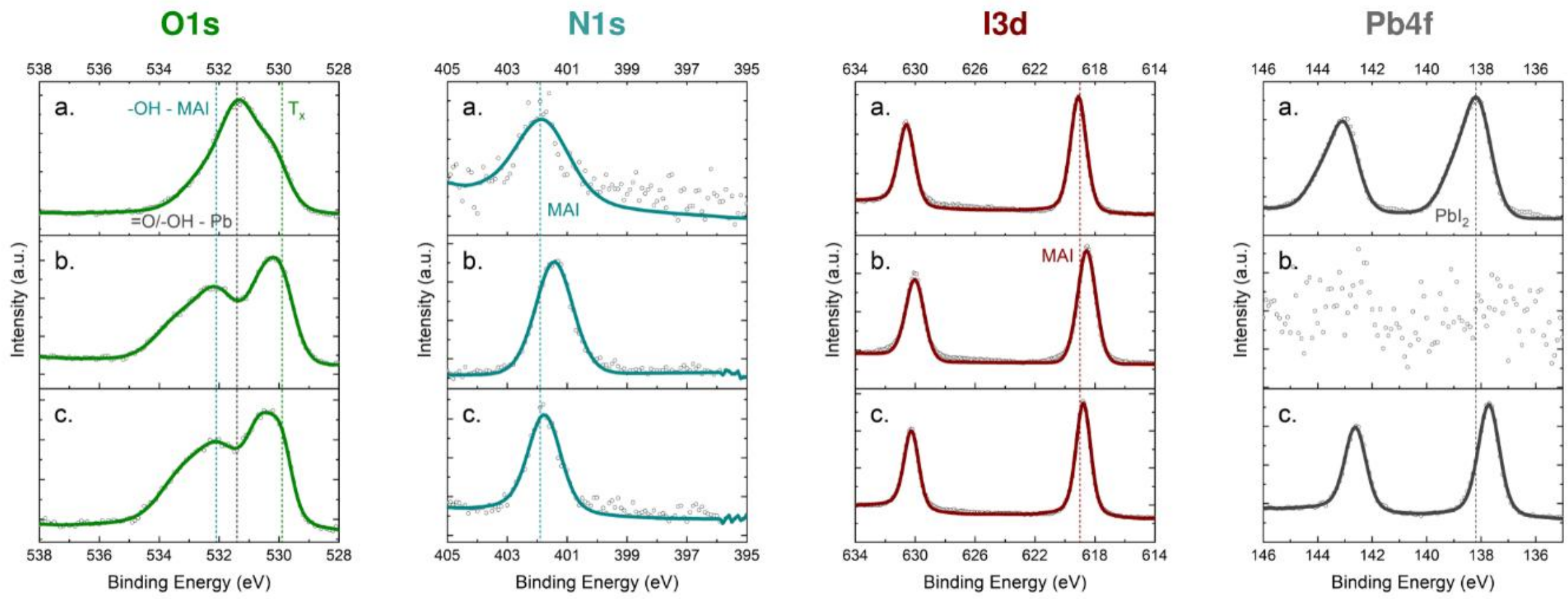

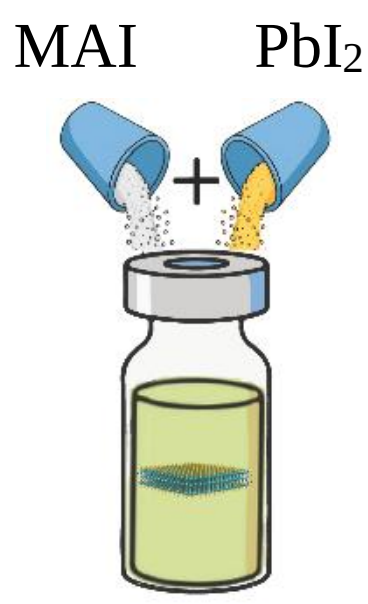

a. **[$Ti_3C_2T_x$ + MAI + $PbI_2$]** **(1 week)**
(ink 3-MP from the main text)

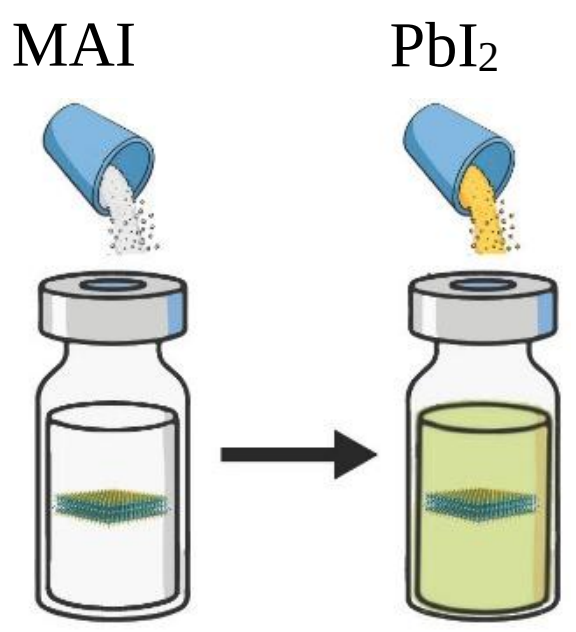

b. **[$Ti_3C_2T_x$ + MAI]** **(1 day)**
c. **[$Ti_3C_2T_x$ + MAI]** **(1 day)** → **[ + $PbI_2$]** **(1 hour)**

**Figure S9.** O1s, N1s, I3d and Pb4f XPS core spectra of $Ti_3C_2T_x$ MXenes after exposure to (a) ink 3-MP (MAI+$PbI_2$) for 1 week (copy of the result from the main text, given here for comparison), (b) MAI for 1 day., and (c) after consequent exposure to MAI (1 day) and $PbI_2$ (for another 1 hour). The typical position referring to: C – Ti – O ($T_x$), nitrogen and iodine in MAI, and lead in $PbI_2$ are marked with green, cyan, dark-brown and gray vertical dashed lines, respectively. Extra peaks that we associate with the formation of the –OH–I and =O/–OH–Pb bonds in the O1s spectra are additionally marked with the cyan and gray lines, respectively.

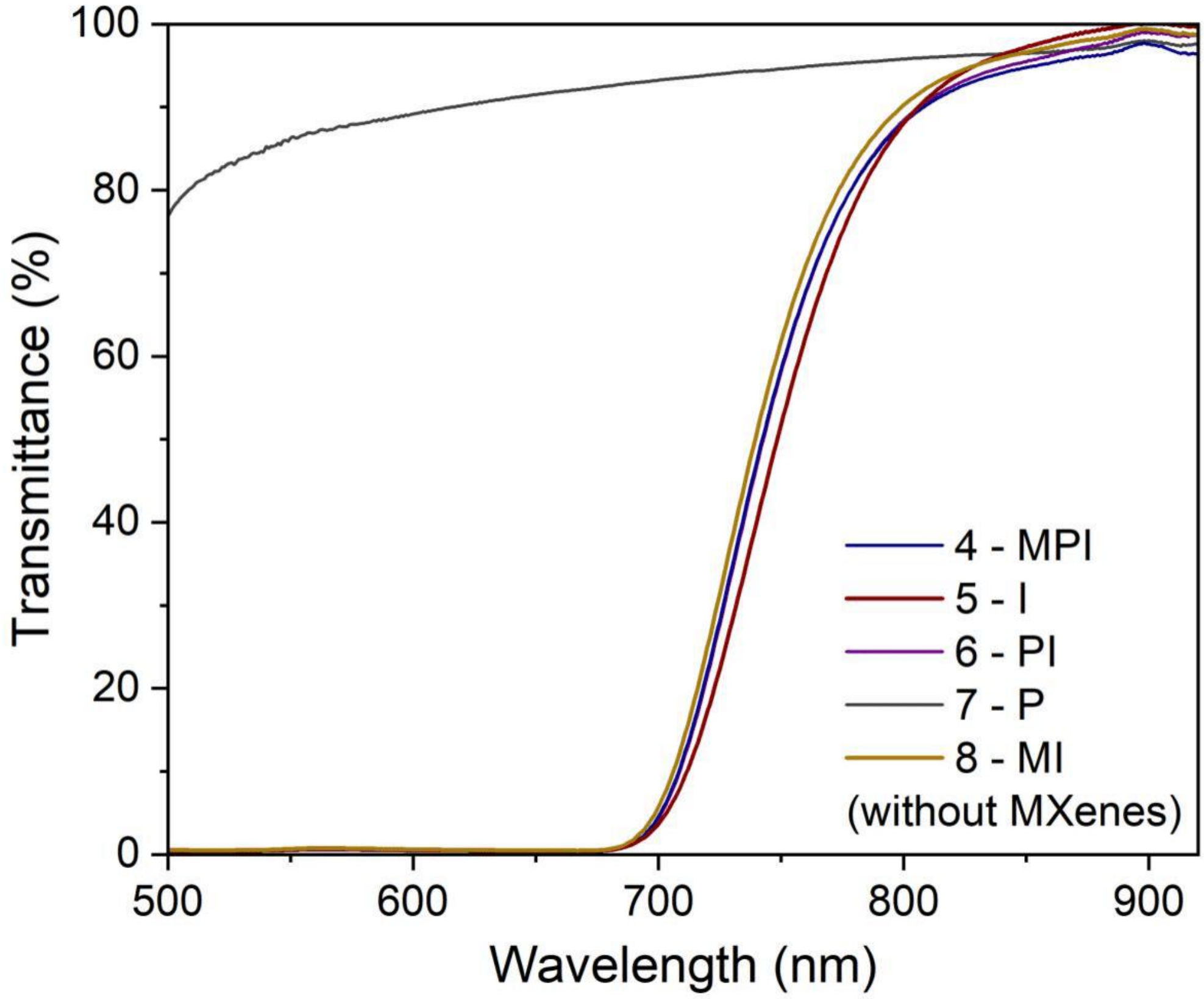

**Figure S10. Transmission control experiment.** Optical transmission spectra of the different reference solutions (4-8), without $Ti_3C_2T_x$ MXene addition, showing the baseline transmission spectrum from, where we can determine that at 920 nm there is no absorption coming from the chemical species that are not MXenes.

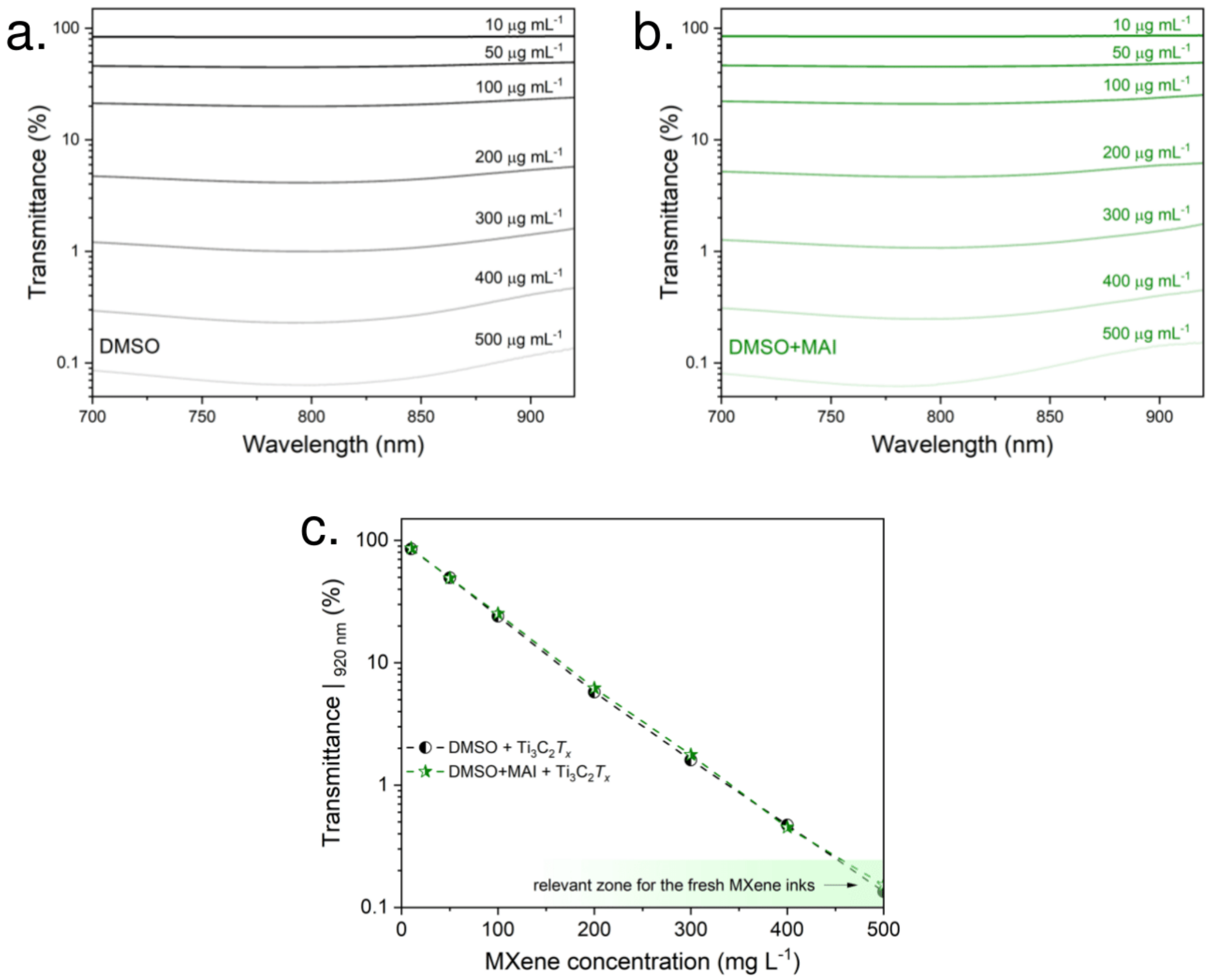

**Figure S11. Transmission–Concentration calibration experiment.** UV-Visible optical transmission spectra of (a) DMSO, and (b) DMSO+MAI solutions with known concentrations of $Ti_3C_2T_x$ MXenes: 10, 50, 100, 200, 300, 400, and 500 µg $mL^{-1}$; (c) transmittance at the wavelength of 920 nm for the reference DMSO, and DMSO+MAI solutions with known concentrations of $Ti_3C_2T_x$ MXenes (calibration curves) as a function of $Ti_3C_2T_x$ MXene concentration in 10-500 µg $mL^{-1}$ range. The green zone in *c* represents the relevant calibration region in our decay experiments that are discussed in the main text.

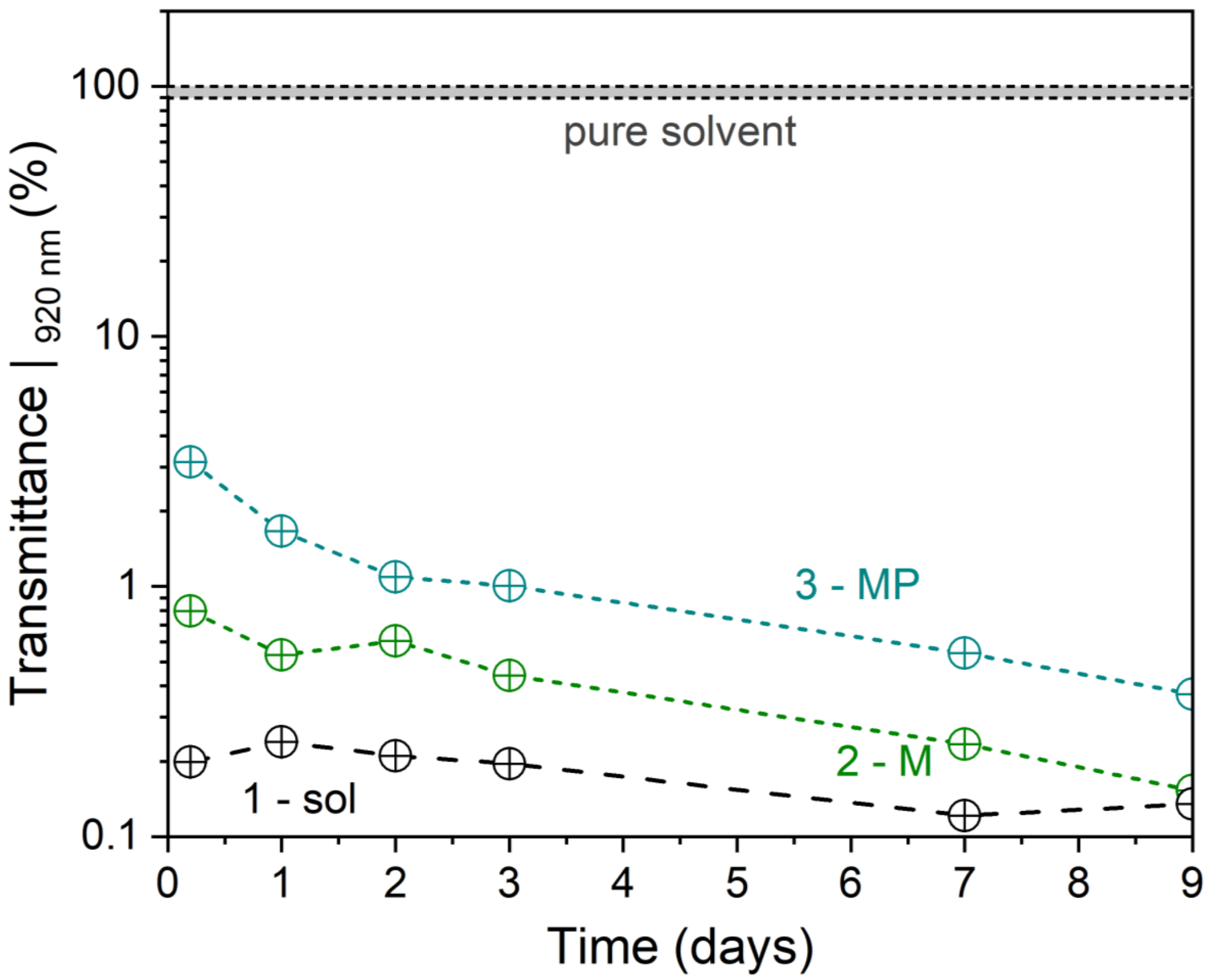

**Figure S12.** Time dependences of the transmittance at the wavelength of 920 nm for the stable $Ti_3C_2T_x$ MXene-containing solutions (1-sol, 2-M and 3-MP) and the reference solution (without $Ti_3C_2T_x$ MXene addition).

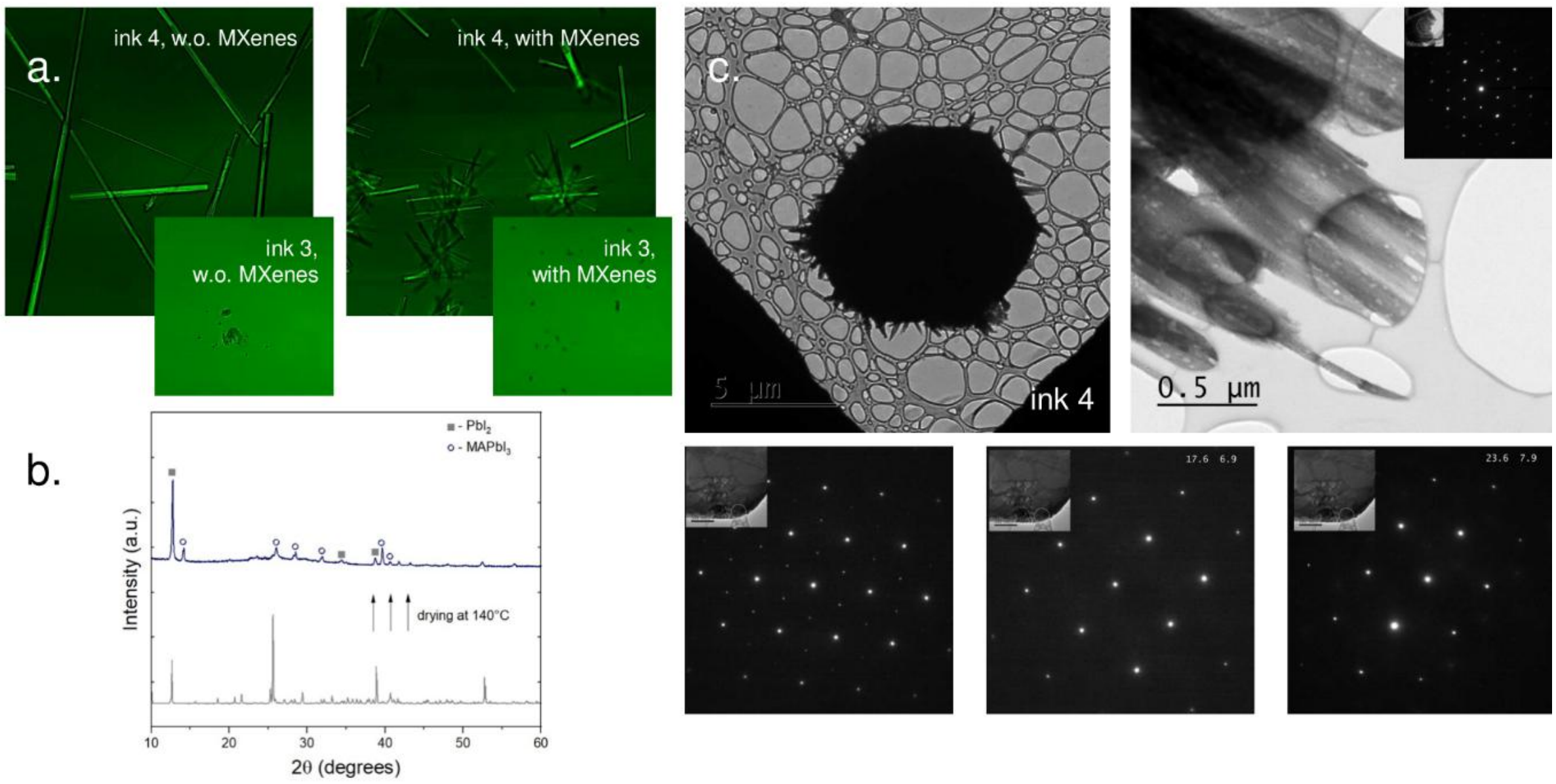

**Figure S13.** (a) confocal microscopy images of solutions 3-MP and 4-MPI with and without $Ti_3C_2T_x$ MXenes; (b) XRD pattern of precipitates from 4-MPI before and after heating them at 140°C; (c) TEM images of precipitates extracted from 4-MPI with respective SAED patterns at various tilt angles.

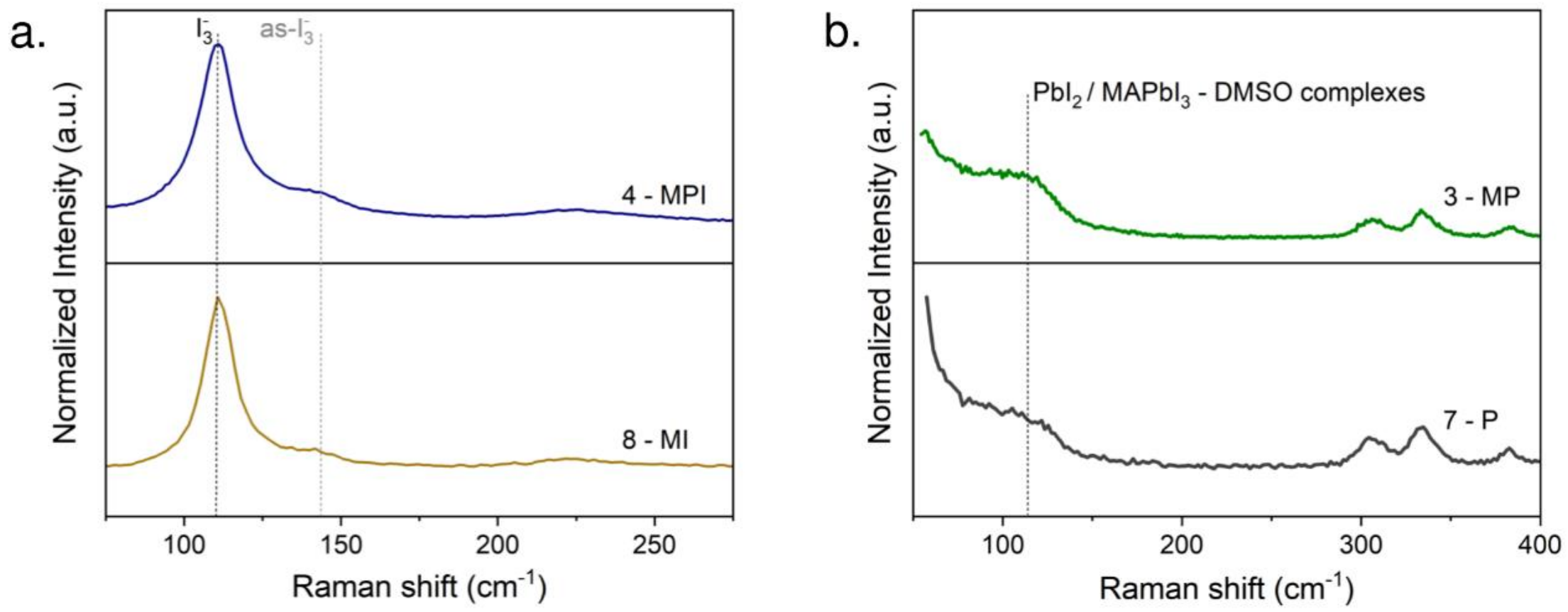

**Figure S14.** Supplementary Raman spectra of the solutions (a) 4-MPI and 8-MI, both containing mixtures of MAI and elemental $I_2$, and (b) 3-MP and 7-P, both containing $PbI_2$, all without $Ti_3C_2T_x$ MXenes addition.

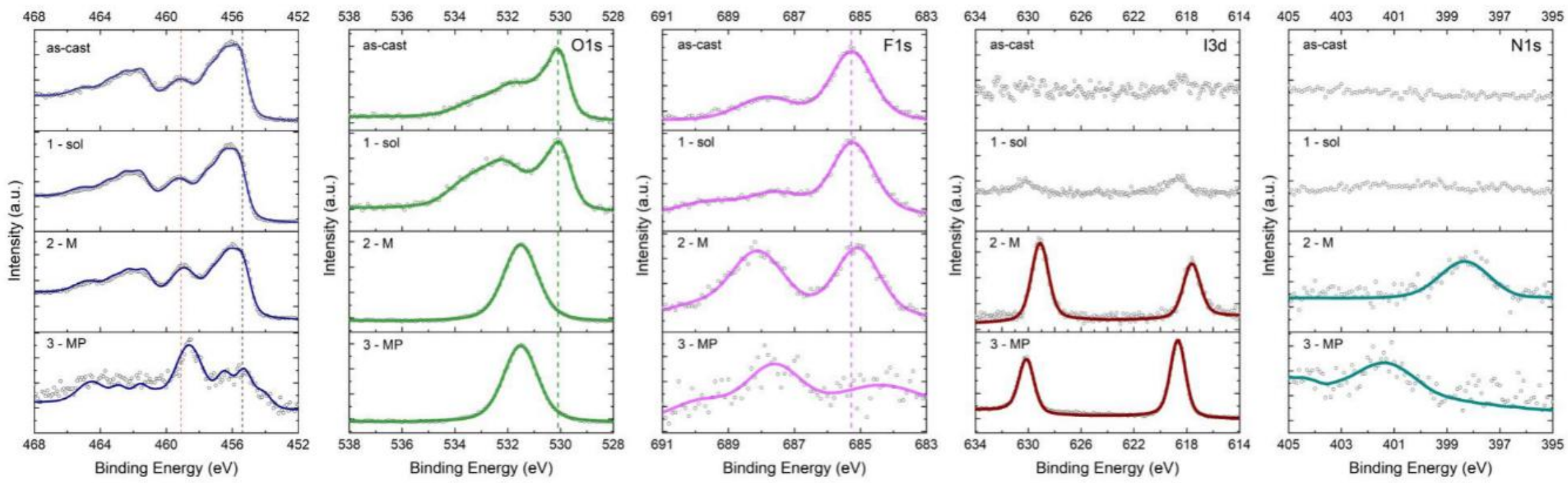

**Figure S15. Changing the MXene preparation protocol: $Ti_3C_2T_x$ → HF-$Ti_3C_2T_x$.** Ti2p, O1s, F1s, I3d and N1s XPS core spectra for as-cast HF-$Ti_3C_2T_x$ MXenes and HF-MXenes extracted from MXene-containing solutions 1-3 after 1 week of exposure.

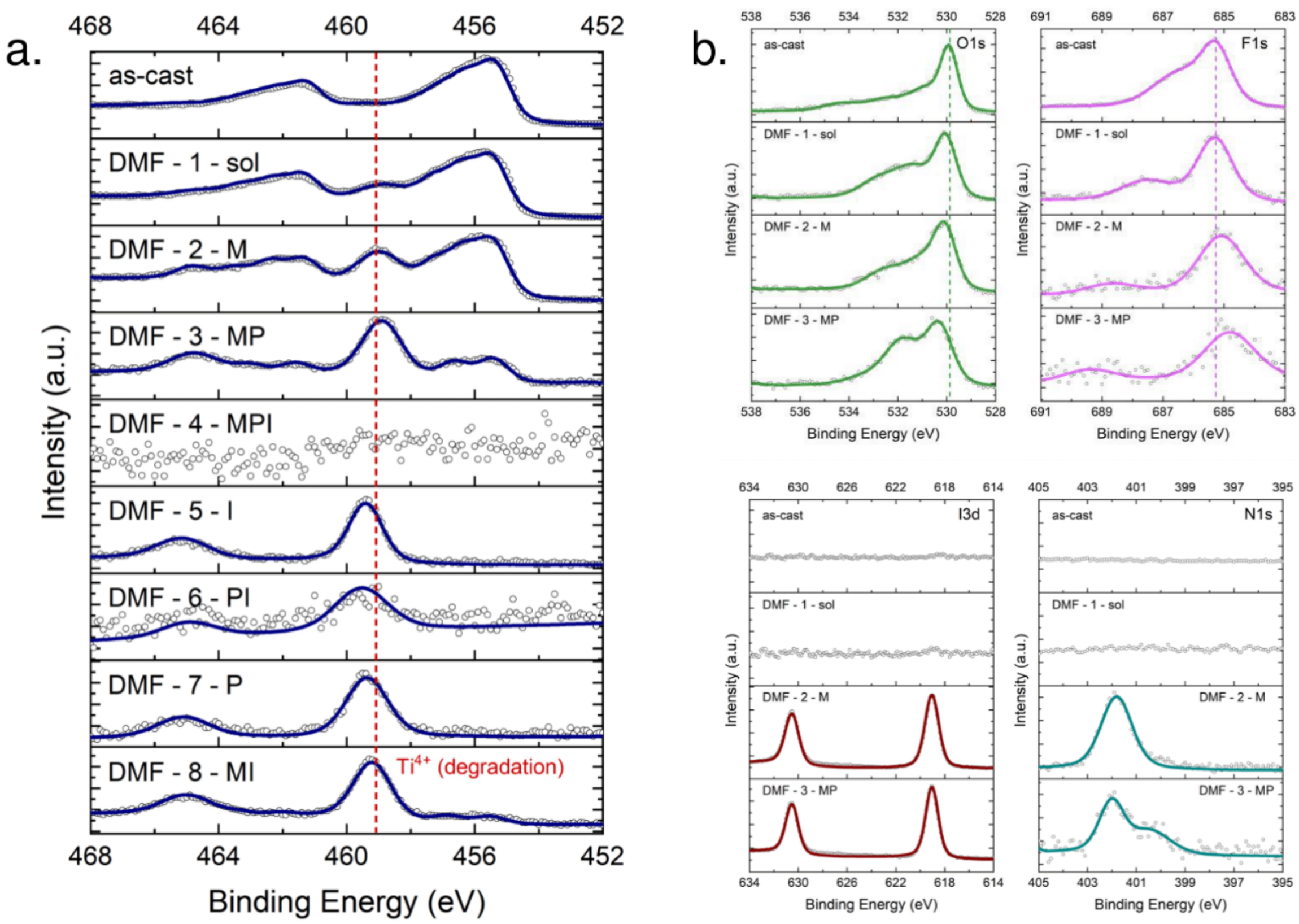

**Figure S16. Changing the solvent: DMSO → DMF.** (a) Ti2p XPS core spectra of as-cast $Ti_3C_2T_x$ MXenes and MXenes extracted from solutions 1-8, with DMF as a solvent; (b) O1s, F1s, I3d and N1s XPS core spectra of as-cast $Ti_3C_2T_x$ MXenes and MXenes extracted from stable MXene-containing solutions 1-3, with DMF as a solvent. MXene extraction was performed after 1 week of exposure.

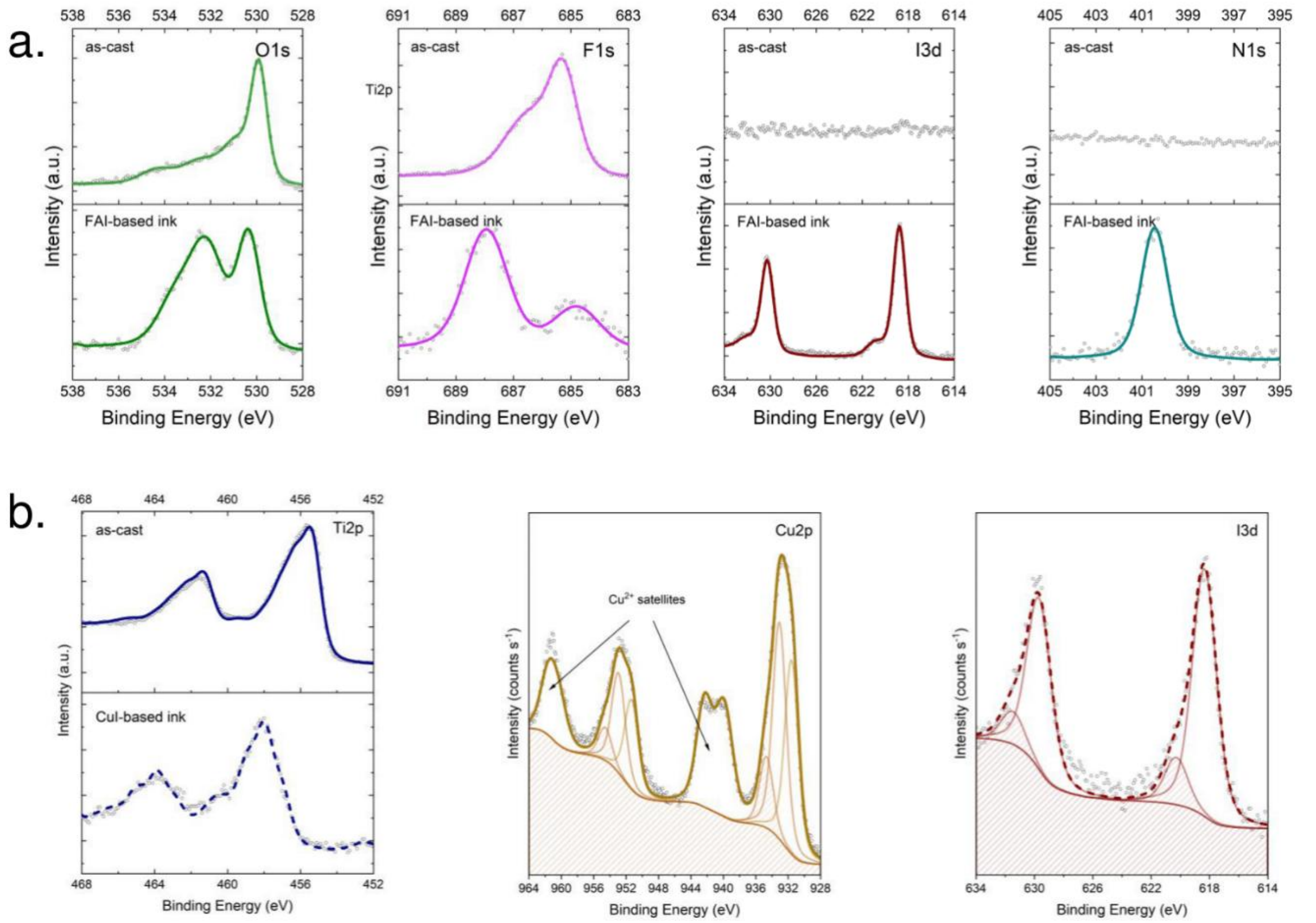

**Figure S17. Changing the organic- and metal-halide species: MAI → FAI and $PbI_2$ → CuI.** (a) O1s, F1s, I3d and N1s XPS core spectra of as-cast $Ti_3C_2T_x$ MXenes and MXenes extracted from DMSO+FAI solution (similar to 2-M); (b) Ti2p, Cu2p and I3d XPS core spectra of as-cast $Ti_3C_2T_x$ MXenes and MXenes extracted from DMSO+CuI solution (similar to 7-P). MXene extraction was performed after 1 week of exposure.

## Supplementary Reference List